\documentclass[a4paper,11pt]{article}
\pdfoutput=1 % if your are submitting a pdflatex ({\it i.e.}~, if you have
\usepackage[symbol]{footmisc} % DO NOT move this below \usepackage{jheppub}. It will mess with the footnote hyperlinking.
\usepackage{jheppub} % for details on the use of the package, please
\usepackage{color, colortbl}
\usepackage[T1]{fontenc}
\usepackage[utf8]{inputenc}
\usepackage{epsfig,latexsym}
\usepackage{amsmath}
\usepackage[dvipsnames]{xcolor}
\usepackage{adjustbox}
\usepackage{tensor}
\usepackage{graphicx}
\usepackage{verbatim}
\usepackage{mathrsfs}
\usepackage{amssymb}
\usepackage{multirow}
\usepackage{epsfig}
\usepackage{color,colordvi}
\usepackage{appendix}
\usepackage{slashed}
\usepackage{array}
\usepackage{cancel}
\usepackage{float}
\usepackage[font=footnotesize]{caption}
\usepackage{epsf}
\usepackage{amsmath}
\usepackage{tikz}
\usetikzlibrary{external,shapes.geometric,arrows,arrows.meta,calc}
\usepackage{physics}
\usepackage{MnSymbol}
\usepackage{rotating}
\usepackage{multirow}
\usepackage[normalem]{ulem}
\usepackage{tcolorbox}
\usepackage{bbold}
\usepackage{bbm}
\usepackage{changepage}
\usepackage{framed}
\usepackage{xspace}
\usepackage{ulem}

\newcommand\numberthis{\addtocounter{equation}{1}\tag{\theequation}}
\definecolor{shadecolor}{rgb}{.9, .9, .9}

\allowdisplaybreaks

\DeclareMathAlphabet{\mathpzc}{OT1}{pzc}{m}{it}

\renewcommand{\theequation}{\thesection.\arabic{equation}} \csname
@addtoreset\endcsname{equation}{section}

\newcolumntype{x}[1]{>{\centering\arraybackslash\hspace{0pt}}p{#1}}

\newcommand{\beq}{\begin{equation}}
\newcommand{\eeq}{\end{equation}}
\renewcommand{\[}{\left[}
\renewcommand{\]}{\right]}
\renewcommand{\(}{\left(}
\renewcommand{\)}{\right)}

\newcommand{\be}{\begin{eqnarray}}
\newcommand{\ee}{\end{eqnarray}}
\newcommand{\bea}{\begin{eqnarray}}
\newcommand{\eea}{\end{eqnarray}}
\newcommand{\bi}{\begin{itemize}}
\newcommand{\ei}{\end{itemize}}
\newcommand{\ben}{\begin{enumerate}}
\newcommand{\een}{\end{enumerate}}
\def\bes{\begin{equation*}}
\def\ees{\end{equation*}}
\def\bead{\begin{aligned}}
\def\eead{\end{aligned}}
\def\bmat{\left(\begin{matrix}}
\def\emat{\end{matrix}\right)}

\def\Re{\text{Re}}
\def\Im{\text{Im}}
\def\diag{\text{diag}}

\def\PL{\text{PL}}
\def\PE{\text{PE}}

\def\cB{{\cal B}}

\def\cD{{\cal D}}
\def\cE{{\cal E}}

\def\cH{{\cal H}}
\def\cI{{\cal I}}
\def\cJ{{\cal J}}
\def\cL{{\cal L}}
\def\cN{{\cal N}}
\def\cO{{\cal O}}

\def\cS{{\cal S}}

\def\CKM{\text{CKM}}
\def\PMNS{\text{PMNS}}

\newcommand{\nuSM}{$\nu$SM\xspace}

\usepackage{xparse}
\usepackage{regexpatch}
\ExplSyntaxOn
\NewDocumentCommand{\neat}{m}
 {
  \tl_set:Nn \l_tmpa_tl { #1 }
  \regex_replace_all:nnN { (.*?)\_\{(.*?)\}\^\{(.*?)\} } { \c{sbt}\cB\{\1\cE\}\cB\{\2\cE\}\cB\{\3\cE\} } \l_tmpa_tl
  \tl_use:N \l_tmpa_tl
 }
\ExplSyntaxOff

\title{The flavor invariants of the \nuSM}

\author[a,b,c]{Christophe Grojean,}
\emailAdd{christophe.grojean@desy.de}

\author[a,b]{Jonathan Kley,}
\emailAdd{jonathan.kley@desy.de}

\author[d]{Damien Leflot,}
\emailAdd{damien.leflot@ens-lyon.fr}

\author[a,e]{Chang-Yuan Yao}
\emailAdd{chang.yuan.yao@desy.de}

\affiliation[a]{Deutsches Elektronen-Synchrotron DESY, Notkestr. 85, 22607 Hamburg, Germany}
\affiliation[b]{Institut für Physik, Humboldt-Universität zu Berlin, 12489 Berlin, Germany}
\affiliation[c]{Theoretical Physics Department, CERN, 1211 Geneva 23, Switzerland}
\affiliation[d]{LAPTh, CNRS et Université Savoie Mont-Blanc, 9 Chemin de Bellevue, F-74941 Annecy, France}
\affiliation[e]{School of Physics, Nankai University, Tianjin 300071, China}

\abstract{Sixty years after the experimental discovery of CP violation in the quark sector, the existence of a similar CP violation in the lepton sector is still to be established. Actually, the structure of such a violation depends crucially on the origin of the neutrino masses. In an attempt at categorizing the leptonic sources of CP violation, we studied the $\nu$SM, the Standard Model extended with three generations of sterile neutrinos, that can  interpolate continuously between the Dirac and Majorana scenarios of neutrino masses. In particular, we perform a classification of the Jarlskog-like flavor invariants entering CP-violating observables and we study their suppression with the heavy Majorana mass in the seesaw limit of the model. To simplify the construction of the invariants, we introduce a graph-based method. With the guidance of the Hilbert series and plethystic logarithm of the theory, we construct the \emph{generating} and \emph{primary} sets of invariants for the $\nu$SM for the first time. Unlike in the Standard Model and some other theories, we find that the numbers of generating invariants and the syzygies among them cannot immediately be read off from the plethystic logarithm, but require a more careful examination. Our analysis reveals that the \emph{generating} set contains 459 invariants, out of which 208 are CP-even and 251 are CP-odd. In the seesaw limit of the $\nu$SM, we show that all parameters of the UV theory can be captured in the effective theory with a certain suppression with the heavy Majorana mass, while these parameters can only appear in a \emph{flavor-invariant} way with a \emph{higher} mass suppression. Furthermore, we discuss how the necessary and sufficient conditions for CP violation can be captured by utilizing these invariants. Along the way, we present useful algorithms to enumerate and build the flavor invariants.}

\begin{document}
\begin{flushright}
CERN-TH-2024-076\\
DESY-24-021 \\
HU-EP-24/14
\end{flushright}
\maketitle
\flushbottom

%%%%%%%%%%%%%%%%%%%%%%%%%%%%%%%%%%%%%%%%%%%%%%%%%%%%%%%%%%%%%%%%%%%%
\section{Introduction}\label{sec:Intro}

The smallness of the observed neutrino masses (needed to explain neutrino oscillation experiments~\cite{Super-Kamiokande:1998qwk,SNO:2001kpb})  is yet another consequence/success of the chiral nature of the Standard Model (SM) gauge structure. In the absence of, yet to be confirmed, right-handed neutrino at \textit{low} energy, these neutrino masses cannot  emerge from marginal interactions among the SM degrees of freedom, like in the case for quarks and charged leptons, but require irrelevant interactions~\cite{Weinberg:1979sa} possibly stemming from interactions with new physics degrees of freedom or new strong interactions (for a review, see Ref.~\cite{ParticleDataGroup:2022pth}). A particularly compelling scenario is the type-I seesaw mechanism with right-handed sterile neutrinos endowed with a \textit{large} SM gauge-invariant Majorana mass and coupled via Yukawa interactions to the left-handed active neutrinos~\cite{Minkowski:1977sc,Yanagida:1979as,Gell-Mann:1979vob,Glashow:1979nm,Mohapatra:1979ia}. The value of the scale of this Majorana mass, which also  measures the breaking of an accidental symmetry of the SM - the conservation of the total number of leptons, remains unknown and largely unconstrained.

In this paper, we will focus on the renormalizable theory of the SM extended with $n_N=3$ flavors of right-handed sterile neutrinos $N$, often referred to as \nuSM or $\nu$MSM~\cite{Asaka:2005an,Asaka:2005pn,Asaka:2006ek,Asaka:2006nq,Asaka:2006rw}. Depending on the mass we assign to these new particles, our discussion will capture both a type-I seesaw scenario and a scenario with light right-handed sterile neutrinos. Just like through the CKM matrix in the quark sector of the SM, adding more parameters to the theory by introducing a mass term opens the possibility to violate CP through physical phases that cannot be removed from the theory using field redefinitions. These new sources of CP violation (CPV) could help generating the matter-antimatter asymmetry via a leptogenesis mechanism~\cite{Davidson:2008bu}.

The purpose of this paper is the characterization of the flavorful parameters of the \nuSM using the language of flavor invariants and in particular classifying the new sources of CPV that are introduced through the neutrino masses with Jarlskog-like flavor invariants~\cite{Jarlskog:1985cw,Jarlskog:1985ht,Bernabeu:1986fc}. In this analysis we will make use of the fact that -- as physical quantities -- all observables should in principle be expressible in terms of a finite number of flavor invariant objects. Such a set of invariants is referred to as the \textit{generating} set or the \textit{basic} set in the literature. This set enables us to express all observables as functions of them, which will allow us to learn about the structure of flavorful CP violation in the theory; for instance, at which order in the couplings certain phases can appear in observables. It is also straightforward to differentiate between the Dirac and Majorana nature of the mass term for the light sterile neutrinos in our formulation by taking the appropriate limits.

Furthermore, the set of invariants can be used to study the seesaw limit of the \nuSM and make statements about the order in the effective theory at which certain UV parameters -- in particular those which are CP-odd -- can appear. For instance, we can show that while all parameters present in the effective theory of the \nuSM appear suppressed with only one or two powers of the heavy Majorana mass, they can only appear in a flavor-invariant way with a suppression of 2 and 4 powers of the Majorana mass.

As a main result of our paper, we will construct the \emph{generating} set of flavor invariants, which is the maximal set capturing all physical effects of the theory. For what concerns CPV effects, we will also endeavor to address the question of what is the minimal set that can capture the necessary \emph{and} sufficient conditions of CP conservation (CPC) in all special spectra (for instance degenerate and vanishing masses) in the theory. In contrast to the quark sector of the SM, where all CPV effects are captured by a single CP-odd Jarlskog invariant, the \nuSM introduces hundreds of CP-odd invariants in the \emph{generating} set. This considerably complicates the analysis, making the identification of CPC conditions in the \nuSM an ongoing challenge yet to be resolved.

In order to simplify the process of finding a \emph{generating} set of flavor invariants, we will make use of tools from invariant theory that have been developed in recent years in the context of theoretical particle physics. Invariants have been used both in UV complete theories~\cite{Branco:1986pr,Branco:1986gr,Botella:1994cs,Branco:2001pq,Lebedev:2002wq,Gunion:2005ja,Davidson:2005cw,Jenkins:2009dy,Trautner:2018ipq,Yu:2019ihs,Wang:2021wdq,Yu:2021cco,Yu:2020gre,Bento:2023owf,Darvishi:2023ckq}
and effective field theories (EFTs) to characterize the parameters of the theory with respect to CP~\cite{Bonnefoy:2021tbt,Bonnefoy:2023bzx,Yu:2022nxj,Yu:2022ttm,Bedi:2024wqg,Darvishi:2024cwe} and other symmetries of EFTs~\cite{Bonnefoy:2022rik}. Some problems which are closely related to our analysis in this paper have been previously investigated in the literature. In particular, the \nuSM with only two generations of charged leptons and neutrinos has been studied in Ref.~\cite{Jenkins:2009dy} and the case of adding two generations of right-handed sterile neutrinos to the SM has been treated in Ref.~\cite{Yu:2021cco}. There, the authors also use the flavor invariants to formulate the necessary and sufficient conditions for CPV in the model. The Hilbert series of the \nuSM with three generations has previously been reported on in Refs.~\cite{Hanany:2010vu,Yu:2022ttm}. As we will see throughout this paper, generalizing the discussion from two to three generations is not an easy task. The more complex flavor group structure leads to a significantly more complicated algebraic ring structure of the invariants in the set, a so-called non-complete intersection ring. We will find that the number of generating invariants in the theory jumps from 38 in Ref.~\cite{Yu:2021cco} for two generations of sterile neutrinos to 459 for three generations of sterile neutrinos, and there is an enormous amount of polynomial relations among them. As a consequence of the complicated algebraic structure of the non-complete intersection ring, there are non-trivial cancellations in the plethystic logarithm which is often used to count invariants given a set of building blocks and their transformation properties. Hence, one should take care when solely relying on these tools from invariant theory to build a complete basis of flavor invariants.

The paper is structured as follows. In Sec.~\ref{sec:TheoryIntro}, we will define more precisely the setup for neutrino masses that we consider in this paper, and introduce all tools from invariant theory we will need to build a \emph{generating} set of flavor invariants. In Sec.~\ref{sec:nuSMinvs}, we will present our results for 3 generations of right-handed neutrinos including the Hilbert series, the \emph{generating} set and the \emph{primary} set. In Sec.~\ref{sec:seesaw_limit}, we will analyze the seesaw limit of the \nuSM, where we discuss how the physical parameters of the full theory map to those in the effective theory. In Sec.~\ref{sec:CPV_condition}, we will study the minimal set of CP-odd invariants used to capture the CPC conditions. We emphasize that this set should be analyzed without assumptions about the spectrum of the parameters. We show the results for the theory with 2 generations of fermions. The case of 3 generations turns out to be too hard to solve, but we nonetheless discuss an approach towards finding such a set. Finally, in Sec.~\ref{sec:Conclusions}, we draw some conclusions. In the Apps.~(\ref{App:parameterization}--\ref{App:CPC2Gen}), we provide additional information and supporting materials that complement the main body of the paper.

%%%%%%%%%%%%%%%%%%%%%%%%%%%%%%%%%%%%%%%%%%%%%%%%%%%%%%%%%%%%%%%%%%%%
\section{The \texorpdfstring{\nuSM}{nuSM} and invariant theory} \label{sec:TheoryIntro}
\subsection{SM extended with right-handed neutrinos} \label{sec:NeutrinoMass}

\begin{table}[t]
        \centering
        \begin{tabular}{c|c|c|c||c}
                & $SU(3)_L \times U(1)_L$ & $SU(3)_e \times U(1)_e$ & $SU(3)_N\times U(1)_N$ & $U(1)_{L+e+N}$\\ \hline
            $Y_e$ &$\mathbf{3}_{+1}$ &$\mathbf{\bar{3}}_{-1}$ &$\mathbf{1}_0$ & 0 \\[0.1cm]
            $Y_N$ &$\mathbf{3}_{+1}$ &$\mathbf{1}_0$ &$\mathbf{\bar{3}}_{-1}$ & 0 \\[0.1cm]
            $M_N$ &$\mathbf{1}_0$ &$\mathbf{1}_0$ &$(\mathbf{\bar{3}}\otimes_s\mathbf{\bar{3}})_{-2}$ & $-2$\\[0.1cm]\hline
            $X_e=Y_e Y_e^{\dagger}$ & $(\mathbf{3}\otimes\mathbf{\bar{3}})_{0}$ & $\mathbf{1}_0$ & $\mathbf{1}_0$ & 0\\[0.1cm]
            $X_N=Y_N Y_N^{\dagger}$ & $(\mathbf{3}\otimes\mathbf{\bar{3}})_{0}$ & $\mathbf{1}_0$ & $\mathbf{1}_0$ & 0\\[0.1cm]
            $X_M=M_N M_N^{*}$ & $\mathbf{1}_0$ & $\mathbf{1}_0$ & $(\mathbf{\bar{3}}\otimes\mathbf{3})_{0}$ & 0\\[0.1cm]
        \end{tabular}
        \caption{The flavor transformation properties of the relevant Yukawa matrices and Majorana mass matrix treated as spurions. The subscripts of the $SU(3)$ representations denote the charge under the $U(1)$ part of the flavor symmetry group. Furthermore, $\otimes_s$ denotes the symmetric tensor product of the simple representations. The charges of all spurions under Abelian lepton number transformations are indicated in the last column. We also show the transformation properties of $X_e$, $X_N$ and $X_M$, which will be used in following sections.}
        \label{tab:flavorTrafo}
\end{table}

Before we start with the analysis, we have to precisely define the theory we will work with. There are several ways to introduce a mass terms for neutrinos at low energies. In this paper, we will extend the SM particle spectrum by adding 3 generations of right-handed sterile neutrinos $N$, which in the literature usually goes under the name of \nuSM.\footnote{In principle 2 generations of sterile neutrinos are enough to generate the observed neutrino masses at low energies~\cite{Ibarra:2003up}. As mentioned before this case has been treated in Ref.~\cite{Yu:2021cco}.} The most general renormalizable Lagrangian that can be built with these fields and the symmetries of the SM gauge group is
\begin{eqnarray} \label{eq:L4}
\mathcal{L}_{\nu\text{SM}}&=&
\sum_{\psi}\bar{\psi}i \slashed{D}\psi
-\left[\frac{1}{2}\left(NCM_NN\right)+\bar{L}Y_NN\widetilde{H}+\bar{L}Y_e e H
+\bar{Q}Y_u u \widetilde{H}+\bar{Q}Y_d d H +\mbox{H.c.}\right]\nonumber\\
&-&\frac{1}{4}G^a_{\mu\nu}G^{a\mu\nu}-\frac{1}{4}W^I_{\mu\nu}W^{I\mu\nu} -\frac{1}{4}B_{\mu\nu}B^{\mu\nu}+(D_\mu H)^\dagger(D^\mu H)
-\lambda\left(H^\dagger H-\frac{v^2}{2}\right)^2,
\end{eqnarray}
where $\psi$ represents all fermion fields $\{Q,u,d,L,e,N\}$, $H$ is the Higgs doublet with the vacuum expectation value (vev) $v$, $\widetilde H_i=\epsilon_{ij}H^*_j$, and $G^a_{\mu\nu},~W^I_{\mu\nu},~B_{\mu\nu}$ are the field strengths of the SM gauge fields and $i, j$ are $SU(2)$ fundamental indices, $a=1,\dots,8$ is a color index, and $I=1,\dots,3$ is a weak isospin index. The gauge symmetries allow for the introduction of Majorana masses for the right-handed neutrinos $N$, which corresponds to a symmetric matrix $M_N$.\footnote{If lepton number is an exact symmetry of the classical Lagrangian (the symmetry is anomalous in the SM and the \nuSM), the Majorana mass term is forbidden, and the neutrino will have a Dirac mass. Then, the flavor structure of the lepton sector will have exactly the same form as the quark sector in the SM.} The $Y_{u,d,e,N}$ are the $3\times 3$ Yukawa couplings with the Higgs field. $C$ is the charge-conjugation matrix, and $D_\mu$ is the gauge covariant derivative.

One can easily check that the fermion kinetic term in $\cL_{\nu\text{SM}}$ is invariant under unitary $U(3)$ flavor transformations of the fermion fields. Assuming that this flavor symmetry is only softly broken by the Yukawa couplings and Majorana mass term, we promote all flavorful couplings to spurions under this symmetry, making the Lagrangian formally invariant. The corresponding transformation properties of the spurions under the non-Abelian part of the flavor group can be found in Tab.~\ref{tab:flavorTrafo}. 

One immediate consequence of this assignment is that the presence of $M_N$ breaks lepton number, as the transformation properties of $M_N$ do not allow for non-trivial rephasings of $N$. Therefore, there will be additional physical Majorana-type phases in the spectrum of the theory. These phases can be removed in the SM because all interactions enjoy a symmetry under respectively rephasing the lepton and quark fields. 

Later on in the paper it will prove useful to have an explicit parameterization for the flavorful matrices. One physical parameterization is given by\footnote{The detail about this physical parameterization can be found in Ref.~\cite{Jenkins:2009dy}. However, they use left-handed Weyl fields to construct the theory, thus leading to a different convention compared to ours. Therefore, we provide the details focusing on the new convention in App.~\ref{App:parameterization}. In addition, we introduce new parameterizations that can be more conveniently used for the study of algebraic properties of the polynomial rings.}
\begin{equation}\label{eq:Parameterization}
    Y_e = \diag\(y_e,y_{\mu},y_{\tau}\) \, , \quad Y_N = V \cdot \diag\(y_1,y_2,y_3\) \cdot W^{\dagger}, \quad M_N = \diag\(m_1,m_2,m_3\) \, ,
\end{equation}
where
\begin{equation}
\begin{split}
V =& U(\theta_{12},\theta_{13},\theta_{23},\delta) \cdot \diag\(1,e^{i \phi_1},e^{i \phi_2}\),\  
W = \diag\(1,e^{i \phi_1^{\prime}},e^{i \phi_2^{\prime}}\) \cdot U(\theta_{12}^{\prime},\theta_{13}^{\prime},\theta_{23}^{\prime},\delta^{\prime}) \, ,
\end{split}
\end{equation}
and $U(\theta_{12},\theta_{13},\theta_{23},\delta)$ has been defined in Eq.~\eqref{eq:ckm}, which is a CKM-like matrix with a phase $\delta\in [0,2\pi)$ and three mixing angles $\theta_{ij}\in [0,\pi/2]$. $\phi_{1,2}\in [0,2\pi)$ and $\phi'_{1,2}\in [0,\pi)$ are additional phases. This parameterization correctly captures the 9 mass parameters, 6 mixing angles and 6 phases of the theory. A detailed counting from a symmetry perspective can be found in Tab.~\ref{tab:Counting}.

In the Dirac limit, the Majorana mass term $M_N$ is set to zero. Consequently, the mixing matrix $W$ and the phases $\phi_{1,2}$ become unphysical, as they can be absorbed by redefinition of the right-handed neutrinos. The flavor structure of the lepton sector then has a similar form as the quark sector. In this scenario, the Pontecorvo-Maki-Nakagawa-Sakata (PMNS) mixing matrix, which describes the mixing between different neutrino flavors, arises in analogy to the CKM matrix in the quark sector of the SM. In the Dirac case, only one \emph{Dirac} CP phase is physical and characterizes the CP-violating effects in the lepton sector.

In the seesaw limit, the heavy Majorana fermions are integrated out, resulting in the emergence of light Majorana neutrino masses for the left-handed SM neutrinos at low energies. The mismatch between the diagonalization matrices in the charged lepton sector and the neutrino sector gives rise to a PMNS matrix. Due to the existence of the Majorana mass term, some rephasings are not allowed, and the two additional \emph{Majorana} CP phases become physical.

The distinction between the Dirac and Majorana nature of neutrinos has important implications for the flavor structure and CP-violating phenomena in the lepton sector, which can be probed experimentally. While neutrino oscillation experiments, such as T2K and NOvA, have made significant progress in measuring the Dirac CP phase, the precise value of this phase is still not known with high accuracy~\cite{Esteban:2020cvm}. Moreover, the existence of Majorana phases remains experimentally elusive. Neutrino oscillation experiments are primarily sensitive to the Dirac phase and do not probe the Majorana phases. The most promising avenue to probe the Majorana nature of neutrinos is through the search for neutrinoless double beta ($0\nu\beta\beta$) decay~\cite{Dolinski:2019nrj,Bilenky:2014uka}, although it is important to note that even if observed, $0\nu\beta\beta$ decay would not directly measure the Majorana phases themselves.

\begin{table}[t]
        \centering
        \begin{tabular}{c|c|c|c}
         & Real Parameters & Imaginary Parameters & Total \\ \hline
        $Y_e,Y_N,M_N$ & $2\times 9+6$ & $2\times 9+3$ & 45 \\ 
        $U_L,U_e,U_N$ & $3\times 3$ & $3\times 6-3$ & 24 \\ \hline
        Difference & 15 & 6 & 21
        \end{tabular}
        \caption{Number of physical parameters in the generic vevs of the flavor spurions $Y_e,Y_N,M_N$ of the \nuSM presented in Tab.~\ref{tab:flavorTrafo}. Note here, that $U_L,U_e,U_N \in SU(3)$ because the lepton numbers are broken by the presence of the vev of $M_N$ as discussed in the main text and cannot be used to remove imaginary parameters, hence the `$-3$' in the second column.}
        \label{tab:Counting}
\end{table}

\subsection{The Hilbert series and plethystic logarithm} \label{sec:PlethysticProgram}

In this section, we will briefly review useful tools from invariant theory, developed in Refs.~\cite{Feng:2007ur,Lehman:2015via,Henning:2017fpj,Xiao:2019uhh} for operator bases and flavor invariants, that simplify the building process and characterization of the low-energy flavor invariants. We will mostly follow the notation introduced in Refs.~\cite{Henning:2017fpj,Wang:2021wdq} here. Our goal in this paper is to find a minimal set of flavor invariants that allows us to parameterize all observables in the theory in terms of those invariants.

The central object of this paper will be flavor invariants $\cI$ that are combinations of Lagrangian parameters of the theory invariant under the maximal possible flavor group of the renormalizable Lagrangian defined in Eq.~\eqref{eq:L4}. In a first step, we want to find a set of invariants that allows us to express all remaining invariants in the theory as a polynomial of the invariants in the set. As mentioned above, this set of invariants is called the \emph{generating set} or \emph{basic set} $\cS_{gen}$, and always has finite cardinality for reductive groups~\cite{Sturmfels:2008Inv,Derksen:2015Inv}.

The \emph{generating} set is a set of invariants, in which no invariant can be expressed as a polynomial of all other invariants in the \emph{generating} set, i.e.,
\begin{equation}
\label{eq:genset}
\cI \neq P(\cS_{gen}\backslash \{\cI\})\,, \quad \forall \cI\in \cS_{gen}\,.
\end{equation}
If an invariant can be expressed as a polynomial of the generating invariants, it will be termed as an \emph{explicit relation} in this paper.\footnote{Later on in this paper, we will also introduce the terms \emph{explicit dependency} and \emph{explicit redundancy} when such a relation is observed.}
But the invariants in the \emph{generating} set may still be \emph{algebraically dependent}, i.e., there could exist relations between them in the form of
\begin{equation}
    P\(\cI_1,\dots,\cI_m\) = 0\,,
\end{equation}
which are called \emph{syzygies} in the invariant literature. We want to stress here that these kind of polynomial relations do not have a linear term in any of the invariants. Otherwise, it would contradict the definition of the \emph{generating} set provided in Eq.~\eqref{eq:genset}.

Among the \emph{generating} set of invariants there exists a set of invariants which are furthermore \emph{algebraically independent}, they are the so-called \emph{primary} invariants.\footnote{Note that this set is not unique and many different choices are viable as long as the set is algebraically independent. Once a set of primary and secondary invariants -- which unlike the primary invariants still have algebraic relations among them -- is chosen, all invariants in the theory can be expressed as a polynomial of them following the so-called Hironaka decomposition (see App.~\ref{app:Hironaka} for more details).} The fact that they are algebraically independent implies that there exists no syzygy only comprising of invariants from the \emph{primary} set. Another interesting result is that the number of physical parameters, i.e., the minimal number of parameters that are left after all transformations allowed by the symmetry group of the theory are used, is equal to the number of invariants in the \emph{primary} set~\cite{Jenkins:2009dy,Derksen:2015Inv, Sturmfels:2008Inv}.

A useful guide to construct those invariants is the so-called \emph{Hilbert series} (HS)
\begin{equation} \label{eq:HSdef}
    \cH(q) = \sum_{i=0}^{\infty} c_i q^i\,,
\end{equation}
which enumerates the number $c_i$ of all possible invariants that can be built from the given set of Lagrangian parameters labeled by $q$ at a given order $i$.

It can be shown that the HS can always be written as a fraction of two polynomials~\cite{Derksen:2015Inv}
\begin{equation} \label{eq:NumDen}
    \cH(q) = \frac{\cN(q)}{\cD(q)} \, ,
\end{equation}
where the numerator is of palindromic form, i.e., $\cN(q) = q^p \cN(1/q)$ with $p$ the highest power of $q$ in $\cN(q)$, and all terms in $\cN(q)$ come with a positive sign. The denominator is of the form $\cD(q) = \prod_{i=1}^{m} (1-q^{d_i})$, where the total number of factors $m$ counts the number of primary invariants corresponding to the physical parameters in the theory, while the exponents $d_i$ in each factor give the power of the spurion in the invariant. If the numerator is trivial, i.e., $\cN(q)=1$ and the complete \emph{generating} set is given by the set of primary invariants, the ring is called a \emph{free ring}.

Of course, most theories contain more than one coupling, and it can be convenient to count each coupling with its own spurion to simplify the identification of the invariants in the HS. For $n$ independent couplings in the theory that are used to build invariants, one defines the \emph{multi-graded} HS
\begin{equation}
    \cH(q_1,\dots,q_n) = \sum_{i_1=0}^{\infty}\dots\sum_{i_n=0}^{\infty} c_{i_1 \dots i_n} q_1^{i_1} \dots q_n^{i_n} \, ,
\end{equation}
where the coefficient $c_{i_1 \dots i_n}$ now count the number of invariants containing the spurions $\(q_1,\dots,q_n\)$ to the power $\(i_1,\dots,i_n\)$. We call these powers the \emph{degrees} of the invariant while we call the sum of the degrees the \emph{order} of the invariant.

Note that the multi-graded HS is no longer guaranteed to come in the form of Eq.~\eqref{eq:NumDen}, which has a palindromic property in numerator with positive terms and the denominator counting the number of primary invariants. To still obtain this information, one can always take the single-graded limit of the HS,  $\cH(q_1,\dots,q_n) \to \cH(q,\dots,q)$, where all spurions in the theory are counted with the same parameter.

After describing the properties of the HS, we now introduce the mathematical methods for calculating it. One convenient way to calculate the HS for reductive Lie groups is the so-called Molien-Weyl formula, which for a single coupling transforming in the representation $R$ of the group $G$ is defined as
\begin{equation}
    \cH(q) = \int d\mu_G \ \exp\(\sum_{k=1}^{\infty} \frac{q^k \chi_R\(z_1^k,\dots,z_d^k\)}{k} \) \equiv \int d\mu_G \PE\[\chi_R(z_1,\dots,z_d);q\] \, ,
\end{equation}
where $d\mu_G$ is the Haar measure of the group, $\chi_R(z_1,\dots,z_d)$ is the character of the representation $R$ of the group $G$ of rank $d$, and we have defined the \emph{plethystic exponential} (PE) in the last step. There is a straightforward generalization of the Molien-Weyl formula for a multi-graded HS in a theory with several couplings transforming in different representations $R_i$
\begin{equation} \label{eq:MolienWeyl}
    \cH(q_1,\dots,q_n) = \int d\mu_G \prod_{i=1}^n \PE\[\chi_{R_i}(z_1,\dots,z_d);q_i\] \,.
\end{equation}
To study some of the properties of the ungraded HS of a theory with several couplings, we can take the single-graded limit $q_i \to q$.

Another useful function is the so-called \emph{plethystic logarithm} (PL) which is the inverse function of the PE that we just defined, i.e., $\PE^{-1}(f(x)) = \PL(f(x))$ and is defined as follows
\begin{equation} \label{eq:PL}
    \PL\[f\(x_1,\dots,x_N\)\] = \sum_{n=1}^{\infty} \frac{\mu(n)}{n} \log\[f\(x_1^n,\dots,x_N^n\)\] \, ,
\end{equation}
where $\mu(n)$ is the so-called Möbius function.\footnote{The Moebius function is defined as 
\begin{equation}
    \mu(n) = 
    \begin{cases}
        0 & \text{n has repeated prime factors} \\
        1 & n=1 \\
        (-1)^j & \text{$n$ is product of $j$ distinct prime numbers} \\
    \end{cases} \, .
\end{equation}
}
In most cases the PL proves extremely helpful because we can simply read off the number of generating invariants and syzygies from the coefficients of the spurions at a given order in the spurions.\footnote{\label{foot:SMInvs}For instance, the PL of the quark sector of the SM can be easily calculated with the HS shown in Ref.~\cite{Jenkins:2009dy}, and is given by $\PL(q)=2 q^2+3 q^4+4 q^6+q^8+q^{12}-q^{24}$. From the positive terms, it can be read off that there are two order-2 invariants, three order-4 invariants, four order-6 invariants, one order-8 invariant and one order-12 invariant, while the negative term shows there is a syzygy at order 24. Among these 11 generating invariants, 10 of them are algebraically independent, which map to the ten physical parameters of the quark sector, with an additional invariant capturing the sign of the CP phase.} In the case of a \emph{complete intersection ring},\footnote{A ring is classified as a complete intersection if the difference between the number of generating invariants and the number of syzygies is equal to the Krull dimension (the Krull dimension being the maximal number of algebraically-independent invariants). Otherwise, it is categorized as a non-complete intersection~\cite{Wang:2021wdq,Derksen:2015Inv}.} where the PL is just a polynomial in the spurions, the positive terms can be identified with the generating invariants of the theory, and the negative terms correspond to the syzygies that exist among them~\cite{Benvenuti:2006qr,Hanany:2010vu}. However, for \emph{non-complete intersection rings}, the PL becomes a non-terminating series, and it has been noted in the literature~\cite{Benvenuti:2006qr,Hanany:2010vu,Wang:2021wdq} that the \emph{leading positive} terms, i.e., all positive terms up to the first term with a negative sign in the PL,\footnote{The terms in the PL is usually sorted according to their powers in the spurions. Refer to footnote~\ref{foot:SMInvs} for an example illustrating the PL of the quark sector.} can be identified with the generating invariants, and the \emph{leading negative} terms, i.e., the first negative terms that appear after the leading positive terms correspond to syzygies.\footnote{All other terms in the non-terminating PL of a non-complete intersection ring after the leading negative terms have -- to our knowledge -- no meaning for the construction of a \emph{generating} set beyond the fact that they appear in a special form of the HS, the so-called Euler form~\cite{Benvenuti:2006qr}. In this form the HS can be written as $\cH(q) = \prod_{n=1}^{\infty}(1-q^n)^{-b_n}$, where it can be shown that the $b_n$ are exactly the coefficients in the PL $\PL\[\cH\(q\)\] = \sum_{n=1}^{\infty} b_n q^n$.\label{foot:EulerForm}}

We will see later for the \nuSM with 3 generations of sterile neutrinos, that this does not necessarily have to hold true, and the interpretation of the positive and negative terms in the PL have to be slightly changed for more complicated invariant rings. This has also been pointed out in the literature~\cite{Wang:2021wdq}. As we will see in the next section, this is also the main difference between the SM quark sector and the \nuSM. While, the representations that the spurions of the SM quark sector transform in are still sufficiently simple to generate a complete intersection ring with a terminating PL, this is no longer true for the \nuSM. Here, the representation the Majorana mass $M_N$ lives in complicates the ring structure significantly, leading to a non-complete intersection ring for both two and three generations of sterile neutrinos. This non-complete intersection ring can also simply arise when increasing the number of generations of fields in a theory. For instance, the complete intersection ring in the quark sector becomes non-complete when the quark has 4 generations~\cite{Hanany:2010vu}.

%%%%%%%%%%%%%%%%%%%%%%%%%%%%%%%%%%%%%%%%%%%%%%%%%%%%%%%%%%%%%%%%%%%%
\section{Building an invariant basis for the \texorpdfstring{\nuSM}{nuSM}}\label{sec:nuSMinvs}
\subsection{Hilbert series of the \texorpdfstring{\nuSM}{nuSM}}
\label{sec:HS}
Before we start building invariants, we will first compute the HS and PL to set our expectations for the \textit{generating} and \textit{primary} set of invariants. We use the Molien--Weyl formula introduced in Eq.~\eqref{eq:MolienWeyl} to calculate the HS for the \nuSM with the spurions $Y_e, Y_N$ and $M_N$.\footnote{To get a dimensionless quantity we take $M_N$ to be divided by the only other mass scale in the problem, the Higgs vev $v$. Only then, one can compare invariants with a different number of insertions of $M_N$.} For that, we need the characters of the fundamental and anti-fundamental representations and the Haar measure of $U(3)$ which are given by~\cite{Henning:2017fpj}
\begin{equation}
    \begin{split}
        \chi_{U(3)}^{\mathbf{3}} & = z_1 + z_2 + z_3 \,, \\
        \chi_{U(3)}^{\mathbf{\bar{3}}} & = z_1^{-1} + z_2^{-1} + z_3^{-1} \,,  \\
        d\mu_{U(3)} & = \frac{1}{6!} \(\prod_{i=1}^3 \frac{dz_i}{2\pi i z_i} \) \(- \frac{\(z_2-z_1\)^2\(z_3-z_1\)^2\(z_3-z_2\)^2}{z_1^2 z_2^2 z_3^2} \) \,.
    \end{split}
\end{equation}
From these, we can construct the characters for the representations of the flavorful Lagrangian parameters of the \nuSM following Tab.~\ref{tab:flavorTrafo}. For instance, the character for $Y_N$ is given by 
\begin{equation}
    \chi_{Y_N} = \chi_{U(3)_L}^{\mathbf{3}}(z_1,z_2,z_3) \chi_{U(3)_N}^{\mathbf{\bar{3}}}(z_4,z_5,z_6) = \(z_1 + z_2 + z_3 \) \(z_4^{-1} + z_5^{-1} + z_6^{-1}\)  \,.
\end{equation} 
The characters for all other spurions can be obtained in the same manner. Using the expression for the Molien--Weyl formula in Eq.~\eqref{eq:MolienWeyl} with the same grading for all spurions, we can calculate the ungraded HS. The calculation involves the integral over the six variables $z_1,\dots,z_6$ over the contour $|z_i|=1$, which can be obtained by calculating the residues. The same calculation has been presented before, please refer to Refs.~\cite{Hanany:2010vu,Yu:2022ttm} for details. We find for the numerator of the ungraded HS
\begin{eqnarray}
\cN(q) &=& 1+q^{4}+5 q^{6}+9 q^{8}+22 q^{10}+61 q^{12}+126 q^{14}+273
   q^{16}+552 q^{18}+1038 q^{20}\nonumber\\
   &&+1880 q^{22}+3293 q^{24}+5441 q^{26}
   +8712
   q^{28}+13417 q^{30}+19867 q^{32}+28414 q^{34}+39351 q^{36}\nonumber\\
   &&+52604
   q^{38}+68220 q^{40}+85783 q^{42}+104588 q^{44}+123852 q^{46}
   +142559
   q^{48}+159328 q^{50}\nonumber\\
   &&+173201 q^{52}+183138 q^{54}+188232 q^{56}+188232
   q^{58}+183138 q^{60}+173201 q^{62}+159328 q^{64}\nonumber\\
   &&
   +142559 q^{66}+123852
   q^{68}+104588 q^{70}+85783 q^{72}+68220 q^{74}+52604 q^{76}+39351
   q^{78}\nonumber\\
   &&+28414 q^{80}+19867 q^{82}
   +13417 q^{84}+8712 q^{86}+5441
   q^{88}+3293 q^{90}+1880 q^{92}+1038 q^{94}\nonumber\\
   &&+552 q^{96}+273 q^{98}+126
   q^{100}+61 q^{102}+22 q^{104}+9 q^{106}+5 q^{108}+q^{110}+q^{114},
   \label{eq:num}
\end{eqnarray}
which has a palindromic form.\footnote{\label{foot:ambiguity} Note that in the numerator Eq.~\eqref{eq:num}, a term $(1+q^2)$ can be factorized, and the same factor also appears in the denominator Eq.~\eqref{eq:deno}.  If this factor were to be simplified, the HS would take a rational form with a numerator featuring some negative terms, in contradiction with the positivity requirement announced earlier. However, one can always multiply a factor $(1+q^k)^m$ in both numerator and denominator if there exists a factor $(1-q^k)^n$ (where $m\leq n$) in the denominator. This multiplication removes a factor of $(1-q^k)^m$ from the denominator while introducing a new factor of $(1-q^{2k})^m$. The total number of factors in the denominator does not change, and the numerator keeps its palindromic form with positive terms. This freedom indicates that there is ambiguity in determining the form of the HS if there is no further requirement of the HS. This is also reflected in the Hironaka decomposition, where the set of primary and secondary invariants are not uniquely determined. Please refer to the App.~\ref{app:Hironaka} for details.}
The denominator is
\begin{eqnarray}
\cD(q) &=&   \left(1-q^2\right)^3 \left(1-q^4\right)^4 \left(1-q^6\right)^4
   \left(1-q^8\right)^2 \left(1-q^{10}\right)^2 \left(1-q^{12}\right)^3
   \left(1-q^{14}\right)^2 \left(1-q^{16}\right).
   \label{eq:deno}
\end{eqnarray}
As expected, the powers of the factors in the denominator add up to 21, the number of physical parameters in the \nuSM which is also the cardinality of the \emph{primary} set. Our result of the ungraded HS is consistent with those found in Refs.~\cite{Hanany:2010vu,Yu:2022ttm}. We have furthermore calculated the multi-graded HS with different parameters $\{e,m,n\}$ counting the degrees of the couplings $\{Y_e,M_N,Y_N\}$, which we only show in App.~\ref{App:GradedHSPL} due to its length. To obtain the results in this paper, we have developed our own \texttt{Mathematica} code that can efficiently calculate the Hilbert series. The code will shortly be published as a \texttt{Mathematica} package under the name \texttt{CHINCHILLA}~\cite{Grojean:2024}.

Plugging the ungraded HS in Eq.~\eqref{eq:PL} to calculate the PL, we find furthermore
\begin{equation} \label{eq:HS_PL}
\begin{split}
  \PL[\cH(q)] =& 3 q^2 + 5 q^4 + 9 q^6 + 10 q^8 + 19 q^{10} + 40 q^{12} + 66 q^{14} + 92 q^{16} + 70 q^{18} - 124 q^{20} \\
  &- 703 q^{22} - 2039 q^{24} - 4391 q^{26} - 7472 q^{28} - 8522 q^{30} + 590 q^{32} + O(q^{34}) \, .
\end{split}
\end{equation}
We only show the PL up to order 32. For higher orders, both positive and negative terms will appear repeatedly in an infinite series, which implies that the \nuSM has an algebraic structure of a \textit{non-complete intersection} ring.

The usual interpretation of the PL suggests that the leading positive terms indicate a total of 314 generating invariants. However, our analysis reveals that this number is incorrect. Discrepancies begin to arise at order 16 in the PL, where the count of generating invariants exceeds 92. This is due to the non-complete intersection nature of the ring, resulting in non-trivial cancellations between the number of generating invariants and the number of syzygies. Concrete examples will be provided in the following sections to clarify this matter. It is worth noting that a similar cancellation was observed in a low-energy neutrino model in Ref.~\cite{Wang:2021wdq}, which also corresponds to a non-complete intersection ring.
There is no rigorous proof for this cancellation, however we can try to understand it as follows.

For a sufficiently simple invariant ring, the orders in the PL corresponding to the appearance of syzygies are well-separated from the orders corresponding to the appearance of the generating invariants. When the ring becomes more complicated, more invariants are needed to describe the full algebraic structure of the ring, hence there exist more generating invariants at higher orders. If syzygies containing the lower order generating invariants still appear at a similar order as in less complicated rings, there will be an overlap between the regions of positive and negative terms. This overlap will result in cancellations between the number of generating invariants and the number of syzygies. Therefore, one should be cautious when using the PL to count the number of generating invariants and the number of syzygies in a non-complete intersection ring. Observing a negative term in the PL does not necessarily imply the absence of generating invariants, but rather indicates the presence of more syzygies than generating invariants. Hence, \emph{the coefficient in the PL is the difference between the number of generating invariants and the number of syzygies}. 

Moreover, the coefficients in the ungraded PL at a specific order can be subject to cancellations from terms which have a different grading for the same total order in the multi-graded PL but cancel once the ungraded limit is taken. In this sense, we can not naively assume that the leading positive terms in Eq.~\eqref{eq:HS_PL} can capture all generating invariants as we expected for the theory with a complete intersection ring. Hence, the multi-graded PL (see App.~\ref{App:GradedHSPL}) is our main guide to check if we have found the correct number of generating invariants and syzygies at a given order in the spurions. It is also conjectured in Ref.~\cite{Yu:2021cco} that the generating invariants are all captured by the terms prior to the pure negative order\footnote{In the multi-graded PL, we sort the terms according to the order (total degrees) of $[emn]$~(c.f. Eq.~\eqref{eq:gradedPL}). In a given order, if all terms are negative, it is called a \emph{pure negative} order.} in multi-graded PL. The pure negative terms occur at order 26 in Eq.~\eqref{eq:gradedPL} in our theory, so the \emph{generating} set should already be found within order 24. However, to test the conjecture, we also construct invariants up to order 26 to see that indeed no generating invariant can be found at this pure negative order. As we will see below, our analysis supports this conjecture.

\subsection{Constructing the invariants} \label{sec:cons_inv}

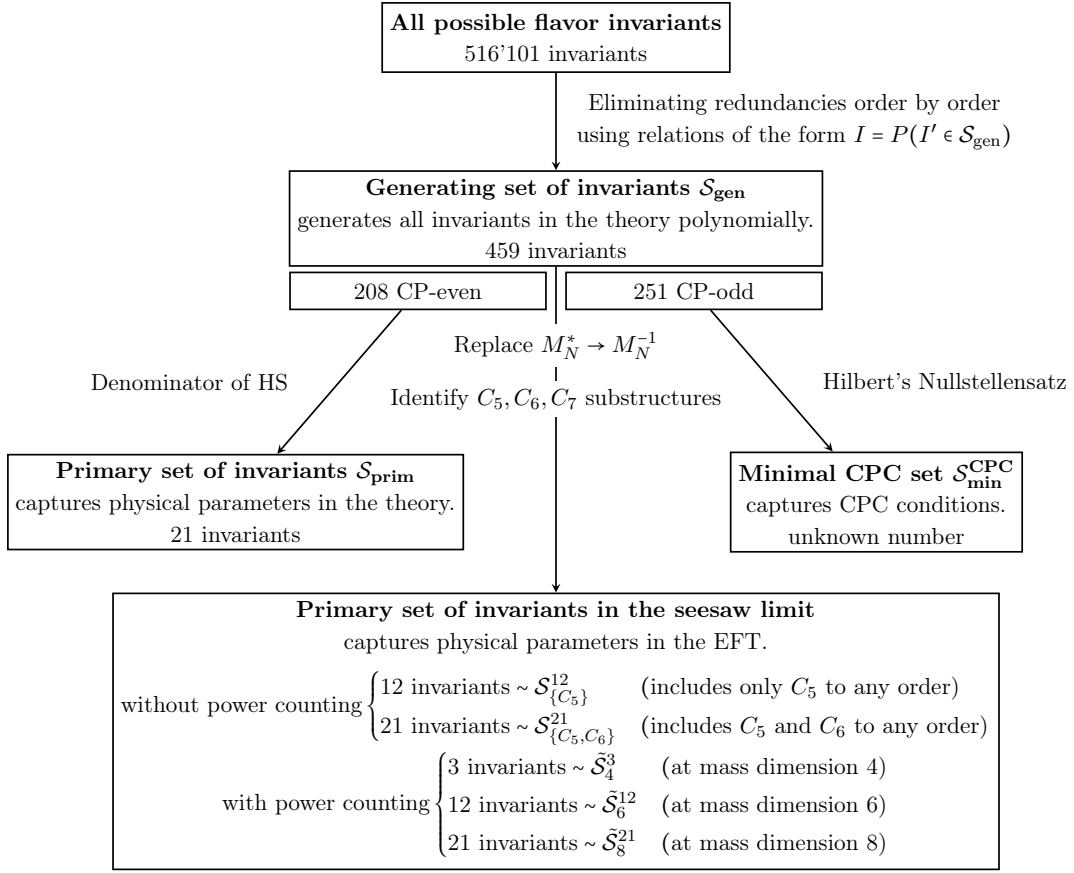
\begin{figure}[tb]
\centering
\scalebox{0.8}{
  \tikzset{
  box/.style={rectangle, draw=black, thick, fill=white, text centered, minimum height=4em, align = center},
  label/.style={text centered,align = center},
  arrow/.style={thick,->,>=stealth}
}
\begin{tikzpicture}

\node (box1) [box, minimum height=3em] {\textbf{All possible flavor invariants} \\ 516'101 invariants};

\node (box3) [box, below of=box1, yshift=-2cm] {  \textbf{Generating set of invariants $\cS_{\text{gen}}$} \\ generates all invariants in the theory polynomially.\\ 459 invariants};

\node (box3a) [box, minimum width=11.0em, minimum height=1.6em, below of=box3, yshift=-0.2cm, xshift=-2.27cm] {208 CP-even};
\node (box3b) [box, minimum width=11.em, minimum height=1.6em, below of=box3, yshift=-0.2cm, xshift=2.27cm] {251 CP-odd};

\node (box4a) [box, below of=box3a, xshift=-3.cm, yshift=-2.5cm] {\textbf{Primary set of invariants $\cS_{\text{prim}}$} \\ captures physical parameters in the theory.\\ 21 invariants};

\node (box4b) [box, below of=box3b, xshift=3.cm, yshift=-2.5cm] {\textbf{Minimal CPC set $\cS_{\text{min}}^{\text{CPC}}$} \\ captures CPC conditions. \\ unknown number};

\node (box5) [box, below of=box3, xshift=0.cm, yshift=-7.5cm] {\textbf{Primary set of invariants in the seesaw limit}\\
captures physical parameters in the EFT.\\[0.2cm]
$\text{without power counting} \begin{cases}
    \text{12 invariants}\sim\cS_{\{C_5\}}^{12} & (\text{includes only $C_5$ to any order)} \\
    \text{21 invariants}\sim\cS_{\{C_5,C_6\}}^{21} & (\text{includes $C_5$ and $C_6$ to any order)}
\end{cases}$\\
$\text{with power counting} \begin{cases}
    \text{3 invariants}\sim\tilde{\cS}_{4}^{3} & (\text{at mass dimension 4)} \\
    \text{12 invariants}\sim\tilde{\cS}_{6}^{12} & (\text{at mass dimension 6)}\\
    \text{21 invariants}\sim\tilde{\cS}_{8}^{21} & (\text{at mass dimension 8)}
\end{cases}$
};
\draw [arrow] (box1) -- (box3) node[midway,right] [label,xshift=0.2cm] {Eliminating redundancies order by order \\ using relations of the form $I = P(I^\prime \in \cS_{\text{gen}})$};
\draw [arrow] (box3a) -- (box4a) node[midway,left,xshift=-0.7cm] [label] {Denominator of HS};
\draw [arrow] (box3b) -- (box4b) node[midway,right,xshift=0.7cm] [label] {Hilbert's Nullstellensatz};
\draw [arrow] (box3) -- (box5) node[midway,fill=white,yshift=1.4cm] {Replace $M_N^*\to M_N^{-1}$} node[midway,fill=white,yshift=0.5cm] {Identify $C_5,C_6, C_7$ substructures};
\end{tikzpicture}
}
\caption{Flow graph of different invariant sets that appear in the analysis alongside the algorithms that are used to obtain one set from another. The number in the top box correspond to all single trace invariants up to a total order of 26 in all spurions. We find 208 CP-even and 251 CP-odd invariants that make up the \emph{generating} set of the ring defined by the Lagrangian parameters and their transformation properties. A complete list of these invariants can be found in App.~\ref{App:InvariantList}. The 21 algebraically independent invariants are selected from the CP-even \emph{generating} set to form the \emph{primary} set. These invariants accurately capture the 21 physical parameters of the theory. To determine the CPC conditions, the CP-odd invariants are selected from the CP-odd \emph{generating} set. However, our program fails to find the minimal set due to the complexity of the theory. Detailed explanations on Hilbert's Nullstellensatz can be found in App.~\ref{App:theorem}. In the seesaw limit, we replace $M_N^*\to M_N^{-1}$ for the generating invariants, and identify invariants with substructures of $C_5, C_6$ and $C_7$ as defined in Eq.~\eqref{eq:SeesawMatching}. The number of primary invariants with and without considering the total suppression of the invariants in the heavy Majorana mass is obtained by calculating the Jacobian rank of the identified invariants. A detailed analysis can be found in Sec.~\ref{sec:seesaw_limit}, where the \emph{primary} sets $\cS$ and $\tilde{\cS}$ are shown explicitly.}
\label{fig:FlowGraphSets}
\end{figure}

Although the PL may provide some clues regarding the number of generating invariants, their specific form remains unknown. While it is possible to construct invariants for some simple models manually, in the case of complex models such as the \nuSM, which involves hundreds of invariants in their \emph{generating} sets, it becomes unfeasible to manually build them. Furthermore, as mentioned at the end of the last section, in a sufficiently complicated ring, the orders at which generating invariants and syzygies appear in the PL might overlap, hence leading to cancellations. To construct a \emph{generating} set, one therefore cannot solely rely on the information provided by the PL. 

Instead,  we will start by constructing all possible invariants up to a given order suggested by the conjecture on the PL, which is done by a graph-based method introduced in this section. The objective is to eliminate redundant invariants with explicit relations and create a \emph{generating} set. This \emph{generating} set will then be used to obtain a \emph{primary} set. Furthermore, we aim to find the minimal set of invariants that can determine the CPC conditions based on the CP-odd generating invariants of the theory. The entire process is summarized in Fig.~\ref{fig:FlowGraphSets}, and we will provide a detailed explanation of each step in the following sections.

\begin{figure}[t]
  \centering
\begin{tikzpicture}[every circle node/.style={draw, fill=blue!10, inner sep=0pt, minimum size=3.5mm},
  every path/.style={-{Stealth[length=2.5mm]}}]
  \node[circle, label={[label distance=0.5mm]$e$}] (e) at (0,0) {{\footnotesize 1}};
  \node[circle, label={[label distance=0.5mm]$L$}] (L) at (2.5,0) {{\footnotesize 2}};
  \node[circle, label={[label distance=0.5mm]$N$}] (N) at (5,0) {{\footnotesize 3}};
  \node[circle, label={[label distance=0.5mm]$N^*$}] (Ns) at (7.5,0) {{\footnotesize 4}};
  \node[circle, label={[label distance=0.5mm]$L^*$}] (Ls) at (10,0) {{\footnotesize 5}};
  \node[circle, label={[label distance=0.5mm]$e^*$}] (es) at (12.5,0) {{\footnotesize 6}};

  \draw (L) to[bend left=-45] node[above]{\textcolor{blue}{$Y_e$}} (e);
  \draw (e) to[bend left=-45] node[above]{\textcolor{blue}{$Y_e^\dagger$}} (L);
  \draw (L) to[bend left=-45] node[above]{\textcolor{blue}{$Y_N$}} (N);
  \draw (N) to[bend left=-45] node[above]{\textcolor{blue}{$Y_N^\dagger$}} (L);
  \draw (Ns) to[bend left=-45] node[above]{\textcolor{blue}{$M_N$}} (N);
  \draw (N) to[bend left=-45] node[above]{\textcolor{blue}{$M_N^*$}} (Ns);
  \draw (Ns) to[bend left=-45] node[above]{\textcolor{blue}{$Y_N^T$}} (Ls);
  \draw (Ls) to[bend left=-45] node[above]{\textcolor{blue}{$Y_N^*$}} (Ns);
  \draw (Ls) to[bend left=-45] node[above]{\textcolor{blue}{$Y_e^*$}} (es);
  \draw (es) to[bend left=-45] node[above]{\textcolor{blue}{$Y_e^T$}} (Ls);

  \node[circle, label={[label distance=0.5mm]$d$}] (d) at (0,-2.5) {{\footnotesize 7}};
  \node[circle, label={[label distance=0.5mm]$Q$}] (Q) at (2.5,-2.5) {{\footnotesize 8}};
  \node[circle, label={[label distance=0.5mm]$u$}] (u) at (5,-2.5) {{\footnotesize 9}};
  \node[circle, label={[label distance=0.5mm]$u^*$}] (us) at (7.5,-2.5) {{\footnotesize 10}};
  \node[circle, label={[label distance=0.5mm]$Q^*$}] (Qs) at (10,-2.5) {{\footnotesize 11}};
  \node[circle, label={[label distance=0.5mm]$d^*$}] (ds) at (12.5,-2.5) {{\footnotesize 12}};

  \draw (Q) to[bend left=-45] node[above]{\textcolor{blue}{$Y_d$}} (d);
  \draw (d) to[bend left=-45] node[above]{\textcolor{blue}{$Y_d^\dagger$}} (Q);
  \draw (u) to[bend left=-45] node[above]{\textcolor{blue}{$Y_u^\dagger$}} (Q);
  \draw (Q) to[bend left=-45] node[above]{\textcolor{blue}{$Y_u$}} (u);
  \draw (Qs) to[bend left=-45] node[above]{\textcolor{blue}{$Y_d^*$}} (ds);
  \draw (ds) to[bend left=-45] node[above]{\textcolor{blue}{$Y_d^T$}} (Qs);
  \draw (us) to[bend left=-45] node[above]{\textcolor{blue}{$Y_u^T$}} (Qs);
  \draw (Qs) to[bend left=-45] node[above]{\textcolor{blue}{$Y_u^*$}} (us);
\end{tikzpicture}
\caption{The flavor invariant graph that can be used to construct all possible single trace flavor invariants in the \nuSM. To any closed walk that follow the arrows, one can associate a single-trace invariant. Note that the graph for the SM would have a ``holomorphic'' structure, i.e.,  it has two separated branches involving separately only fields or only their conjugates. This changes in the \nuSM, where the transformation properties of the Majorana mass $M_N$ connect the holomorphic and anti-holomorphic branches. More details can be found in the main text.}
\label{fig:invariantGraph}
\end{figure}
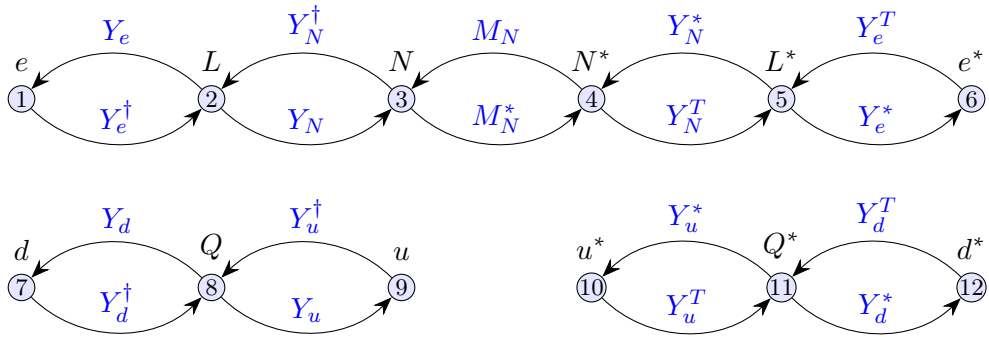

\paragraph{Flavor invariants from closed walks in a graph}

The construction of flavor invariants in our paper heavily relies on the flavor invariant graph, which is inspired by Ref.~\cite{Jenkins:2007ip}. We show the graph of the \nuSM in Fig.~\ref{fig:invariantGraph}, where the lepton sector and quark sector are presented in the top and bottom panels respectively. In a graph, the nodes represent the different fields and their conjugates, while the arrows are labeled with flavor matrices connecting various fields at the nodes. These graphs have the following two main advantages. First, they display the transformation rules for all flavor matrices. Taking the basic elements of the graph as an example,
\begin{center}
\begin{tikzpicture}[every circle node/.style={draw, fill=blue!10, inner sep=0pt, minimum size=3.5mm},
  every path/.style={-{Stealth[length=2.5mm]}}]
  \node[circle, label={[label distance=0.5mm]$F_i$}] (A) at (0,0) {{\footnotesize $i$}};
  \node[circle, label={[label distance=0.5mm]$F_j$}] (B) at (2.5,0) {{\footnotesize $j$}};
  \draw (A) to node[above]{\textcolor{blue}{$Y$}} (B);
\end{tikzpicture}
\end{center}
where the arrow labeled with a flavor matrix $Y$, starting from vertex $i$ to vertex $j$, which are labeled with the fields $F_i$ and $F_j$ respectively, indicating that the flavor matrix $Y$ should transform as $Y\to \neat{U_{F_i}^{} Y U_{F_j}^{\dagger}}$ under the flavor group. The numbers displayed in the vertex are simply for labeling purposes, and $F_i$ should be understood as the field at vertex $i$, not to be confused with the flavor index. For example, for the graph in Fig.~\ref{fig:invariantGraph}, we can read off $Y_e\to U_LY_eU_e^{\dagger}$ and $Y_e^*\to \neat{U_{L^*}^{}Y_{e}^{*}U_{e^*}^{\dagger}=U_{L}^{*}Y_{e}^{*}U_{e}^{T}}$, where we have used the fact that $U_{f^*}=U_{f}^{*}$. Second, following the directional flow of the arrows, one can pass through vertices and edges, creating ``paths''. In mathematical terminology, a path with repeated inclusion of vertices and edges is referred to as a ``walk''. To every walk, one can associate an object with specific transformation under the flavor transformations. For instance, in the graph below, starting from vertex $i$ and following a sequence of arrows until reaching vertex $j$,
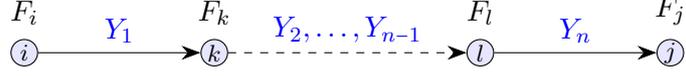
\begin{center}
\begin{tikzpicture}[every circle node/.style={draw, fill=blue!10, inner sep=0pt, minimum size=3.5mm},
  every path/.style={-{Stealth[length=2.5mm]}}]
  \node[circle, label={[label distance=0.5mm]$F_i$}] (A) at (0,0) {{\footnotesize $i$}};
  \node[circle, label={[label distance=0.5mm]$F_k$}] (P) at (2.5,0) {{\footnotesize $k$}};
  \node[circle, label={[label distance=0.5mm]$F_l$}] (Q) at (6,0) {{\footnotesize $l$}};
  \node[circle, label={[label distance=0.5mm]$F_j$}] (B) at (8.5,0) {{\footnotesize $j$}};
  \draw (A) to node[above]{\textcolor{blue}{$Y_1$}} (P);
  \draw[dashed] (P) to node[above]{\textcolor{blue}{$Y_2,\dots,Y_{n-1}$}} (Q);
  \draw (Q) to node[above]{\textcolor{blue}{$Y_n$}} (B);
\end{tikzpicture}
\end{center}
the product of flavor matrices $X\equiv Y_1\dots Y_n$ will transform as $X\to \neat{U_{F_i}^{} X U_{F_j}^{\dagger}}$. We can simply read off an example from Fig.~\ref{fig:invariantGraph} that $\neat{Y_{e}^{\dagger} Y_{N}^{}\to U_{e}^{} Y_{e}^{\dagger} Y_{N}^{} U_{N}^{\dagger}}$. If there is a walk with $i=j$, 
which means the walk is closed and ends at the starting vertex, we have $X\to \neat{U_{F_i}^{} X U_{F_i}^{\dagger}}$. This implies that we can construct a single trace invariant $\Tr(X)$ under the action of the group. As a result, flavor invariants are mapped to the closed walks in the graph, and one can identify a closed walk with each single-trace invariant in the theory.\footnote{One can also find other forms of the flavor invariants, but they can be converted to the single-trace forms. See Ref.~\cite{Bonnefoy:2021tbt} for the discussion.}

For instance, one can find the following correspondences between invariants and the closed walks in the quark and lepton sectors
(the red paths represent closed walks on the graphs denoted by the black arrows connecting the vertices)
\begin{equation}
\label{eq:walk}
\begin{split}
\begin{tikzpicture}[baseline={([yshift=-1ex]current bounding box.center)}, every circle node/.style={draw, fill=blue!10, inner sep=0pt, minimum size=3.5mm},
  every path/.style={-{Stealth[length=2.5mm]}}]
  \node[circle, label={[label distance=0.5mm]95:$d$}] (d) at (0,-2.5) {{\footnotesize 7}};
  \node[circle, label={[label distance=1mm]$Q$}] (Q) at (2.5,-2.5) {{\footnotesize 8}};
  \node[circle, label={[label distance=0.5mm]80:$u$}] (u) at (5,-2.5) {{\footnotesize 9}};
  \coordinate (dl) at ($(d) + (-0.3,0) $);
  \coordinate (Qb) at ($(Q) + (0,-0.3) $);
  \coordinate (ur) at ($(u) + (0.3,0) $);
  \coordinate (Qu) at ($(Q) + (0,0.3) $);
  \draw[rounded corners, red, style=](Qb) to[bend left=-45] (ur) to[bend left=-45] (Qu) to[bend left=-45] (dl) to[bend left=-45] (Qb);
  \draw (Q) to[bend left=-45] node[above=3pt]{\textcolor{blue}{$Y_d$}} (d);
  \draw (d) to[bend left=-45] node[above]{\textcolor{blue}{$Y_d^\dagger$}} (Q);
  \draw (u) to[bend left=-45] node[above=3pt]{\textcolor{blue}{$Y_u^\dagger$}} (Q);
  \draw (Q) to[bend left=-45] node[above]{\textcolor{blue}{$Y_u$}} (u);
\end{tikzpicture}&\sim 8\to 9\to 8\to 7\to 8\\
&\sim \Tr(\neat{Y_{u}^{}Y_{u}^{\dagger}Y_{d}^{}Y_{d}^{\dagger}})\,,\\
\begin{tikzpicture}[baseline={([yshift=-1ex]current bounding box.center)}, every circle node/.style={draw, fill=blue!10, inner sep=0pt, minimum size=3.5mm},
  every path/.style={-{Stealth[length=2.5mm]}}]
  \node[circle, label={[label distance=1.0mm]110:$L$}] (L) at (2.5,0) {{\footnotesize 2}};
  \node[circle, label={[label distance=1.5mm]90:$N$}] (N) at (5,0) {{\footnotesize 3}};
  \node[circle, label={[label distance=0.5mm]70:$N^*$}] (Ns) at (7.5,0) {{\footnotesize 4}};
  \coordinate (Ll) at ($(L) + (-0.3,0)$);
  \coordinate (Nb) at ($(N) + (0,-0.2)$);
  \coordinate (Nsr) at ($(Ns) + (0.3,0)$);
  \coordinate (Nsr2) at ($(Ns) + (0.35,0)$);
  \coordinate (Nu) at ($(N) + (-0.1,0.3)$);
  \coordinate (Nl) at ($(N) + (-0.3,0)$);
  \coordinate (Ll) at ($(L) + (-0.3,0)$);
  \draw[rounded corners, red, style=](Ll) to[bend left=-50] (Nb) to[bend left=-45] (Nsr) to[bend left=-50] (Nl) to[bend left=-60] (Nsr2)to[bend left=-50] (Nu)to[bend left=-45] (Ll);
  \draw (L) to[bend left=-45] node[above]{\textcolor{blue}{$Y_N$}} (N);
  \draw (N) to[bend left=-45] node[above=2pt]{\textcolor{blue}{$Y_N^\dagger$}} (L);
  \draw (Ns) to[bend left=-45] node[above=5pt]{\textcolor{blue}{$M_N$}} (N);
  \draw (N) to[bend left=-45] node[above]{\textcolor{blue}{$M_N^*$}} (Ns);
\end{tikzpicture} &\sim 2\to 3\to 4\to 3\to 4\to 3\to 2\\
&\sim \Tr(\neat{Y_{N}^{}M_{N}^{*}M_{N}^{}M_{N}^{*}M_{N}^{}Y_{N}^{\dagger}})\,.
\end{split}
\end{equation}
Note that the walks are equivalently given as a chain of numbers corresponding to the vertices that are passed through. We only show here simple examples of walks in subsets of the \nuSM graph. Obviously, one can consider more complicated walks involving more vertices and obtain new flavor invariants accordingly. Due to the cyclicity of the trace, the invariant associated to a closed walk is independent of the starting vertex, e.g. $ 8\to 9\to 8\to 7\to 8 = 7 \to 8 \to 9 \to 8 \to 7$ for the first example above. Based on this, we can rotate the numbers in the chain to arrange them in lexicographically smallest order. In addition, since the last vertex in a closed walk is always the same as the first vertex, we can omit the last number in the chain. To further shorten the notation, we can also remove the arrow, resulting in an integer representation of the walk notation. By following this approach, all single trace invariants are uniquely represented as integers. For instance, the two invariants above are represented by the two integers 7898 and 234343, respectively.

The inclusion of the Majorana neutrino mass $M_N$ creates an important difference between the lepton and the quark sectors. 
The quark sector has two separate parts in the graph: the closed walks involve either the fields or their complex conjugate at the vertices, but never mix both. The walks are holomorphic or anti-holomorphic. The set of invariants built from the anti-holomorphic closed walks is equivalent of the one built from the holomorphic walks (see App.~\ref{App:AlgGenInvs} for details).
The introduction of the Majorana neutrino mass term $M_N$ complicates the scenario in the lepton sector. The two conjugate parts are connected through this new flavor matrix, making the invariant structure much more complex, as we will see below.

In the Dirac limit, $M_N\to 0$, the quark and the lepton sectors  obviously have the same flavor structure, and thus the flavor invariants have the same form. In App.~\ref{App:AlgGenInvs}, we show how to construct the flavor invariants in the quark sector based on the graph approach. The generating flavor invariants are shown explicitly, which can be easily mapped to the flavor invariants of the lepton sector in the Dirac limit. 

For non-zero and finite $M_N$, we can systematically enumerate the closed walks corresponding to all single-trace flavor invariants up to arbitrarily high order. Based on the arguments of Ref.~\cite{Yu:2021cco}, we are particularly interested in invariants up to order 26 which is the order of the first pure negative term in the multi-graded PL of the \nuSM, and therefore it should be possible to obtain a set of generating invariants out of them. In App.~\ref{App:AlgGenInvs}, we present the brute-force algorithm we used to construct all the 516'001 single-trace invariants we obtain up to order 26. However, there is still a lot of redundancy among these invariants, and we can immediately reduce the set of invariants using some simple relations.
\begin{itemize}
\item Transpose redundancy

The flavor invariant associated with the walk $W_1\equiv ij\dots kl$ is always accompanied by another walk with primed vertices in reverse order $W_2\equiv l'k'\dots j'i'$, where $v'=7-v$ for the graph in the lepton sector, featuring a left and right mirror symmetry. The invariants generated from these two walks are the same, and they are connected through the trace identity $\Tr \big(A^T\big)=\Tr\big(A\big)$. For instance, $\Tr\Big(Y_eY_e^\dagger\Big)=\Tr\Big(Y_e^*Y_e^T\Big)$.

\item Conjugate redundancy

The walk $W_1$ defined above is also associated with $W_3\equiv i'j'\dots k'l'$. The invariants generated by these two walks are conjugate to each other. Thus, both $\Tr(A)$ and $\Tr(A^*)$ will be generated in our construction. As the CP properties of $\Re\Tr(A)$ and $\Im\Tr(A)$ are more transparent than those of $\Tr(A)$ and $\Tr(A^*)$, we will trade $\Tr(A)$ and $\Tr(A^*)$, which are generated by the graphs, with $\Re\Tr(A)$ and $\Im\Tr(A)$, whenever a complex invariant is found.

\item Cayley--Hamilton theorem

The Cayley--Hamilton theorem, along with its simple variations, enables us to eliminate invariants or establish relations among them. Note that the theorem is not fully utilized, as there are more complex variations that are not easy to employ. Therefore, we will implement a generic numerical algorithm to handle the remaining redundancies. 

\end{itemize}
The details of these redundancies are discussed in App.~\ref{App:AlgGenInvs}. After eliminating these redundancies, we are left with a set of 8'666 invariants, which is still too large to form a set of generating invariants.

\paragraph{Construction of the generating set}
After the set of invariants are pre-reduced with well-known trace relations, we search for explicit relations of invariants at a given order in terms of the other invariants at the previous lower orders. This allows us to identify the \emph{generating} set of invariants which, by definition, does not have any explicit redundancy. To fully remove all {\it explicit} dependence and obtain a {\it generating} set, we introduce a numerical algorithm outlined in App.~\ref{App:AlgGenSet}, which converts the problem of finding polynomial relations to a problem of solving finite system of linear equations. This algorithm has been used in different forms in Refs.~\cite{Trautner:2018ipq,Wang:2021wdq,Bonnefoy:2021tbt,Bonnefoy:2022rik} before. We have improved the algorithm for this work to avoid redundant syzygies\footnote{Redundant syzygies are those which have previously appeared in the algorithm at a lower order in the spurions and are multiplied by another syzygy, which has previously appeared, or some invariant of the \emph{generating} set (or a sum of both), hence reappearing at a higher order. These kind of syzygies evidently do not carry any new information. This is also discussed in Ref.~\cite{Trautner:2018ipq}, where the term ``old relation'' is used.} as we show in detail in App.~\ref{App:AlgGenSet}. By using this method, we are able to generate all possible polynomial relations among the invariants at each degree, which includes both explicit relations and syzygies. Whenever an explicit relation is found, the corresponding invariant is removed from our set. When the algorithm has eliminated all explicit redundancies at a given order, it continues constructing relations at the next order including all the invariants that have survived the algorithm up to this point in its preliminary \emph{generating} set, until the maximum order 26 is reached.

Running this algorithm up to order 26, the invariant list is reduced to a set without explicit dependencies. Our final set includes 459 invariants which form a {\it generating} set of the flavor invariants in the \nuSM. Out of the 459 generating invariants 208 are CP-even and 251 are CP-odd. We want to stress again that all invariants in the theory can be captured by polynomials of these 459 flavor invariants. Hence, any observable in the theory is in principle expressible in terms of these invariants. In the following sections, we will further reduce the {\it generating} set to a {\it primary} set which captures all physical parameters in the theory. Additionally, we aim to reduce the CP-odd \emph{generating} set to a minimal CPC set that captures all necessary and sufficient conditions for CPC in the \nuSM~(c.f. Fig.~\ref{fig:FlowGraphSets}). We provide the full \emph{generating} set in App.~\ref{App:InvariantList}, which is split into CP-even set and CP-odd set. From now on, we will use $\cI_i (\cJ_i)$ to represent the $i$th invariant in the CP-even (CP-odd) set.

\paragraph{Explicit relations and syzygies}
To cross-check if the correct amount of generating invariants was found, we can use the information provided by the multi-graded PL. In order to match the numbers in the PL given by the number of generating invariants minus the number of syzygies at each order, we also have to find all syzygies at a given order. All of this can be achieved by following the process outlined above. Now, we will provide some examples for the polynomial relations we have found, which include both explicit relations and syzygies. Our program scans and checks all terms in the PL from lowest to highest order. Prior to order 12, our reduced invariant set accurately reproduces the terms in the PL. This means that no polynomial relation can be found, and all the invariants in our pre-reduced set are generating invariants. At degree $e^6n^6$ in the spurions, two invariants are found in our pre-reduced set, given by $\cI\equiv\Re\Tr(\neat{X_{N}^{2}X_{e}^{2}X_{N}^{}X_{e}^{}})$ and $\cJ_{10}\equiv \Im\Tr(\neat{X_{N}^{2}X_{e}^{2}X_{N}^{}X_{e}^{}})$, where $X_N$ and $X_e$ are defined in Tab.~\ref{tab:flavorTrafo}. From the multi-graded PL in Eq.~\eqref{eq:gradedPL}, a positive term $+e^6n^6$ is observed. According to the usual understanding of the PL, there should be only one generating invariant at this degree, which means one of the invariants $\cI$ and $\cJ_{10}$ is redundant. This is confirmed by our algorithm, we find that $\cI$ can be written as a polynomial of other lower degree CP-even invariants
\begin{equation}
\begin{split}
6 \cI =\  &\cI_3^3 \cI_1^3-\cI_3 \cI_6 \cI_1^3-3 \cI_3^2 \cI_7 \cI_1^2+3 \cI_3 \cI_{12} \cI_1^2-\cI_3^3 \cI_4 \cI_1+\cI_4 \cI_{11} \cI_1+\\
+&3 \cI_3^2 \cI_{13} \cI_1-3 \cI_3 \cI_{18} \cI_1+\cI_3 \cI_6 \cI_9-\cI_9 \cI_{11}+3 \cI_{12} \cI_{13}+3 \cI_7 \cI_{18}\,.
\end{split}
\end{equation}
With this explicit relation, the CP-even invariant $\cI$ becomes redundant, while no such explicit relation can be found for the CP-odd invariant $\cJ_{10}$; as a result the latter is included in our \emph{generating} set. The same explicit relation was found in the discussion of the quark flavor invariants in Ref.~\cite{Jenkins:2009dy}. The CP-odd invariant $\cJ_{10}$ in the lepton sector is analogous to the Jarlskog invariant $I_{6,6}^{(-)}$ in the quark sector as shown in Eq.~\eqref{eq:quark_inv}, and is the only CP-odd generating invariant in the \nuSM that has no dependence on $M_N$. Note that $I_{6,6}^{(-)}=2i\,\Im\Tr(\neat{X_{u}^{2}X_{d}^{2}X_{u}^{}X_{d}^{}})$, while in our notation the additional factor ``$2i$'' is omitted.

Running our algorithm up to order 14, we get exactly the same number of generating invariants as indicated by the positive terms in the multi-graded PL. However, we run into a mismatch at degree $m^8n^8$. There are 10 invariants from the pre-reduced set, but we only find 9 explicit relations. As a result, we are left with one invariant $\cJ_{76}$ that can not be written as any polynomial of other invariants and by definition, it should be identified as a {\it generating} invariant. But looking at the multi-graded PL in Eq.~\eqref{eq:gradedPL}, one does not find a positive term at degree $m^8 n^8$ and would naively believe that there are no generating invariants at this degree. However, as we mentioned in Sec.~\ref{sec:HS}, there can be non-trivial cancellations between the number of generating invariants and syzygies. Indeed, we also find another relation at degree $m^8 n^8$, which is given by
\begin{align*}
&~~~3 \cI_5^2 \cI_3^4-3 \cI_2 \cI_{10} \cI_3^4-8 \cI_2^3 \cI_8 \cI_3^3+12 \cI_2 \cI_5 \cI_8 \cI_3^3+8 \cI_8 \cI_{10} \cI_3^3+12 \cI_2^2 \cI_{16} \cI_3^3-24 \cI_5 \cI_{16} \cI_3^3+\\
&+30 \cI_2^2 \cI_8^2 \cI_3^2-18 \cI_5 \cI_8^2 \cI_3^2+36 \cI_{16}^2 \cI_3^2-6 \cI_5^2 \cI_6 \cI_3^2+6 \cI_2 \cI_6 \cI_{10} \cI_3^2+8 \cI_2^3 \cI_{14} \cI_3^2-12 \cI_2 \cI_5 \cI_{14} \cI_3^2+\\
&-8 \cI_{10} \cI_{14} \cI_3^2+10 \cI_2^3 \cI_{15} \cI_3^2-18 \cI_2 \cI_5 \cI_{15} \cI_3^2-4 \cI_{10} \cI_{15} \cI_3^2-60 \cI_2 \cI_8 \cI_{16} \cI_3^2-12 \cI_2^2 \cI_{20} \cI_3^2+\\
&+24 \cI_5 \cI_{20} \cI_3^2-24 \cI_2^2 \cI_{21} \cI_3^2+24 \cI_5 \cI_{21} \cI_3^2+24 \cI_2 \cI_{28} \cI_3^2-24 \cI_2 \cI_8^3 \cI_3+8 \cI_2^3 \cI_6 \cI_8 \cI_3+\numberthis\\
&-12 \cI_2 \cI_5 \cI_6 \cI_8 \cI_3-8 \cI_6 \cI_8 \cI_{10} \cI_3-24 \cI_2^2 \cI_8 \cI_{14} \cI_3+24 \cI_5 \cI_8 \cI_{14} \cI_3-48 \cI_2^2 \cI_8 \cI_{15} \cI_3+24 \cI_5 \cI_8 \cI_{15} \cI_3+\\
&+48 \cI_8^2 \cI_{16} \cI_3-12 \cI_2^2 \cI_6 \cI_{16} \cI_3+24 \cI_5 \cI_6 \cI_{16} \cI_3+24 \cI_2 \cI_{14} \cI_{16} \cI_3+48 \cI_2 \cI_{15} \cI_{16} \cI_3-16 \cI_2^3 \cI_{19} \cI_3+\\
&+24 \cI_2 \cI_5 \cI_{19} \cI_3+16 \cI_{10} \cI_{19} \cI_3+48 \cI_2 \cI_8 \cI_{20} \cI_3-48 \cI_{16} \cI_{20} \cI_3+48 \cI_2 \cI_8 \cI_{21} \cI_3-48 \cI_{16} \cI_{21} \cI_3+\\
&+48 \cI_2^2 \cI_{27} \cI_3-48 \cI_5 \cI_{27} \cI_3-48 \cI_8 \cI_{28} \cI_3-48 \cI_2 \cI_{39} \cI_3+3 \cI_5^2 \cI_6^2-6 \cI_2^2 \cI_6 \cI_8^2+6 \cI_5 \cI_6 \cI_8^2+6 \cI_2^2 \cI_{14}^2+\\
&-6 \cI_5 \cI_{14}^2+24 \cI_2^2 \cI_{15}^2-12 \cI_5 \cI_{15}^2-12 \cI_6 \cI_{16}^2+24 \cI_{20}^2+12 \cI_{21}^2-3 \cI_2 \cI_6^2 \cI_{10}+12 \cI_2 \cI_8^2 \cI_{14}+\\
&+24 \cI_2 \cI_8^2 \cI_{15}-10 \cI_2^3 \cI_6 \cI_{15}+18 \cI_2 \cI_5 \cI_6 \cI_{15}+4 \cI_6 \cI_{10} \cI_{15}+12 \cI_2 \cI_6 \cI_8 \cI_{16}-24 \cI_8 \cI_{14} \cI_{16}+\\
&-24 \cI_8 \cI_{15} \cI_{16}+24 \cI_2^2 \cI_8 \cI_{19}-24 \cI_5 \cI_8 \cI_{19}-24 \cI_2 \cI_{16} \cI_{19}-24 \cI_8^2 \cI_{20}+12 \cI_2^2 \cI_6 \cI_{20}-24 \cI_5 \cI_6 \cI_{20}+\\
&-48 \cI_2 \cI_{15} \cI_{20}-24 \cI_2 \cI_{14} \cI_{21}+6 \cI_2^3 \cI_{26}-6 \cI_2 \cI_5 \cI_{26}-12 \cI_{10} \cI_{26}-48 \cI_2 \cI_8 \cI_{27}+48 \cI_{16} \cI_{27}+\\
&+24 \cI_{14} \cI_{28}-24 \cI_2^2 \cI_{38}+24 \cI_5 \cI_{38}+48 \cI_8 \cI_{39}+24 \cI_2 \cI_{55}+12 \cJ_1^2=0\,,
\end{align*}
This polynomial relation indicates a redundancy of the CP-odd invariant squared $\cJ_{1}^2$ -- not an explicit redundancy of an invariant -- and therefore by definition it is a syzygy. Hence, combining the two findings we have
\begin{equation}
    \PL\(e,m,n\) \supset + m^8 n^8 - m^8 n^8 \, ,
\end{equation}
explaining the non-trivial zero in the PL. At higher degrees in the spurions even stronger cancellations appear, that can even generate negative terms at orders where generating invariants may exist. Some examples are
\begin{equation}
\label{eq:cancellation_example}
\begin{split}
    \PL\(e,m,n\) \supset &(21-1)\,e^4 m^4 n^8 + (3-1)\,e^2 m^4 n^{12}
    + (3-6)\,e^2 m^6 n^{10} +(2-6)\,e^2 m^8 n^{8}\, ,
\end{split}
\end{equation}
where we have used $(n_g-n_s)$ as a coefficient to indicate the $n_g$ generating invariants and $n_s$ syzygies at the corresponding degree. 

Although there may be challenges when identifying generating invariants and syzygies, as long as the terms in PL are correctly interpreted, we are able to determine the correct number of the generating invariants and syzygies at each degree. Our algorithm accurately generates the terms in PL up to order 24 based on the updated description of the PL. However, at order 26, there are some mismatches due to ``redundant syzygies'', which are not simply products of syzygies that appeared at a lower order in the spurions. We provide solutions to address these mismatches. For more details, please refer to App.~\ref{App:AlgGenSet}. The order 26 is the first order that only has negative terms in the multi-graded PL. We have also confirmed that there are no further generating invariants at this order, as all the invariants are found to be a polynomial of lower order invariants.

Finally, our program successfully generates the correct number of generating invariants and syzygies at each degree of $[emn]$, reproducing the coefficients of the multi-graded PL shown in Eq.~\eqref{eq:gradedPL}. We have summarized the number of generating invariants and syzygies, as well as the CP-even and CP-odd counting at each degree in Tabs.~\ref{tab:graded_count_1} and \ref{tab:graded_count_2}. In the new interpretation, the PL should be presented in a new form taking into account the non-trivial cancellations. The coefficient of each term should be replaced with $(n_g-n_s)$ as shown and described around Eq.~\eqref{eq:cancellation_example}. According to the new multi-graded PL, one should be able to read off the correct number of generating invariants and syzygies at each degree easily. Due to its length, we will not show it here. Furthermore, we summarize the counting information at each ungraded order in Tab.~\ref{tab:ungraded_count}. With this table, the ungraded PL shown in Eq.~\eqref{eq:HS_PL} can also be revised to the new form
\begin{eqnarray}
\PL[\cH(q)] =&(3-0)\,q^{2}+(5-0)\,q^{4}+(9-0)\,q^{6}+(10-0)\,q^{8}+(19-0)\,q^{10}+(40-0)\,q^{12}\nonumber\\
&+(66-0)\,q^{14}+(94-2)\,q^{16}+(102-32)\,q^{18}+(76-200)\,q^{20}\\
&+(30-733)\,q^{22}+(5-2044)\,q^{24}+(0-4391)\,q^{26}+ O(q^{28})\,.\nonumber
\end{eqnarray}
One can find that up to order 26, the coefficients are exactly the same as those shown in Eq.~\eqref{eq:HS_PL}.

\begin{table}[t]
\centering
\setlength{\tabcolsep}{4pt}
\begin{tabular}{|c|c|ccccccccccccc|c|}
\hline
\multicolumn{2}{|c|}{Order} & $2$ & $4$ & $6$ & $8$ & $10$ & $12$ & $14$ & $16$ & $18$ & $20$ & $22$ & $24$ & $26$ & Total\\\hline\hline
\multirow{2}{*}{Generating} & CP-even & $3$ & $5$ & $9$ & $8$ & $12$ & $17$ & $25$ & $33$ & $41$ & $34$ & $17$ & $4$ & $0$ & $208$
\\\cline{2-16}
& CP-odd & $0$ & $0$ & $0$ & $2$ & $7$ & $23$ & $41$ & $61$ & $61$ & $42$ & $13$ & $1$ & $0$ & $251$ \\\hline
\multicolumn{2}{|c|}{Syz.} & $0$ & $0$ & $0$ & $0$ & $0$ & $0$ & $0$ & $2$ & $32$ & $200$ & $733$ & $2044$ & $4391$ & $7402$ \\\hline
\multicolumn{2}{|c|}{PL$=$Generating$-$Syz.} & $3$ & $5$ & $9$ & $10$ & $19$ & $40$ & $66$ & $92$ & $70$ & $-124$ & $-703$ & $-2039$ & $-4391$ & $-6943$ \\\hline
\end{tabular}
\caption{The number of generating invariants and syzygies from order 2 to order 26, where the generating invariants are split into CP-even and CP-odd in the counting. The difference between the number of generating invariants and number of syzygies precisely aligns with the terms in the ungraded PL shown in Eq.~\eqref{eq:HS_PL}. In the last column, we list the total number of CP-even, CP-odd generating invariants, syzygies, and their difference. In the complete intersection ring, the difference between the number of generating invariants and number of syzygies should be the Krull dimension, which is 21 in our theory. The negative number shown here featuring a non-complete intersection ring.}
\label{tab:ungraded_count}
\end{table}

At higher orders, the coefficients in the PL can no longer be connected to meaningful quantities like the number of generating invariants or the number of syzygies at a given degree~\cite{Benvenuti:2006qr, Wang:2021wdq}. For instance, at order 28, we can find two positive terms
 \begin{equation}
 \label{eq:PL_order28}
    \PL\(e,m,n\) \supset +6 m^{14} n^{14} + 4 m^{16} n^{12} \, .
\end{equation}
However, all invariants constructed by brute force at these two degrees are redundant after applying the Cayley--Hamilton theorem. Therefore, there is no generating invariant, and these two positive terms must be misleading.\footnote{The only meaning we can give to the coefficients of these terms is due to the Euler form we mentioned in footnote \ref{foot:EulerForm}.} The scanning program should terminate at the first pure negative order~\cite{Wang:2021wdq}, and we can be assured that we get a complete and minimal set of generating invariants. It is worth noting that identifying syzygies at each order is not necessary to obtain the \emph{generating} set. The identification of explicit relations is sufficient for this purpose. Counting correct number of syzygies adds much complexity to our analysis, and it is only necessary to verify that we have found the correct number of generating invariants as indicated by the PL.

\subsection{A primary set for the \texorpdfstring{\nuSM}{nuSM}} \label{Sec:PrimSet}
To reduce the {\it generating} set of invariants to a {\it primary} set that captures all physical parameters of the theory, we can use the denominator of the HS as a guide.\footnote{In addition, the Hironaka decomposition (c.f. App.~\ref{app:Hironaka}) documented in mathematical literature proves to be useful. The significance of the Hironaka decomposition lies in its ability to simplify the analysis of the invariants in the ring, particularly in scenarios where primary invariants are CP-even and secondary invariants are CP-odd. This enables a more focused investigation of the linear span of CP-odd secondary invariants, providing necessary and sufficient conditions for CP conservation~\cite{Trautner:2018ipq}. Once a set of primary and secondary invariants is chosen, there will be a unique decomposition of any invariant in the theory in terms of the primary and secondary invariants in the form of Eq.~\eqref{eq:Hironaka}. This decomposition can be used to check if any invariant can be written in the form dictated by it, which indicates if the choice of primary and secondary invariants is a proper choice. However, according to Eq.~\eqref{eq:num} and Eq.~\eqref{eq:HironakaHS}, identifying all secondary invariants requires scanning the invariants up to order 114, which poses a significant challenge in our theory. Therefore, we will not attempt to present the invariants corresponding to the Hironaka decomposition.} The \emph{primary} set is defined as a set of algebraically independent invariants. These invariants will capture all physical parameters in the theory. In order to reduce redundancies from the beginning, we choose the candidate sets with cardinalities equal to 21, which is the number of physical parameters in the case of the \nuSM. One way to check if a candidate set is algebraically independent is to calculate the Jacobian with respect to all parameters in a given parameterization of the Lagrangian, for which we will use the parameterization from Eq.~\eqref{eq:parameterization_3gen_new}. If the rank is equal to the number of physical parameters in the theory, a set of algebraically independent invariants is found. Following this procedure we find the following \emph{primary} set of invariants 
\begin{align*}
\label{eq:PrimSetnuSM}
\cI_{1}&=\,\Tr\left(X_e\right),~\cI_{2}=\,\Tr\left(X_M\right),~\cI_{3}=\,\Tr\left(X_N\right),~\cI_{5}=\,\Tr\left(X_M^2\right),~\cI_{6}=\,\Tr\left(X_N^2\right),\\
\cI_{7}&=\,\Tr\left(X_e X_N\right),~\cI_{8}=\,\Tr\left(Z_{MN}\right),~\cI_{9}=\,\Tr\left(X_e^3\right),~\cI_{12}=\,\Tr\left(X_eX_N^2\right),~\cI_{13}=\,\Tr\left(X_e^2X_N\right),\\
\cI_{15}&=\,\Tr\left(X_N Z_{MN}\right),~\cI_{23}=\,\Re\Tr\left( X_e X_N Z_{MN} \right),~\cI_{25}=\,\Tr\left(X_e^2 Z_{MN}\right),\\
\cI_{34}&=\,\Tr\left(\neat{X_{e}^{2} Y_{N}^{} M_{N}^{*} Y_{N}^{T} Y_{N}^{*} M_{N}^{} Y_{N}^{\dagger}}\right),~\cI_{35}=\,\Tr\left(\neat{X_{e}^{} Y_{N}^{} M_{N}^{*} Y_{N}^{T} X_{e}^{*} Y_{N}^{*} M_{N}^{} Y_{N}^{\dagger}}\right),\numberthis\\
\cI_{47}&=\,\Tr\left(\neat{X_{e}^{2} Y_{N}^{} M_{N}^{*} Y_{N}^{T} X_{N}^{*} Y_{N}^{*} M_{N}^{} Y_{N}^{\dagger}}\right),~\cI_{50}=\,\Re\Tr\left(\neat{X_{e}^{} X_{N}^{} Y_{N}^{} M_{N}^{*} Y_{N}^{T} X_{e}^{*} Y_{N}^{*} M_{N}^{} Y_{N}^{\dagger}}\right),\\
\cI_{54}&=\,\Tr\left(\neat{X_{e}^{2} Y_{N}^{} M_{N}^{*} Y_{N}^{T} X_{e}^{*} Y_{N}^{*} M_{N}^{} Y_{N}^{\dagger}}\right),~\cI_{65}=\,\Re\Tr\left(\neat{X_{e}^{2} X_{N}^{2} Y_{N}^{} M_{N}^{*} Y_{N}^{T} Y_{N}^{*} M_{N}^{} Y_{N}^{\dagger}}\right),\\
\cI_{79}&=\,\Tr\left(\neat{X_{e}^{2} Y_{N}^{} M_{N}^{*} Y_{N}^{T} X_{e}^{*2} Y_{N}^{*} M_{N}^{} Y_{N}^{\dagger}}\right),~\cI_{91}=\,\Re\Tr\left(\neat{X_{e}^{2} X_{N}^{2} Y_{N}^{} M_{N}^{*} M_{N}^{} M_{N}^{*} Y_{N}^{T} Y_{N}^{*} M_{N}^{} Y_{N}^{\dagger}}\right) \, ,
\end{align*}
where $\cI_i$ is the $i$th invariant in the CP-even set shown in App.~\ref{App:InvariantList}, and we have defined $X_e= Y_e Y_e^{\dagger},X_N= \neat{Y_{N}^{} Y_{N}^{\dagger}},X_M=\neat{M_{N}^{} M_{N}^{*}}$ and $Z_{MN} = \neat{Y_{N}^{} M_{N}^{*} M_{N}^{} Y_{N}^{\dagger}}$. We want to stress here that the set of algebraically independent invariants is not unique. In particular, there are many sets that are compatible with the denominator of the ungraded HS and have a Jacobian rank of 21. 

Furthermore, we have only chosen CP-even invariants from the \emph{generating} set to form the \emph{primary} set. However, it is also possible to include CP-odd invariants, provided they are algebraically independent. For instance, some CP-odd invariants are included in the \emph{primary} set of the seesaw effective field theory in Ref.~\cite{Yu:2022ttm}. It might sound strange at first that CP-even invariants can be used to capture the physical phases in the theory. However, the primary invariants only capture the absolute value of the physical parameters. In particular, the sign of the phases in the theory are \emph{not} described by the primary invariants. This is a well-known result in the quark sector of the SM~\cite{Jenkins:2009dy}, where the Jarlskog invariant has to be added to the set of primary invariants to complete the \emph{generating} set of invariants, enabling the expression of all observables in the theory in terms of flavor invariants.\footnote{As already mentioned in footnote~\ref{foot:SMInvs}, the square of the Jarlskog invariant, i.e., the square of the sign of the CKM phase, is in turn expressible in terms of CP-even invariants in agreement with the statement that the square of a sign is trivial.}

In addition, we have chosen the orders of the invariants in our \emph{primary} set to follow those of the denominator of the HS in Eq.~\eqref{eq:deno}. For instance, the first term $(1-q^2)^3$ of the denominator in Eq.~\eqref{eq:deno} indicates that there should be 3 invariants of order 2. However, following our discussion in footnote~\ref{foot:ambiguity}, one can change the numbers in the denominator of the HS by multiplying the numerator and denominator of the HS with the same factor $(1+q^k)^m$. In this case, more algebraically independent subsets of the \emph{generating} set with cardinality 21 can function as a \emph{primary} set.

\section{The seesaw limit}
\label{sec:seesaw_limit}
A phenomenologically interesting limit of the \nuSM is the seesaw limit where $M_N$ is taken to be much larger than the electroweak scale $v$, allowing for an EFT description of the model. In this limit, the heavy neutrinos are integrated out, resulting in a matching between the low energy theory and the UV theory. The seesaw limit can be captured by the invariants in this paper by making the observation that $M_N^{-1}$, appearing in the EFT description, transforms in the same way as $M_N^*$. Hence, by replacing $M_N^* \to M_N^{-1}$ in all invariants, we can analyze this limit. After the replacement there are several structures appearing in the invariants which correspond to different orders in the EFT. In particular, we have~\cite{deBlas:2017xtg,Elgaard-Clausen:2017xkq}
\begin{equation} \label{eq:SeesawMatching}
    \begin{split}
        \frac{C_5}{\Lambda} & \sim \neat{Y_{N}^{} M_{N}^{-1} Y_{N}^{T}}\,, \\
        \frac{C_6}{\Lambda^2} & \sim \neat{Y_{N}^{} (M_{N}^{*} M_{N}^{})^{-1} Y_{N}^{\dagger}} \,, \\
        \frac{C_7}{\Lambda^3} & \sim \neat{Y_{N}^{} (M_{N}^{} M_{N}^{*} M_{N}^{})^{-1} Y_{N}^{T}} \,, \\
        & \qquad\qquad\vdots \\
    \end{split}
\end{equation}
where $C_5$, $C_6$ and $C_7$ are the Wilson coefficients of the Weinberg operator~\cite{Weinberg:1979sa}, the operator $\cO_{Hl}^{(1)}$ in the Warsaw basis~\cite{Grzadkowski:2010es} and the operator $\cO_{lHD}^{(2)}$ in Ref.~\cite{Lehman:2014jma}, respectively.\footnote{Note, that at dimension 6 (and also at higher orders), there are other operators that can appear in the matching. $\cO_{Hl}^{(1)}$ at dimension-6 and $\cO_{lHD}^{(2)}$ at dimension-7 are only examples of an operator that receives a contribution from the matching of the \nuSM to the SMEFT~\cite{Elgaard-Clausen:2017xkq}. However, our choice of Wilson coefficients ensures the lowest overall suppression of the invariants with the Majorana mass.}

As before, it is insightful to construct the invariant graphs to understand more easily which kind of structures can appear in the flavor invariants. The relevant graph for this limit is show in Fig.~\ref{fig:invariantGraphEFT}. Here, we show the graph for two types of the Wilson coefficients $C_{LL^T}$ and $C_{LL^{\dagger}}$ as an example, where $C_{LL^T}\to U_LC_{LL^T}U_L^T$ and $C_{LL^\dagger}\to U_LC_{LL^\dagger}U_L^\dagger$ respectively. For instance, $C_{LL^T}\sim C_5, C_7$ and $C_{LL^{\dagger}}\sim C_6$. This shows that the graph approach serves as a general method for constructing flavor invariants and can be employed effectively in both UV and EFT studies. Furthermore, the graph in the EFT allows for a clearer understanding of the structure of the invariants. For instance, when constructing invariants for the EFT with the Wilson coefficient $C_5$, a closed walk in the graph indicates that $C_5^*$ must also be included. Following this conclusion, it becomes clear that all constructed invariants must adhere to an even power counting.

\begin{figure}[t]
  \centering
\begin{tikzpicture}[every circle node/.style={draw, fill=blue!10, inner sep=0pt, minimum size=3.5mm},  every path/.style={-{Stealth[length=2.5mm]}}]
  \node[circle, label={[label distance=0.5mm]$e$}] (e) at (0,0) {{\footnotesize 1}};
  \node[circle, label={[label distance=0.5mm]$L$}] (L) at (2.5,0) {{\footnotesize 2}};
  \node[circle, label={[label distance=0.5mm]$L^*$}] (Ls) at (5,0) {{\footnotesize 3}};
  \node[circle, label={[label distance=0.5mm]$e^*$}] (es) at (7.5,0) {{\footnotesize 4}};

  \draw (L) to[bend left=-45] node[above]{\textcolor{blue}{$Y_e$}} (e);
  \draw (e) to[bend left=-45] node[above]{\textcolor{blue}{$Y_e^\dagger$}} (L);
  \draw[draw=green!70!black, loop above] (L) to [out=55, in=125, distance=2cm] node[above] {\textcolor{green!70!black}{$C_6$}} (L);
  \draw[draw=green!70!black, loop above] (Ls) to [out=55, in=125, distance=2cm] node[above] {\textcolor{green!70!black}{$C_6^*$}} (Ls);
  \draw[red] (L) to[bend left=-45] node[above]{\textcolor{red}{$C_5,C_7$}} (Ls);
  \draw[red] (Ls) to[bend left=-45] node[above]{\textcolor{red}{$C_5^*,C_7^*$}} (L);
  \draw (Ls) to[bend left=-45] node[above]{\textcolor{blue}{$Y_e^*$}} (es);
  \draw (es) to[bend left=-45] node[above]{\textcolor{blue}{$Y_e^T$}} (Ls);
\end{tikzpicture}
\caption{The flavor invariant graph that can be used to construct all possible single trace flavor invariants in the seesaw limit of the \nuSM, where $C_5$, $C_6$ and $C_7$ are the Wilson coefficients of the Weinberg operator~\cite{Weinberg:1979sa}, the operator $\cO_{Hl}^{(1)}$ in the Warsaw basis~\cite{Grzadkowski:2010es} and the operator $\cO_{lHD}^{(2)}$ in Ref.~\cite{Lehman:2014jma}, respectively, and they transform as $C_{5,7}\to \neat{U_{L}^{} C_{5,7}^{} U_{L}^{T}}$ and $C_6\to \neat{U_{L}^{} C_{6}^{} U_{L}^{\dagger}}$. We show in the main text that it is sufficient to consider $C_5$, $C_6$ and $C_7$ to capture all parameters of the full theory.}
\label{fig:invariantGraphEFT}
\end{figure}

We can make several interesting observations using our generating invariants. Keeping the invariants with only the structures of $C_5$ and $Y_e$ and calculating the Jacobian of the invariants with respect to all parameters in a chosen parameterization, we find that its rank is 12. This corresponds to the six masses, three mixing angles and three phases that appear in the low-energy theory of three flavors of neutrinos with a Majorana mass term $\cL \supset - 1/2 \, \bar{\nu}_L m_{\nu} \nu_L^c + \text{H.c.}$. There are a total of 15 invariants with only $C_5$ and $Y_e$ structure, we show the 12 algebraically independent invariants as follows:
\begin{equation}
\cS_{\{C_5\}}^{12} \equiv \{\cI'_{1},\, \cI'_{4},\, \cI'_{9},\, \cI'_{14},\, \cI'_{22},\, \cI'_{34},\, \cI'_{35},\, \cI'_{54},\, \cI'_{79},\, \cJ'_{101},\, \cJ'_{168},\, \cJ'_{221}\}\,,
\end{equation}
where $\cI'_i(\cJ'_i)$ can be obtained from our generating invariants $\cI_i(\cJ_i)$ (refer to App.~\ref{App:InvariantList}) by the replacement $M_N^*\to M_N^{-1}$. We introduce the notation $\cS_{c}^{r}$ to indicate that the EFT only involves the Wilson coefficients in the list $c$, and the corresponding theory has a rank-$r$ primary invariant set.

By also keeping the invariants with the structures corresponding to $C_6$ in Eq.~\eqref{eq:SeesawMatching}, we find that the Jacobian of the invariants has rank 21, corresponding to the number of physical parameters in the \nuSM. We choose the \emph{primary} set to be
\begin{equation}
\cS_{\{C_5,C_6\}}^{21} \equiv \cS_{\{C_5\}}^{12} \cup \{\cI'_{8},\, \cI'_{17},\, \cI'_{25},\, \cJ'_{20},\, \cJ'_{58},\, \cJ'_{62},\, \cJ'_{124},\, \cJ'_{125},\, \cJ'_{189}\}\,.
\end{equation}
This has previously been noted in Ref.~\cite{Yu:2022ttm}, where this analysis is performed in the effective theory of the seesaw model. They claim that the physical information about the full theory can be completely covered by the EFT with only two operators corresponding to the Wilson coefficient $C_5$ and $C_6$. Here, we come to the same conclusion in the full theory of the \nuSM.

Note that, up to now we have not considered the power counting of the effective theory, i.e., we also allowed for invariants with several insertions of the $C_5$ structure to obtain the ranks of 12 and 21. On the other hand, working in the effective theory by taking the seesaw limit, one should work consistently to a certain order in the power counting of the theory, which in this case is defined by the characteristic scale of the process divided by the Majorana mass. By counting the number of insertions of $M_N$, we find that one cannot reach rank 12 and 21 at dimension 5 and 6, respectively. Instead, without any insertions of $M_N$, one can reach rank three, corresponding to the masses of the charged leptons. The \emph{primary} set is simply formed by three flavor invariants that only involve $Y_e$:
\begin{equation}
\tilde{\cS}_{4}^{3} \equiv \{\cI'_{1},\, \cI'_{4},\, \cI'_{9}\}\,,
\end{equation}
where $\tilde{\cS}_{d}^{r}$ is the rank-$r$ \emph{primary} set of invariants at mass dimension $d$. At mass dimension 6 (with 2 insertions of $M_N$), rank 12 can be reached, and the corresponding \emph{primary} set is given by:
\begin{equation}
\tilde{\cS}_{6}^{12} \equiv \tilde{\cS}_{4}^{3} \cup \{\cI'_{8},\, \cI'_{14},\, \cI'_{17},\, \cI'_{22},\, \cI'_{25},\, \cI'_{34},\, \cI'_{35},\, \cI'_{54},\, \cI'_{79}\}\,.
\end{equation}
At dimension 8 (with 4 insertions of $M_N$), one can reach rank 21, which correspond to the \emph{primary} set:
\begin{equation}
\tilde{\cS}_{8}^{21} \equiv \tilde{\cS}_{6}^{12} \cup \{\cI'_{16},\, \cI'_{21},\, \cI'_{24},\, \cI'_{32},\, \cI'_{37},\, \cI'_{51},\, \cI'_{53},\, \cI'_{78},\, \cI'_{112}\}\,.
\end{equation}
This set also includes invariants with the structure of $C_7$. Hence, even though all the information about the \nuSM is in principle contained in the Wilson coefficients $C_5$ and $C_6$, this information can only be accessed at higher mass dimensions. Considering the power counting of the theory, full rank is first reached at dimension 8 by also considering the structure of $C_7$ in the invariants instead of going to higher mass dimensions beyond dimension 8 by only including the structures of the Wilson coefficients $C_5$ and $C_6$. We have summarized the results in Fig.~\ref{fig:FlowGraphSets}.

%%%%%%%%%%%%%%%%%%%%%%%%%%%%%%%%%%%%%%%%%%%%%%%%%%%%%%%%%%%%%%%%%%%%
\section{Conditions for CP conservation}\label{sec:CPV_condition}

The invariant structures of the \nuSM have been presented in various papers for different numbers of lepton flavors. In Ref.~\cite{Jenkins:2009dy}, the \emph{generating} set for the \nuSM with only two generations of right-handed neutrinos and charged leptons ($n_N=n_f=2$) has been shown, but the discussion of CPC conditions has not been expanded on. In Ref.~\cite{Yu:2021cco}, the minimal seesaw model with two generations of right-handed neutrinos and three generations of charged leptons ($n_N=2, n_f=3$) has been discussed. However, the discussion of CPC conditions is provided only with some assumptions about the parameter spectrum. For the three-generation case ($n_N=n_f=3$), the discussion of CPC conditions is still lacking, as the \emph{generating} set has not been constructed, although the Hilbert series has been calculated in Refs.~\cite{Hanany:2010vu,Yu:2022ttm}. In Ref.~\cite{Branco:2001pq}, without using the Hilbert series and explicit construction of the \emph{generating} set, six CP-odd invariants are provided to characterize CPV effects of the three generation case. Also there, assumptions are made regarding the spectrum of the theory. 

The spirit of using CP-odd flavor invariants to determine CPC conditions is that they can capture all possible CPC conditions of the theory, which not only include the vanishing phases, but also conditions that lead to unphysical phases, such as degenerate masses and texture zeros. This paradigm is well-established in the quark sector of the SM, where the Jarlskog invariant captures all such conditions leading to the CPC of the SM. In this paper, our objective is to identify a set of Jarlskog-like invariants capturing all possible CPC conditions of the \nuSM.

It should be noted, that without the guidance of the invariants, it is not easy to derive all CPC conditions of a theory, as some special conditions may appear that are beyond the conditions as we mentioned above. We show the complete list of CPC conditions for the \nuSM with $n_N=n_f=2$ in App.~\ref{App:CPC2Gen}, where in one particularly interesting case a scenario with {\it pseudo-real} couplings\footnote{By pseudo-real couplings, we denote a set of couplings which have irremovable phases but still conserve CP. This is the case in models with discrete symmetries, when the effect of a CP transformation on the Lagrangian parameters can be undone by a field redefinition, even if there exists no basis where all couplings can be made real. This has been previously noted in the context of Three Higgs Doublet Models~\cite{Ivanov:2015mwl} and toy models with more involved discrete symmetry groups~\cite{Trautner:2016ezn}. We show an explicit example for the \nuSM with two generations in App.~\ref{App:CPC2Gen}.} is found. As a result, studying CPC conditions through the generating CP-odd invariants provides a more reliable approach. In the following, we will define the minimal invariant set that captures the CPC conditions, and propose useful methods to find it. However, due to the complexity of the invariant structure of the \nuSM, finding the proposed minimal set in general can be challenging. The main goal of this section is to establish a framework for studying the CPC conditions, with the final solutions to these sets left for future work.

The CP-odd invariants in the \emph{generating} set $\cS_{\text{gen}}$ act as generators of all CPV observables. When these invariants vanish, it establishes the necessary and sufficient conditions for CP conservation. However, this does not mean that the CP-odd \emph{generating} set is the minimal set characterizing this property. The CP-odd \emph{generating} set is required to generate any value of the CP-odd invariants in a given parameterization, while the set that can determine if CPV exists only captures their roots. In principle, the latter set should be a subset of the CP-odd \emph{generating} set, and from now on, we will call it the CPC set. If the invariants in the CPC set is reduced to the minimal number, the CPC set becomes the minimal CPC set $\cS_{\text{min}}^{\text{CPC}}$.

By definition, the minimal CPC set is sufficient enough to generate all roots corresponding to the CPC conditions, all other CP-odd invariants should vanish automatically on these roots. It is possible to find a minimal CPC set, such that
\begin{equation}\label{eq:CPVsetDef}
    \cJ_{\text{min}} = 0, \, \forall \cJ_{\text{min}} \in \cS_{\text{min}}^{\text{CPC}} \Longrightarrow \cJ = 0, \, \forall \cJ \in \cS_{\text{gen}} \backslash \cS_{\text{min}}^{\text{CPC}}.
\end{equation}
The straightforward method one can try is to solve the common zeros of polynomials in a candidate minimal CPC set,\footnote{We have done this for the \emph{generating} set of the \nuSM ($n_N=n_f=2$) with the package \texttt{Macaulay2}~\cite{M2}, for which the complete set of CPC conditions are listed in App.~\ref{App:CPC2Gen}. However, the common zeros of the subset of the generating invariants are still difficult to solve, thus we can not determine the minimal CPC set.} and subsequently apply the solutions to the other CP-odd invariants to check whether they will vanish. However, this is not practical for complicated polynomials. Without directly solving the polynomial equations, one can also use the syzygies to determine whether other invariants are automatically zero given that all invariants in a minimal set are set to zero. This approach was e.g. followed in Ref.~\cite{Trautner:2018ipq}, where the author found some syzygies that can determine the minimal CPC set in the 2HDM. In this section, we will show that there is a specific form of the syzygy that can help to determine whether an invariant will vanish.

The general problem concerning the common zeros of polynomials is closely connected to Hilbert's Nullstellensatz~\cite{eisenbud2013commutative,atiyah2018introduction,Cox:2015ode}, a theorem that establishes a fundamental relationship between geometry and algebra. We have presented this theorem and relevant mathematical terms in App.~\ref{App:theorem}. In this section, we will employ Hilbert's Nullstellensatz to reframe the problem of identifying the minimal CPC set. In the invariant ring, or more generally in the polynomial ring $R:=\mathbb{Q}[x_1,\dots,x_n]$, where $x_{1,\dots,n}$ are the parameters in the theory. The CP-odd generating invariants are polynomials in this ring. Let $\cJ_s,\cJ_1,\dots,\cJ_m\in R$, Hilbert's Nullstellensatz says that if an invariant $\cJ_s$ vanishes on all the common zeros of the $\cJ_{1,\dots,m}$, then there exist some integer $t$, such that $\cJ_s^t$ is a subset of the ideal $I:=(\cJ_1,\dots,\cJ_m)$, i.e.,
\begin{equation}
  \label{eq:jpower}
  \cJ_s^t=f_1 \cJ_1+\dots+f_m \cJ_m\,,\quad f_i\in R\,,
\end{equation}
where $f_i$ are the ring elements, i.e., they are polynomials of the parameters. However, since $\cJ_s$ and $\cJ_{1,\dots,m}$ are elements in the invariant ring, $f_i$ should also be invariants. They can be parameterized by the generating invariants in a polynomial form
\begin{equation}
\label{eq:jpower_new}
\cJ_s^t=\sum_{i=1}^{m}P_i(\cJ_l,\cI_k)\cJ_i\,,\quad i\neq s\,.
\end{equation}
Therefore, Hilbert's Nullstellensatz tells us that if a CP-odd invariant $\cJ_s$ is redundant in the presence of a given CPC set, there must exist a syzygy of some power of $\cJ_s$ that can be used to eliminate this invariant. This theorem is quite helpful when using the syzygy approach. For example, in Ref.~\cite{Trautner:2018ipq}, the syzygies are used to identify the CPC conditions in the 2HDM, one can find that the syzygies regarding the vanishing invariant $\cJ_s$ can be obtained at order $\cJ_s^2$, there is no need to discuss the spectrum of the syzygies as presented in their analysis.

It is possible to come up with an elimination algorithm based on the Hilbert's Nullstellensatz to find the minimal CPC set. However, the problem of finding a syzygy like Eq.~\eqref{eq:jpower_new}, or in more mathematical language, determining whether an ideal $\cJ_s^t$ is a subset of another ideal $I:=(\cJ_1,\dots,\cJ_m)$ highly relies on the calculation of the Gr{\"o}bner basis, which is computationally quite expensive in complicated polynomial rings. The undetermined power $t$ also introduces a lot of complexity in this problem. There are different methods and packages that are suitable for this problem. As outlined in App.~\ref{App:AlgGenSet}, the syzygy problem can also be converted to a finite system of linear equations, which can be addressed using standard linear algebra techniques. However, some high-degree syzygies lead to very complex linear systems that can not be easily solved. Some software systems are devoted to studying the algebraic geometry and commutative algebra, such as \texttt{Macaulay2}~\cite{M2} and \texttt{Singular}~\cite{DGPS}. They have proved helpful when exploring the algebraic structures of simple theories. In addition, there is the \texttt{Mathematica} function \textbf{PolynomialReduce}, that can also be used to solve these problems. However, it is also based on the expensive Gr{\"o}bner basis calculation. Despite the capabilities of these packages, they failed to generate results within a reasonable time frame for the three-generation case of the \nuSM.

Although we were unable to construct the minimal CPC set for the theory with three generations of fermions, we did identify some example of syzygy that follows the form shown in Eq.~\eqref{eq:jpower_new}. For instance,
\begin{equation}
\begin{split}
2 \cJ_{14}^2 &= \cJ_1 \left(2 \cJ_{34} \cI_2-2 \cJ_{76}-\cJ_{11} \cI_2^2-\cJ_{11} \cI_5\right) + \cJ_3 \left(4 \cJ_{36}- 4 \cJ_{13} \cI_2 + \cI_2^2\cJ_3+\cI_5\cJ_3\right)\\
&+2\cJ_4 \left(\cJ_{11} \cI_2 - \cJ_{34}\right)-2 \cJ_{11} \cJ_{15} + 2 \cJ_{13}^2\,,
\end{split}
\end{equation}
one can find that if $\{\cJ_1,\cJ_3,\cJ_4,\cJ_{11},\cJ_{13}\}$ is set to zero, then $\cJ_{14}=0$ automatically, which means if we include the list of five invariants in the CPC set, the inclusion of $\cJ_{14}$ becomes unnecessary.

In addition, we also find that some CP-odd invariants must be added to the minimal CPC set. Specifically, if only one CP-odd invariant $\cJ_s$ in the generating set is non-zero under a specific spectrum, it would not be possible to establish Eq.~\eqref{eq:jpower_new}. Consequently, $\cJ_s$ cannot be eliminated by any CPC set unless it is included in the set. We have found 3 such cases where only one CP-odd invariant is non-zero. The first is the limit $M_N \to 0$ where the theory is reduced to a copy of the SM quark sector. Then the analogue of the Jarlskog invariant $\mathcal{J}_{10}=\,\Im\Tr(\neat{X_{N}^{2} X_{e}^{2} X_{N}^{} X_{e}^{}})$ is the only non-vanishing CP-odd invariant and has to be included in the minimal CPC set. The other 2 cases are found under the spectrum $\{M_N \to m_N \mathbb{1}, Y_e \to 0\}$ and $\{M_N \to m_N \mathbb{1},Y_N \to y_N \mathbb{1}\}$ which force us to add $\mathcal{J}_{74}=\,\Im\Tr(\neat{Y_{N}^{} Y_{N}^{\dagger} Y_{N}^{} M_{N}^{*} Y_{N}^{T} Y_{N}^{*} M_{N}^{} Y_{N}^{\dagger} Y_{N}^{} M_{N}^{*} Y_{N}^{T} Y_{N}^{*} Y_{N}^{T} Y_{N}^{*} M_{N}^{} Y_{N}^{\dagger}})$ and \newline $\mathcal{J}_{251}=\,\Im\Tr(\neat{X_{e}^{2} Y_{N}^{} M_{N}^{*} Y_{N}^{T} X_{e}^{*2} Y_{N}^{*} M_{N}^{} Y_{N}^{\dagger} X_{e}^{} Y_{N}^{} M_{N}^{*} Y_{N}^{T} X_{e}^{*} Y_{N}^{*} M_{N}^{} Y_{N}^{\dagger}})$ to the minimal CPC set respectively. Here, as before, we define $X_{e,N} = \neat{Y_{e,N}^{}Y_{e,N}^{\dagger}}$. It's interesting that we have to rely on the highest-order invariant $\mathcal{J}_{251}$ to establish the CPC conditions of the theory. Notably, $\mathcal{J}_{251}$ is the last invariant in the CP-odd generating set, as shown in App.~\ref{App:InvariantList}.

We want to emphasize one point here. The parameters of the neutrino sector are not measured well enough today to even exclude that one of the neutrinos is massless. Therefore, having a set of flavor invariants which determines the CPC conditions without assumptions on the spectrum is important to make general statements about the theory that hold true for all possible experimental results. 

In conclusion, we introduced Hilbert's Nullstellensatz as a reliable mathematical language to address the problem of the CPC condition. However, the problem is inherently complex and cannot be solved with our current efforts. It deserves further exploration as it requires larger computational resources to be successfully resolved. Additionally, we have solved the complete CPC conditions of the \nuSM for the two-generation case, as detailed in App.~\ref{App:CPC2Gen}, and some special conditions corresponding to the pseudo-real couplings are found, which deserve further investigation.

%%%%%%%%%%%%%%%%%%%%%%%%%%%%%%%%%%%%%%%%%%%%%%%%%%%%%%%%%%%%%%%%%%%%
\section{Conclusions}\label{sec:Conclusions}
In this paper, we have analyzed the algebraic structure of the \nuSM, the SM extended with three generations of sterile neutrinos, by constructing its set of generating flavor invariants. In the quark sector, the theory exhibits a complete intersection ring, and the flavor invariants can be easily generated with the guidance of the Hilbert series. However, in the \nuSM, the theory corresponds to a non-complete intersection ring, introducing many complexities. The analysis of the Hilbert series in this context requires a more careful examination to extract correct information about the number of generating invariants and syzygies. We introduce a novel graph-based approach for constructing flavor invariants. This method maps all single-trace invariants to walks on the graph, allowing for a systematic construction of invariants. Notably, this method can be extended to construct flavor invariants (or covariants) in broader theoretical frameworks beyond the scope of our current work. The resulting invariants are then reduced to a \emph{generating} set by removing redundant invariants. The number of generating invariants and syzygies are carefully compared with the plethystic logarithm and we find agreement at each order in the flavor spurions. The \emph{generating} set comprises of in total 459 invariants, out of which 208 are CP-even and 251 are CP-odd. We have furthermore reduced this set to a \emph{primary} set of flavor invariants capturing the 21 physical parameters of the renormalizable Lagrangian.

We can show that the \emph{generating} set is able to capture the physical degrees of freedom of the seesaw limit. Considering structures in the flavor invariants that are suppressed with one and two powers of the heavy Majorana mass in the seesaw limit, we showed that the 12 and 21 physical parameters are captured by our \emph{generating} set of invariants at the respective orders. Here, the number 12 accounts for the number of masses, mixing angles and phases in the low-energy theory of neutrinos and charged leptons. By also taking the total suppression of the invariants with the heavy Majorana mass in the effective theory into account, it can be seen in a straightforward way, that the 12 and 21 physical parameters can only be accessed with a suppression of two and four powers of the Majorana mass respectively.

In the spirit of the Jarlskog invariant, we have defined a minimal set of CP-odd invariants that can determine the CPC conditions of the theory, which is called the minimal CPC set. This set is defined to be assumption-free regarding the spectrum of the parameters, making it available to determine whether CPV exists in the theory for all upcoming experimental results. The reduction of the CP-odd \emph{generating} set to a minimal CPC set turns out to be a hard task and we only managed to reduce the set of CP-odd flavor invariants to a more minimal set in the case of two generations of fermions, while the reduction of the three generation case was too complicated to complete by our means. Along the way, we present useful algorithms which are essential to our analysis and hopefully prove to be useful for other analyses of flavor invariants in the future. It is worth noting that the flavor structures of the type-I and type-III seesaws are identical to that of the \nuSM, while the couplings of the type-II seesaw cannot be identified with those of the \nuSM due to the different particle content. Instead, the only flavorful parameter in the theory has the same transformation properties as those studied in Ref.~\cite{Jenkins:2009dy,Wang:2021wdq}. Therefore, the flavor invariants identified in this paper can be directly applied to the type-I and type-III seesaw models by renaming the flavor matrices.

An obvious next step for the flavor invariants is to connect them to phenomenological applications to study, for instance, CPV observables in a consistent way with the help of flavor invariants.

%%%%%%%%%%%%%%%%%%%%%%%%%%%%%%%%%%%%%%%%%%%%%%%%%%%%%%%%%%%%%%%%%%%%
\section*{Acknowledgments}
We thank Quentin Bonnefoy, Gabriele Dian, Emanuele Gendy, Gabriel Massoni Salla, Paula Pilatus and Pablo Qu\'ilez for useful discussions. DL thanks the DESY theory group for their hospitality during a long-term visit, where part of this work was completed. He also thanks the ENS de Lyon for providing the scholarship that made this visit possible. This work is supported by the Deutsche Forschungsgemeinschaft under Germany’s Excellence Strategy EXC 2121 “Quantum Universe” -- 390833306, as well as by the grant 491245950. This project also has received funding from the European Union’s Horizon Europe research and innovation programme under the Marie Skłodowska-Curie Staff Exchange grant agreement No 101086085 - ASYMMETRY. C.Y.Y. is supported in part by the Grants No. NSFC-11975130, No. NSFC-12035008, No. NSFC-12047533, the National Key Research and Development Program of China under Grant No. 2017YFA0402200, the China Postdoctoral Science Foundation under Grant No. 2018M641621 and the Helmholtz-OCPC International Postdoctoral Exchange Fellowship Program.

%%%%%%%%%%%%%%%%%%%%%%%%%%%%%%%%%%%%%%%%%%%%%%%%%%%%%%%%%%%%%%%%%%%%
\section*{Appendices}
In App.~\ref{App:parameterization}, we present useful parameterizations for the flavorful couplings that we use throughout this paper. In App.~\ref{App:GradedHSPL}, we show our results for the multi-graded Hilbert series and the multi-graded plethystic logarithm that we do not show in the main text due to their length. In App.~\ref{App:Algorithms}, we present the algorithms used in this paper, which include the algorithm to generate flavor invariants and the algorithm to reduce the invariants to a \emph{generating} set. In App.~\ref{App:InvariantList}, we list all invariants in the \emph{generating} set along with their CP parities. In App.~\ref{app:Hironaka}, we introduce the Hironaka decomposition that we use in some parts of the paper. In App.~\ref{App:theorem}, we introduce Hilbert's Nullstellensatz~(a fundamental theorem in algebraic geometry stating the relationship between polynomial equations and the points where they vanish) and present the relevant concepts that are useful for obtaining a minimal set of CP-odd invariants that capture the necessary and sufficient conditions for CPC. Finally, in App.~\ref{App:CPC2Gen}, we present the complete list of CPC conditions for the two generation case obtained through the ideal-related method.

%%%%%%%%%%%%%%%%%%%%%%%%%%%%%%%%%%%%%%%%%%%%%%%%%%%%%%%%%%%%%%%%%%%%
\appendix

\section{Parameterization of flavor matrices}
\label{App:parameterization}

\subsection{Standard parameterization}
The parameterizations of the flavor matrices in the lepton and quark sectors have been discussed in detail in Refs.~\cite{Jenkins:2007ip, Jenkins:2009dy}. We use a slightly different convention in this paper and thus, we will follow a similar discussion here to introduce the parameterization in this new convention. The fermionic part of the Lagrangian is given by
\begin{equation}
\mathcal{L}_4
\supset\sum_{\Psi}\bar{\Psi}i \slashed{D}\Psi
-\left[\frac{1}{2}\left(NCM_NN\right)+\bar{L}Y_NN\widetilde{H}+\bar{L}Y_e e H
+\bar{Q}Y_u u \widetilde{H}+\bar{Q}Y_d d H +\mbox{H.c.}\right],
\end{equation}
where the kinetic term sums all fermion fields $\Psi=\{Q, L, u, d, e, N\}$.
The flavor transformations of the fermion fields are given by
\begin{equation}
  \begin{split}
    L& \to U_L\, L\,, \quad  e \to U_e\, e\,, \quad  N \to U_N\, N\,, \\
    Q& \to U_Q\, Q\,, \quad  u \to U_u\, u\,, \quad  d \to U_d\, d\,. \\
  \end{split}
\end{equation}
The corresponding Yukawa matrices and the Majorana mass matrix transform as
\begin{equation}
\label{eq:flavor_transformation}
  \begin{split}
    Y_e& \to U_L Y_e U_e^{\dagger}\,, \quad Y_N \to U_L Y_N U_N^{\dagger}\,,\quad M_N \to U_N^* M_N U_N^{\dagger}\,, \\
    Y_u& \to U_Q Y_u U_u^{\dagger}\,, \quad Y_d \to U_Q Y_d U_d^{\dagger} \,.\\
  \end{split}
\end{equation}
The Yukawa matrices $Y_{N,e,u,d}$ are general complex matrices, while the Majorana mass matrix $M_N$ is symmetric, which can be diagonalized as follows
\begin{equation}
  \label{eq:diagonalization}
  \begin{split}
    Y_e&=V_e\,\widehat{Y}_e\,W_e^{\dagger}\,,\quad
    Y_N=V_N\,\widehat{Y}_N\,W_N^{\dagger}\,,\quad
    M_N=V_N'\,\widehat{M}_N\,V_N'^{T}\,,\\
    Y_u&=V_u\,\widehat{Y}_u\,W_u^{\dagger}\,,\quad
    Y_d=V_d\,\widehat{Y}_d\,W_d^{\dagger}\,,\\
  \end{split}
\end{equation}
where $\widehat{Y}_{N,e,u,d}$ and $\widehat{M}_N$ are diagonal matrices with real and non-negative entries. $V_{f},W_{f}$ with $f=N,e,u,d$ and $V'_N$ are unitary matrices. We can choose specific flavor transformations in Eq.~\eqref{eq:flavor_transformation} to get a fixed mass basis. In this paper, we will work on the charged lepton diagonal basis in the lepton sector and up basis in the quark sector. This can be achieved by setting $U_L=V_e^{\dagger}, U_e=W_e^{\dagger}, U_N=V_N'^{T}, U_Q=V_u^{\dagger}, U_u=W_u^{\dagger}, U_d=W_d^{\dagger}$. The flavor matrices will be fixed to the following forms,
\begin{equation}
  \label{eq:basis}
  \begin{split}
    Y_e&=\widehat{Y}_e\,,\quad
    Y_N=V_L\,\widehat{Y}_N\,W^{\dagger}\,,\quad
    M_N=\widehat{M}_N\,,\\
    Y_u&=\widehat{Y}_u\,,\quad
    Y_d=V_\CKM\,\widehat{Y}_d\,,\\
  \end{split}
\end{equation}
where $V_\CKM=V_u^{\dagger}V_d$ is the CKM matrix, corresponding to the mismatch between the diagonalization matrices of the up and down sectors. $V_{L}=V_e^{\dagger}V_N$ is a similar matrix in the lepton sector, describing the mismatch between the diagonalization matrices of lepton Yukawa matrices. The existence of the Majorana mass matrix in the lepton sector introduces another mixing matrix $W=V_N'^{T}W_N$, which describes the mismatch between the diagonalization matrices of $Y_N$ and $M_N$. If the Majorana mass term is forbidden in the theory, the neutrino will have a Dirac mass and we can simply choose $U_N=W_N$, which will lead to $W=\mathbbm{1}$. The lepton sector will have the same structure as the quark sector and $V_L$ can be identified with the PMNS matrix, phenomenologically.

From Eq.~\eqref{eq:diagonalization}, if we assume that the flavor matrices have no degenerate or vanishing eigenvalues,\footnote{If there is a degenerate or vanishing mass spectrum, there will be a larger degrees of freedom to redefine the mixing matrix, some parameters in our current parameterization could be unphysical, and they can be removed by those redefinition. Under this enlarged symmetry, we should adopt a new parameterization, the number of parameters should exactly match the number of physical observables. On the other hand, if we adopt the most general parameterization, and set the special spectrum afterwards, some parameters in the parameterization will be redundant. In some cases, the redundant parameters could be canceled in the expression, and they never appear in the flavor invariants. In other cases, some parameters will appear together as a single polynomial in different invariants. In the latter case, the polynomial that correlates them should be considered as a single parameter in the parameterization, which should remove the redundancies.} the diagonalization matrices will at least have a column phase redefinition freedom,  $V_{f}\rightarrow V_f e^{i\widehat{\Phi}_f}\,, W_{f}\rightarrow W_f e^{i\widehat{\Phi}_f}, f=\{N,e,u,d\}$ and $V_N'\rightarrow V_N'\eta_N$. This can be expressed as the a rephasing invariance of the diagonal matrices 
\begin{equation}
\begin{aligned}
    e^{i \widehat{\Phi}_f} \widehat{Y}_f e^{-i \widehat{\Phi}_f} & = \widehat{Y}_f, \quad f=N,e,u,d\,, \\
    \eta_N \widehat{M}_N \eta_N & = \widehat{M}_N \,,
\end{aligned}
\end{equation}
where $\widehat{\Phi}_{f}\equiv \diag(\phi_{f_1}, \phi_{f_2}, \phi_{f_3})\,, f=N,e,u,d$ are diagonal complex phase matrices, and $\eta_N$ is a diagonal matrix with eigenvalues $\pm 1$. Under these rephasings, the mixing matrices transform as
\begin{equation}
  \label{eq:phase_redefinition}
  V_\CKM \rightarrow e^{-i \widehat{\Phi}_u}V_\CKM e^{i\widehat{\Phi}_d}\,,\quad
  V_L \rightarrow e^{-i \widehat{\Phi}_e}V_L e^{i \widehat{\Phi}_N}\,,\quad
  W \rightarrow  \eta_N W e^{i \widehat{\Phi}_N}\,,
\end{equation}
The $3\times 3$ unitary matrix can be parameterized as
\begin{equation}
  \label{eq:unitary_matrix}
  U_3= e^{i \varphi}e^{i \widehat{\Psi}}U(\theta_{12},\theta_{13},\theta_{22},\delta)e^{i \widehat{\Phi}}\,,
\end{equation}
where $\varphi$ is a overall phase, $\widehat{\Psi}=\diag(0,\psi_1,\psi_2)$ and $\widehat{\Phi}=\diag(0,\phi_1,\phi_2)$, and $U(\theta_{12},\theta_{13},\theta_{23},\delta)$ takes the standard form
\begin{eqnarray}
  \label{eq:ckm}
&& U(\theta_{12},\theta_{13},\theta_{23},\delta) \equiv
\left( \begin{array}{ccc}
1 & 0 & 0 \\
0 & c_{23} & s_{23} \\
0 & -s_{23} & c_{23} \end{array} \right)
\left( \begin{array}{ccc}
c_{13} & 0 & s_{13}e^{-i \delta} \\
0 & 1 & 0 \\
- s_{13}e^{i \delta} & 0 & c_{13} \end{array} \right)
  \left( \begin{array}{ccc}
c_{12} & s_{12} & 0 \\
-s_{12} & c_{12} & 0\\
0 & 0 & 1 \end{array} \right)\,,
\end{eqnarray}
where $s_{ij}\equiv\sin\theta_{ij}$, $c_{ij}\equiv\cos\theta_{ij}$, and $\theta_{ij}\in [0,\pi/2], \delta\in [0,2\pi)$.

In the quark sector, by introducing the phase redefinition in Eq.~\eqref{eq:phase_redefinition}, the phases in the unitary matrix $V_\CKM$ can be  absorbed, it takes the standard form as given in Eq.~\eqref{eq:ckm}. The number of parameters are summarized as follows
\begin{eqnarray*}
  \begin{array}{c|ccc}
\hline
\text{Matrices}& \text{Masses} & \text{Angles} & \text{Phases} \\
\hline
\widehat{Y}_u &  3 & 0 & 0  \\
\widehat{Y}_d &  3 & 0 & 0  \\
V_{\CKM} & 0 & 3 & 1\\[5pt]\hline
\text{Total} & 6 &  3 & 1\\
  \hline
\end{array}
\end{eqnarray*}
There are a total of 10 parameters, consisting of 6 quark masses, 3 mixing angles, and 1 CP phase. 

In the lepton sector, we can parameterize $V_L$ and $W$ as
  \begin{equation}
    \begin{split}
      \label{eq:VW}
  V_L=&\, e^{i \varphi}e^{i \widehat{\Psi}}U(\theta_{12},\theta_{13},\theta_{23},\delta)e^{i \widehat{\Phi}}\,,\\
   W=&\, e^{i \varphi'}e^{i \widehat{\Psi}'/2}U(\theta_{12}',\theta_{13}',\theta_{23}',\delta')e^{i \widehat{\Phi}'}\,.
    \end{split}
  \end{equation}
Since $V_L$ and $W$ share the same rephasing matrix $\widehat{\Phi}_N$, we can use this freedom to remove either $\widehat{\Phi}$ or $\widehat{\Phi}'$ in Eq.~\eqref{eq:VW}, depending on what observable we are interested in. If the flavor invariant only depends on $Y_N$ and $Y_e$, then the mixing matrix $W$ can be simply set to $W=\mathbbm{1}$, since it is not an observable. We can use the phase matrices $\widehat{\Phi}_e$ and $\widehat{\Phi}_N$ to remove the phases of the unitary matrix $V_L$, then it takes the same form as given in Eq.~\eqref{eq:ckm}. The mixing matrices in this case take the following form
  \begin{equation}
    V_L=U(\theta_{12},\theta_{13},\theta_{23},\delta), \quad W=\mathbbm{1}\,.
  \end{equation}
If the flavor invariant only depends on $Y_N$ and $M_N$, then the mixing matrix $V_L$ is not an observable, which can be set to identity matrix. We can use the rephasing matrix $\widehat{\Phi}_N$ to remove the phases $\varphi'$ and $\widehat{\Phi}'$ in Eq.~\eqref{eq:VW}, the effect of $\eta_N$ is indicated by the factor of two in the phase matrix $e^{i\widehat{\Psi}'/2}$, which limits the phases in the range of $\psi'_i/2\in [0,\pi)$ if by convention $\psi'_i\in [0,2\pi)$. In this case the mixing matrices will take the following form
  \begin{equation}
    V_L = \mathbbm{1}\,,\quad W=\, e^{i \widehat{\Psi}'/2}U(\theta_{12}',\theta_{13}',\theta_{23}',\delta')\,.
  \end{equation}
If the flavor invariant depends on the three flavor matrices $Y_e, Y_N$ and $M_N$,\footnote{The flavor invariants can not be constructed by only $Y_e$ and $M_N$, since they are disconnected objects if $Y_N$ is not included, as shown in Fig.~\ref{fig:invariantGraph}.} we can use the rephasing phases $\widehat{\Phi}_e$ and $\widehat{\Phi}_N$ to remove the phases in $V_L$ and the phase $\varphi'$ in $W$ in Eq.~\eqref{eq:VW}, without loss of generality. Then, the mixing matrices take the following form
\begin{equation}
    V_L = U(\theta_{12},\theta_{13},\theta_{23},\delta)\,, \quad W=\, e^{i \widehat{\Psi}'/2}U(\theta_{12}',\theta_{13}',\theta_{23}',\delta')e^{i \widehat{\Phi}'}\,.\quad
  \end{equation}
We can also use $\widehat{\Phi}_e$ and $\widehat{\Phi}_N$ to remove the phases $e^{i\varphi}e^{i\widehat{\Psi}}$ and $e^{i\varphi'}e^{i\widehat{\Phi}'}$ in $V_L$ and $W$ in Eq.~\eqref{eq:VW}, respectively, then the mixing matrices are given by
\begin{equation}
\label{eq:param_lepton}
    V_L = U(\theta_{12},\theta_{13},\theta_{23},\delta)e^{i\widehat{\Phi}}\,, \quad W=\, e^{i \widehat{\Psi}'/2}U(\theta_{12}',\theta_{13}',\theta_{23}',\delta')\,.\quad
  \end{equation}
The third case corresponds to the most general parameterization of the mixing matrices $V_L$ and $W$ and we can work in this basis without loss of generality. To be specific, we use the following parameterization
\begin{equation}
\label{eq:parameterization_3gen}
Y_e = \diag\(y_e,y_{\mu},y_{\tau}\),\ Y_N = V_L \cdot \diag\(y_{1},y_{2},y_{3}\) \cdot W^{\dagger},\  M_N = \diag\(m_1,m_2,m_3\) \, ,
\end{equation}
with $V_L$ and $W$ defined in Eq.~\eqref{eq:param_lepton}. We summarize the the number of parameters in each matrix in Tab.~\ref{tab:parameters}.

For a more comprehensive discussion, please refer to Ref.~\cite{Jenkins:2009dy}. In their paper, the cases for $n_N=n_f=2$ and $n_N=2,n_f=3$ in the lepton sector are also discussed. For $n_N=n_f=2$, all flavor matrices are $2\times 2$ matrices, the $2\times2$ unitary matrix is given by
\begin{equation}
U_2=e^{i\varphi} \diag(1,e^{i\psi})\cdot
\begin{pmatrix}
\cos\theta & \sin\theta\\
-\sin\theta & \cos\theta
\end{pmatrix}\cdot\diag(1,e^{i\phi})\,.
\end{equation}
Follow a similar discussion, we can easily find the following parameterization,
\begin{equation}
\label{eq:parameterization_2gen}
    Y_e = \diag\(y_e,y_{\mu}\), \qquad Y_N = V_L \cdot \diag\(y_{1},y_{2}\) \cdot W^{\dagger}, \qquad M_N = \diag\(m_1,m_2\) \, ,
\end{equation}
with 
\begin{equation}
\label{eq:mixing_matrix_2gen}
    V_L =\begin{pmatrix} \cos\theta & \sin\theta \\ -\sin\theta & \cos\theta \end{pmatrix} \cdot \diag\(1,e^{i \phi}\), \quad W = \diag\(1,e^{i \varphi}\) \cdot \begin{pmatrix} \cos\alpha & \sin\alpha \\ -\sin\alpha & \cos\alpha \end{pmatrix} \, .
\end{equation} 

For $n_N=2,n_f=3$, $Y_N$ is a $3\times 2$ matrix, and $M_N$ is a $2\times 2$ matrix. Similarly, the flavor matrices can be parameterized as
\begin{equation}
\label{eq:parameterization_3x2gen}
Y_e = \diag\(y_e,y_{\mu},y_{\tau}\),\quad Y_N = V_L \cdot 
\begin{pmatrix}
    y_1 & 0\\
    0 & y_2\\
    0 & 0
\end{pmatrix} \cdot W^{\dagger},\quad  M_N = \diag\(m_1,m_2\) \, ,
\end{equation}
with $V_L$ and $W$ defined as follows
\begin{equation}
\label{eq:mixing_matrix_3x2gen}
V_L =U(\theta_{12},\theta_{13},\theta_{23},\delta)\cdot \diag \(1,e^{i \phi},1\)\,, \quad 
W = \diag\(1,e^{i \varphi}\) \cdot \begin{pmatrix} \cos\alpha & \sin\alpha \\ -\sin\alpha & \cos\alpha \end{pmatrix} \,,
\end{equation} 
which is analogous to the parameterization in Eq.~\eqref{eq:param_lepton}, however the last phase in $\hat{\Phi}$ is unphysical, since it vanishes when multiplied with the zeros in the last row of the $3\times2$ block diagonal matrix.

\begin{table}[!ht]
\centering
\begin{eqnarray*}
{\arraycolsep=5pt
\begin{array}{c|ccc}
    \hline
    \text{Matrices}& \text{Masses} & \text{Angles} & \text{Phases} \\ \hline
\widehat{M}_N &  3\,[2]\,(2) & 0 & 0  \\
\widehat{Y}_N &  3\,[2]\,(2) & 0 & 0  \\
\widehat{Y}_e &  3\,[2]\,(3) & 0 & 0  \\
V_L & 0 & 3\,[1]\,(3) & 3\,[1]\,(2) \\
W & 0 & 3\,[1]\,(1) & 3\,[1]\,(1)  \\
  \hline
\text{Total} & 9\,[6]\,(7) &  6\,[2]\,(4) & 6\,[2]\,(3)\\\hline
\end{array}
}
\end{eqnarray*}
\caption{The number of masses, mixing angles and phases in the parameterization of the lepton sector for the case of $n_N=n_f=3\,[n_N=n_f=2]\,(n_N=2,n_f=3)$.}
\label{tab:parameters}
\end{table}

In the Dirac limit, $M_N$ is set to zero, and $W$ is not relevant, we can simply set $W=\mathbbm{1}$. In the case of $n_N=n_f=3$, the matrix $V_L$ will become the PMNS matrix $V_{\PMNS}$, which takes the same standard form as $V_{\CKM}$. The parameterization is given as
\begin{equation}
\label{eq:parameterization_3gen_Dirac1}
Y_e = \diag\(y_e,y_{\mu},y_{\tau}\),\quad Y_N = U(\theta_{12},\theta_{13},\theta_{23},\delta) \cdot \diag\(y_{1},y_{2},y_{3}\)\, .
\end{equation}
Similarly, for $n_N=n_f=2$, it can be parameterized as
\begin{equation}
\label{eq:parameterization_2gen_Dirac1}
    Y_e = \diag\(y_e,y_{\mu}\), \qquad Y_N = \begin{pmatrix} \cos\theta & \sin\theta \\ -\sin\theta & \cos\theta \end{pmatrix} \cdot \diag\(y_{1},y_{2}\)\, ,
\end{equation}
where the phase of $V_L$ in the Majorana case can be removed by the rephasing of the right-handed neutrino $N$. For $n_N=2, n_f=3$, the parameterization is given by
\begin{equation}
\label{eq:parameterization_3x2gen_Dirac1}
Y_e = \diag\(y_e,y_{\mu},y_{\tau}\),\quad Y_N = U(\theta_{12},\theta_{13},\theta_{23},\delta) \cdot 
\begin{pmatrix}
    y_1 & 0\\
    0 & y_2\\
    0 & 0
\end{pmatrix}\, ,
\end{equation}
where the phase $e^{i\phi}$ in Eq.~\eqref{eq:mixing_matrix_3x2gen} can be absorbed by the field $N$ in the Dirac case. The parameters in the Dirac limit are summarized in Tab.~\ref{tab:parameters_Dirac}.
\begin{table}[!ht]
\centering
\begin{eqnarray*}
  \begin{array}{c|ccc}
    \hline
    \text{Matrices}& \text{Masses} & \text{Angles} & \text{Phases} \\
    \hline
\widehat{Y}_N &  3[2](2) & 0 & 0  \\
\widehat{Y}_e &  3[2](3) & 0 & 0  \\
V_{\PMNS}  & 0 & 3[1](3) &1[0](1) \\
  \hline
\text{Total} & 6[4](5) & 3[1](3) & 1[0](1)\\\hline
\end{array}
\end{eqnarray*}
\caption{The number of masses, mixing angles and phases in the parameterization of the lepton sector for the case of $n_N=n_f=3\,[n_N=n_f=2]\,(n_N=2,n_f=3)$ in the Dirac limit.}
\label{tab:parameters_Dirac}
\end{table}

\subsection{Algebraic parameterization}
The parameterization described above is favored for its phenomenological relevance, as Yukawa matrices are factorized into eigenvalues and mixing matrices, which aligns with experimental observables. However, when exploring the algebraic structures of invariants, the inclusion of trigonometric functions introduces complexity. The use of sine and cosine as objects in the polynomial expansion of flavor invariants can complicate the exploration of these structures.  Consequently, alternative parameterizations that are more suitable for polynomial expressions are needed.

One possible solution is to parameterize the trigonometric functions. A frequently used parameterization for the unit circle is provided as follows:
\begin{equation}
x(t) = \frac{1-t^2}{1+t^2},\quad y(t) = \frac{2t}{1+t^2}\,,\quad \text{with~} t\in(-\infty,+\infty).
\end{equation}
However the point $(-1,0)$ on the unit circle can only be obtained in the limit $t\to\infty$. Another parameterization that can cover the whole circle is given by
\begin{equation}
x(t) = \frac{1-6t^2+t^4}{1+2t^2+t^4},\quad y(t) = \frac{4t-4t^3}{1+2t^2+t^4}\,,\quad \text{with~} t\in (-1,1]\,.
\end{equation}
Thus the sine and cosine functions in the mixing matrices of our above parameterization can be replaced with $x(t)$ and $y(t)$ respectively. However, this parameterization may introduce new complexities, as it leads to rational polynomials.

Following the parameterization in Eq.~\eqref{eq:parameterization_3gen}, it is possible to work with the diagonal basis of both $Y_e$ and $M_N$ while leaving $Y_N$ undiagonalized. The parameterization is given as follows
\begin{equation}
\label{eq:parameterization_3gen_new}
Y_e = \diag\(y_e,y_{\mu},y_{\tau}\),\quad Y_N = \begin{pmatrix}
r_{11} & c_{12} & c_{13}\\
r_{21} & c_{22} & c_{23}\\
r_{31} & c_{32} & c_{33}\\
\end{pmatrix},\quad  
M_N = \diag\(m_1,m_2,m_3\) \, ,
\end{equation}
where $r_{ij}$ and $c_{ij}$ are general labels for real and complex parameters respectively, and $c_{ij}$ can be easily written as real and imaginary parts by introducing two real parameters. The phases of the first column of $Y_N$ are absorbed by the rephasing of charged lepton fields.\footnote{It is not necessary for the rephasing degree of freedom to target the first column, any phase in each row of $Y_N$ can be eliminated.} In this parameterization all invariants are expressed as polynomials of simple variables, making it easier to analyze the algebraic structures of the theory. It is easy to see that the number of parameters in this parameterization is still 21, which is the same as the physical parameterization in Eq.~\eqref{eq:parameterization_3gen}. Similarly, for $n_N=n_f=2$, we can parameterize it as 
\begin{equation}
\label{eq:parameterization_2gen_new}
Y_e = \diag\(y_e,y_{\mu}\), \qquad Y_N = \begin{pmatrix}
r_{11} & c_{12}\\
r_{21} & c_{22}   
\end{pmatrix}, \qquad
M_N = \diag\(m_1,m_2\) \, .
\end{equation}
For $n_N=2, n_f=3$, the parameterization is given by
\begin{equation}
\label{eq:parameterization_3x2gen_new}
Y_e = \diag\(y_e,y_{\mu},y_{\tau}\),\quad Y_N =
\begin{pmatrix}
    r_{11} & c_{12}\\
    r_{21} & c_{22}\\
    r_{31} & c_{32}
\end{pmatrix},\quad  M_N = \diag\(m_1,m_2\) \, .
\end{equation}
The mapping between these two parameterizations can be found by solving equations built from the entries of $Y_N$, which will not be shown here.

Under the Dirac limit, the Majorana mass term $M_N$ is set to 0, the parameters in Yukawa matrix $Y_N$ can be further reduced by the field redefinition of right-handed neutrino $N$. Starting from the $Y_N$ in Eq.~\eqref{eq:parameterization_3gen_new}, we can absorb the phases in $c_{12}$ and $c_{13}$ by rephasing of $N$, these two real parameters can be further set to 0 by two rotations of $N$, i.e., $N\to R_{12}(\theta_2)R_{23}(\theta_1)N$, where $R_{ij}$ is the rotation matrix acting on the $(i,j)$ entries of $N$. The rotation $R_{12}$ can make $r_{21}$ and $r_{31}$ complex, but their phases can always be absorbed by the rephasing of charged lepton fields. With these two leading zeros in the second and third columns, the rephasing of $N$ can be used again to remove the phases in $c_{22}$ and $c_{23}$, the corresponding real parameters $r_{22}$ and $r_{23}$ can be mixed by the rotation $N\to R_{23}(\theta_3)N$. By setting proper value of $\theta_3$, the parameter $r_{23}$ can be set to zero. The leading two zeros in the third column making it possible to remove the phase in $c_{33}$ by rephasing of $N$, resulting in a real parameter $r_{33}$. The whole transformation process can be presented as follows
\begin{equation}
Y_N\to \begin{pmatrix}
 r_{11} & r_{12} & r_{13} \\
 r_{21} & c_{22} & c_{23} \\
 r_{31} & c_{32} & c_{33} \\
\end{pmatrix}\to 
\begin{pmatrix}
 r_{11} & 0 & 0 \\
 r_{21} & c_{22} & c_{23} \\
 r_{31} & c_{32} & c_{33} \\
\end{pmatrix} \to 
\begin{pmatrix}
 r_{11} & 0 & 0 \\
 r_{21} & r_{22} & r_{23} \\
 r_{31} & c_{32} & c_{33} \\
\end{pmatrix}\to 
\begin{pmatrix}
 r_{11} & 0 & 0 \\
 r_{21} & r_{22} & 0 \\
 r_{31} & c_{32} & c_{33} \\
\end{pmatrix}\to
\begin{pmatrix}
 r_{11} & 0 & 0 \\
 r_{21} & r_{22} & 0 \\
 r_{31} & c_{32} & r_{33} \\
\end{pmatrix}\,.
\end{equation}
There is no further field redefinition that can be used to reduce the number of parameters, as a result, in the Dirac limit with $n_N=n_f=3$, we can parameterize the flavor matrices as
\begin{equation}
\label{eq:parameterization_3gen_Dirac}
Y_e = \diag\(y_e,y_{\mu},y_{\tau}\),\quad Y_N = \begin{pmatrix}
r_{11} & 0 & 0\\
r_{21} & r_{22} & 0\\
r_{31} & c_{32} & r_{33}\\
\end{pmatrix}\, .
\end{equation}
We find there are exactly 10 real parameters in this parameterization, as expected. Similarly, for $n_N=n_f=2$ the flavor matrices are parameterized as 
\begin{equation}
\label{eq:parameterization_2gen_Dirac}
Y_e = \diag\(y_e,y_{\mu}\), \qquad Y_N = \begin{pmatrix}
r_{11} & 0\\
r_{21} & r_{22}   
\end{pmatrix}\, .
\end{equation}
For $n_N=2, n_f=3$ the parameterization is given by
\begin{equation}
\label{eq:parameterization_3x2gen_Dirac}
Y_e = \diag\(y_e,y_{\mu},y_{\tau}\),\quad Y_N =
\begin{pmatrix}
    r_{11} & 0\\
    r_{21} & r_{22}\\
    r_{31} & c_{32}
\end{pmatrix}\, .
\end{equation}

%%%%%%%%%%%%%%%%%%%%%%%%%%%%%%%%%%%%%%%%%%%%%%%%%%%%%%%%%%%%%%%%%%%%
\section{Results for multi-graded Hilbert series and plethystic logarithm}\label{App:GradedHSPL}

\subsection{Model with \texorpdfstring{$n_N=n_f=3$}{nN=nf=3}}\label{App:HS3x3}

In Eqs.~\eqref{eq:num} and \eqref{eq:deno}, we have presented the ungraded Hilbert series, where the same grading is used for all spurions. However, the information encoded in the ungraded HS is not enough for analyses, where the PL is used to count the number of generating invariants and syzygies at each order. Therefore, in this section, we present the multi-graded HS, where the single spurion $q$ in Eqs.~\eqref{eq:num} and \eqref{eq:deno} is traded for multiple spurions $e,m$ and $n$, corresponding to the spurions of the flavor matrices $Y_e,M_N$ and $Y_N$ respectively. The denominator of the multi-graded HS is given as follows
\begin{equation}\label{eq:gradedHS_den}
  \begin{split}
    \cD&(e,m,n)=(1-e^2)(1-m^2)(1-n^2)(1-e^4)(1-m^4)(1-n^4)(1-e^2 n^2)^2(1-m^2 n^2)\\
    &(1-e^6)(1-m^6)(1-n^6)(1-e^2 n^4)(1-e^4 n^2)(1-m^2 n^4)^2(1-m^4 n^2)(1-e^2 m^2 n^2)\\
    &(1-m^2 n^6)(1-m^4 n^4)(1-e^2 m^2 n^4)(1-e^2 m^4 n^2)(1-e^4 m^2 n^2)(1-m^2 n^8)\\
    &(1-e^2 m^2 n^6)(1-e^4 m^2 n^4)(1-e^4 m^4 n^2)(1-e^4 m^2 n^6)(1-e^4 m^4 n^4)(1-e^6 m^2 n^4)\\
    &(1-e^4 m^2 n^8)(1-e^8 m^2 n^4)(1-e^8 m^4 n^4)\,,
  \end{split}
\end{equation}
while there are 6582 terms in the numerator, which goes up to order $\cO([emn]^{196})$ in total powers of spurions. Due to its length we only show the terms up to $\cO([emn]^{26})$ below
\begin{align*}
\cN&(e,m,n) = 1 - e^2 n^2 + 2 m^4 n^4 + e^4 n^4 + 2 e^2 m^2 n^4 + 2 m^4 n^6 + 2 m^6 n^4 + 4 e^2 m^2 n^6\\ 
&+ 4 e^2 m^4 n^4 + 3 e^4 m^2 n^4 + 3 m^4 n^8 + 3 m^6 n^6 + m^8 n^4 + 5 e^2 m^2 n^8 + 7 e^2 m^4 n^6 + 3 e^2 m^6 n^4\\
&+ 4 e^4 m^2 n^6 + 5 e^4 m^4 n^4 + e^6 m^2 n^4 + m^4 n^{10} + 3 m^6 n^8 + m^8 n^6 + 3 e^2 m^2 n^{10} + 9 e^2 m^4 n^8\\
& + 6 e^2 m^6 n^6 + e^2 m^8 n^4 + 4 e^4 m^2 n^8 + 10 e^4 m^4 n^6 + 5 e^4 m^6 n^4 + 3 e^6 m^2 n^6 + 3 e^6 m^4 n^4\\
& + m^4 n^{12} + m^6 n^{10} + 3 m^8 n^8 + e^2 m^2 n^{12} + 5 e^2 m^4 n^{10} + 9 e^2 m^6 n^8 + 2 e^2 m^8 n^6 + e^4 m^2 n^{10}\\
& + 16 e^4 m^4 n^8 + 10 e^4 m^6 n^6 + 2 e^4 m^8 n^4 + 4 e^6 m^2 n^8 + 8 e^6 m^4 n^6 + 3 e^6 m^6 n^4 + 2 e^8 m^2 n^6\\
& + e^8 m^4 n^4 + m^8 n^{10} + m^{10} n^8 + e^2 m^4 n^{12} + 6 e^2 m^6 n^{10} + 7 e^2 m^8 n^8 - e^2 m^{10} n^6\numberthis\\
& + 10 e^4 m^4 n^{10} + 19 e^4 m^6 n^8 + 4 e^4 m^8 n^6 + e^6 m^2 n^{10} + 14 e^6 m^4 n^8 + 8 e^6 m^6 n^6 + e^6 m^8 n^4\\
& + e^8 m^2 n^8 + 5 e^8 m^4 n^6 + 2 e^8 m^6 n^4 + e^{10} m^2 n^6 - m^6 n^{14} + 2 m^{10} n^{10} + m^{12} n^8 - 3 e^2 m^4 n^{14}\\
& + 2 e^2 m^6 n^{12} + 5 e^2 m^8 n^{10} + 2 e^2 m^{10} n^8 - e^2 m^{12} n^6 - e^4 m^2 n^{14} + 6 e^4 m^4 n^{12} + 15 e^4 m^6 n^{10}\\
& + 12 e^4 m^8 n^8 - 2 e^4 m^{10} n^6 - e^6 m^2 n^{12} + 9 e^6 m^4 n^{10} + 16 e^6 m^6 n^8 + 2 e^6 m^8 n^6 - 2 e^8 m^2 n^{10}\\
& + 9 e^8 m^4 n^8 + 4 e^8 m^6 n^6 + e^8 m^8 n^4 - 2 e^{10} m^2 n^8 + 2 e^{10} m^4 n^6 - m^6 n^{16} - 2 m^8 n^{14} + m^{12} n^{10}\\
& - 2 e^2 m^4 n^{16} - 4 e^2 m^6 n^{14} + 3 e^2 m^8 n^{12} + 2 e^2 m^{10} n^{10} - 2 e^4 m^4 n^{14} + 6 e^4 m^6 n^{12} + 7 e^4 m^8 n^{10}\\
& - e^4 m^{10} n^8 - 2 e^4 m^{12} n^6 - 2 e^6 m^2 n^{14} + 2 e^6 m^4 n^{12} + 13 e^6 m^6 n^{10} + 8 e^6 m^8 n^8 - 3 e^6 m^{10} n^6\\
& - 3 e^8 m^2 n^{12} + 3 e^8 m^4 n^{10} + 8 e^8 m^6 n^8 + 2 e^8 m^8 n^6 - 2 e^{10} m^2 n^{10} - 3 e^{10} m^4 n^8 - e^{12} m^2 n^8\\
& - m^8 n^{16} - m^{10} n^{14} - e^2 m^4 n^{18} - 3 e^2 m^6 n^{16} - 6 e^2 m^8 n^{14} - e^2 m^{10} n^{12} - 3 e^2 m^{12} n^{10}\\
& + e^4 m^4 n^{16} - 10 e^4 m^6 n^{14} - 5 e^4 m^8 n^{12} - 7 e^4 m^{10} n^{10} - 3 e^4 m^{12} n^8 - 7 e^6 m^4 n^{14} + 2 e^6 m^6 n^{12}\\
& - e^6 m^8 n^{10} - 2 e^6 m^{10} n^8 - 2 e^6 m^{12} n^6 - 2 e^8 m^2 n^{14} - 4 e^8 m^4 n^{12} + 2 e^8 m^6 n^{10} + 3 e^8 m^8 n^8\\
& - 2 e^8 m^{10} n^6 - 2 e^{10} m^2 n^{12} - 6 e^{10} m^4 n^{10} - 3 e^{10} m^6 n^8 - 3 e^{12} m^4 n^8 - 3 m^{10} n^{16} - m^{12} n^{14}\\
& - m^{14} n^{12} - e^2 m^6 n^{18} - 7 e^2 m^8 n^{16} - 10 e^2 m^{10} n^{14} - 5 e^2 m^{12} n^{12} - 2 e^2 m^{14} n^{10} - 15 e^4 m^6 n^{16}\\
& - 33 e^4 m^8 n^{14} - 22 e^4 m^{10} n^{12} - 13 e^4 m^{12} n^{10} - e^4 m^{14} n^8 - 2 e^6 m^4 n^{16} - 21 e^6 m^6 n^{14}\\
& - 21 e^6 m^8 n^{12} - 16 e^6 m^{10} n^{10} - 4 e^6 m^{12} n^8 - 11 e^8 m^4 n^{14} - 15 e^8 m^6 n^{12} - 15 e^8 m^8 n^{10}\\
& - 6 e^8 m^{10} n^8 - 2 e^8 m^{12} n^6 - e^{10} m^2 n^{14} - 9 e^{10} m^4 n^{12} - 8 e^{10} m^6 n^{10} - 5 e^{10} m^8 n^8 - e^{10} m^{10} n^6\\
& - 3 e^{12} m^6 n^8 - e^{14} m^4 n^8 + \cO([emn]^{28})\,.
\end{align*}
It is worth noting that the multi-graded HS lacks certain properties of the ungraded HS, such as a matching number of factors in the denominator with physical observables and a palindromic form in the numerator. In addition, as already mentioned in footnote~\ref{foot:ambiguity}, there is ambiguity when determining the form of the ungraded HS, similar ambiguity can arise for the multi-graded HS. For instance, it is possible to introduce a common factor $(1+e^2n^2)$ to both the numerator and denominator, then the factors $(1-e^2n^2)^2$ in denominator become $(1-e^2n^2)(1-e^4n^4)$, the numerator will also change accordingly. The form of the HS is less constrained in the multi-graded case. This ambiguity prevents us from further exploring the invariant structures of the theory. Instead of decoding the HS, analyzing the PL is more helpful in our case, since it is unique for both ungraded and multi-graded HS. 

The invariants in our theory form a non-complete intersection ring making the PL a non-terminating series as a result. According to Eq.~\eqref{eq:PL}, the PL can only be calculated up to some given order in the spurions. However, as we discussed in Sec.~\ref{sec:cons_inv}, we are only interested in the terms in the PL up to the first purely negative order, which we find to occur at order $\cO([emn]^{26})$. The PL up to this order is given by
\begin{align*}
\label{eq:gradedPL}
    & \PL(e,m,n)= \( e^2+m^2+n^2\) + \( e^4+m^4+n^4+e^2 n^2+m^2 n^2\) + \( e^6+m^6+n^6+e^2 n^4 \right. \\
    & \left. + e^4 n^2+2 m^2 n^4+m^4 n^2+e^2 m^2 n^2 \) + \( e^4 n^4+m^2 n^6+3 m^4 n^4+3 e^2 m^2 n^4 + e^2 m^4 n^2  \right. \\
    & \left. + e^4 m^2 n^2 \) + \( m^2 n^8+2 m^4 n^6+2 m^6 n^4+5 e^2 m^2 n^6+4 e^2 m^4 n^4 + 4 e^4 m^2 n^4+e^4 m^4 n^2 \) \\
    & + \( e^6 n^6+3 m^4 n^8+3 m^6 n^6+m^8 n^4+5 e^2 m^2 n^8 + 9 e^2 m^4 n^6+3 e^2 m^6 n^4 +7 e^4 m^2 n^6 \right. \\
    & \left. +6 e^4 m^4 n^4+2 e^6 m^2 n^4\) + \( m^4 n^{10} + 3 m^6 n^8+m^8 n^6+3 e^2 m^2 n^{10}+11 e^2 m^4 n^8 \right. \\
    & \left. +8 e^2 m^6 n^6+e^2 m^8 n^4+9 e^4 m^2 n^8+14 e^4 m^4 n^6+5 e^4 m^6 n^4+6 e^6 m^2 n^6+3 e^6 m^4 n^4 \right. \\
    & \left. +e^8 m^2 n^4 \)+ \( m^4 n^{12}+m^6 n^{10}+e^2 m^2 n^{12}+8 e^2 m^4 n^{10}+8 e^2 m^6 n^8+3 e^2 m^8 n^6 \right. \\
    & \left. +6 e^4 m^2 n^{10}+20 e^4 m^4 n^8+13 e^4 m^6 n^6+2 e^4 m^8 n^4+8 e^6 m^2 n^8 +13 e^6 m^4 n^6 \right. \\
    & \left. +3 e^6 m^6 n^4+3 e^8 m^2 n^6+2 e^8 m^4 n^4 \) + \(2 e^2 m^4 n^{12}+3 e^4 m^2 n^{12}+11 e^4 m^4 n^{10} \right. \\
    & \left. +11 e^4 m^6 n^8+5 e^4 m^8 n^6+5 e^6 m^2 n^{10} +18 e^6 m^4 n^8+13 e^6 m^6 n^6+e^6 m^8 n^4+4 e^8 m^2 n^8 \right. \\
    & \left. +8 e^8 m^4 n^6+2 e^8 m^6 n^4 +e^{10} m^2 n^6-3 m^8 n^{10}-3 m^{10} n^8-3 e^2 m^6 n^{10}-4 e^2 m^8 n^8 \right. \\
    & \left. -e^2 m^{10} n^6 \) + \( 2 e^6 m^6 n^8 + 4 e^6 m^8 n^6 + 9 e^8 m^4 n^8 + 7 e^8 m^6 n^6 + e^8 m^8 n^4 + 3 e^{10} m^4 n^6 \right. \\
    & \left. - m^6 n^{14} - 9 m^8 n^{12} - 8 m^{10} n^{10} - 4 m^{12} n^8 - 2 e^2 m^4 n^{14} - 21 e^2 m^6 n^{12} - 32 e^2 m^8 n^{10} \right. \\
    & \left. - 14 e^2 m^{10} n^8 - e^2 m^{12} n^6 - 9 e^4 m^4 n^{12} - 28 e^4 m^6 n^{10} - 18 e^4 m^8 n^8 - 2 e^4 m^{10} n^6 \right. \\
    & \left. - e^6 m^4 n^{10} \) + \( 3 e^8 m^8 n^6 + 2 e^{10} m^6 n^6 - m^6 n^{16} - 10 m^8 n^{14} - 18 m^{10} n^{12} - 9 m^{12} n^{10} \right. \\
    & \left. - 2 m^{14} n^8 - 2 e^2 m^4 n^{16} - 34 e^2 m^6 n^{14} - 76 e^2 m^8 n^{12} - 55 e^2 m^{10} n^{10} - 12 e^2 m^{12} n^8 \right. \\
    & \left. - 27 e^4 m^4 n^{14} - 103 e^4 m^6 n^{12} - 109 e^4 m^8 n^{10}- 39 e^4 m^{10} n^8 - 2 e^4 m^{12} n^6 - 2 e^6 m^2 n^{14} \right. \\
    & \left. - 45 e^6 m^4 n^{12} - 83 e^6 m^6 n^{10} - 35 e^6 m^8 n^8 - 3 e^6 m^{10} n^6 - 6 e^8 m^2 n^{12} - 21 e^8 m^4 n^{10} \right.\numberthis \\
    & \left. - 9 e^8 m^6 n^8 - 4 e^{10} m^2 n^{10} - e^{10} m^4 n^8 \) + \( e^{10} m^8 n^6 - e^{12} n^{12} - 11 m^8 n^{16} - 20 m^{10} n^{14} \right. \\
    & \left. - 21 m^{12} n^{12} - 5 m^{14} n^{10} - m^{16} n^8 - e^2 m^4 n^{18} - 33 e^2 m^6 n^{16} - 110 e^2 m^8 n^{14} - 116 e^2 m^{10} n^{12} \right. \\
    & \left.- 48 e^2 m^{12} n^{10} - 5 e^2 m^{14} n^8 - 31 e^4 m^4 n^{16} - 174 e^4 m^6 n^{14} - 284 e^4 m^8 n^{12} - 162 e^4 m^{10} n^{10} \right. \\
    & \left. - 33 e^4 m^{12} n^8 - e^6 m^2 n^{16} - 87 e^6 m^4 n^{14} - 261 e^6 m^6 n^{12} - 226 e^6 m^8 n^{10} - 59 e^6 m^{10} n^8 \right. \\
    & \left. - 2 e^6 m^{12} n^6 - 8 e^8 m^2 n^{14} - 94 e^8 m^4 n^{12} - 134 e^8 m^6 n^{10} - 48 e^8 m^8 n^8 - 2 e^8 m^{10} n^6 \right. \\
    & \left. - 10 e^{10} m^2 n^{12} - 32 e^{10} m^4 n^{10}- 15 e^{10} m^6 n^8 - 3 e^{12} m^2 n^{10} - 2 e^{12} m^4 n^8 \) + \(-5 m^8 n^{18} \right. \\
    & \left. - 19 m^{10} n^{16} - 16 m^{12} n^{14} - 7 m^{14} n^{12}- m^{16} n^{10} - 22 e^2 m^6 n^{18} - 105 e^2 m^8 n^{16} \right. \\
    & \left. - 149 e^2 m^{10} n^{14} - 88 e^2 m^{12} n^{12} - 20 e^2 m^{14} n^{10} - e^2 m^{16} n^8 - 21 e^4 m^4 n^{18} - 192 e^4 m^6 n^{16} \right. \\
    & \left. - 424 e^4 m^8 n^{14} - 371 e^4 m^{10} n^{12} - 130 e^4 m^{12} n^{10} - 13 e^4 m^{14} n^8 - 96 e^6 m^4 n^{16} - 448 e^6 m^6 n^{14} \right. \\
    & \left. - 603 e^6 m^8 n^{12} - 303 e^6 m^{10} n^{10} - 45 e^6 m^{12} n^8 - 5 e^8 m^2 n^{16} - 161 e^8 m^4 n^{14} - 417 e^8 m^6 n^{12} \right. \\
    & \left. - 314 e^8 m^8 n^{10} - 70 e^8 m^{10} n^8 - 2 e^8 m^{12} n^6 - 12 e^{10} m^2 n^{14} - 108 e^{10} m^4 n^{12} - 140 e^{10} m^6 n^{10} \right. \\
    & \left. - 40 e^{10} m^8 n^8 - e^{10} m^{10} n^6 - 8 e^{12} m^2 n^{12} - 23 e^{12} m^4 n^{10} - 9 e^{12} m^6 n^8 - e^{14} m^2 n^{10} - e^{14} m^4 n^8 \) \\
    & + \mathcal{O}\(\[emn\]^{28}\)\,,
\end{align*}
where the terms are grouped by parentheses at each order. We can see that the terms in $\cO([emn]^{26})$ are all negative.

Under the Dirac limit, the HS can be obtained by setting $m\to 0$, which will have a very simple form as has been found for the quark sector in Ref.~\cite{Jenkins:2009dy}
\begin{equation}
\cH(e,n)=\frac{1+e^6 n^6}{\left(1-e^2\right) \left(1-e^4\right) \left(1-e^6\right) \left(1-n^2\right)
   \left(1-n^4\right) \left(1-n^6\right) \left(1-e^2 n^2\right) \left(1-e^4 n^2\right) \left(1-e^2
   n^4\right) \left(1-e^4 n^4\right)}\,.
\end{equation}
The ungraded HS is given by
\begin{equation}
\cH(q)=\frac{1+q^{12}}{\left(1-q^2\right)^2 \left(1-q^4\right)^3 \left(1-q^6\right)^4 \left(1-q^8\right)}\,.
\end{equation}
The Dirac case corresponds to a complete intersection ring, and the multi-graded PL has finite number of terms, which are given as follows
\begin{equation}
\PL(e,n)=e^2+e^4+e^6+n^2+n^4+n^6+e^2 n^2+e^4 n^2+e^2 n^4+e^4 n^4+e^6 n^6-e^{12} n^{12}\,.
\end{equation}
The corresponding ungraded PL can be obtained by setting $e,n\to q$, which has the following form
\begin{equation}
\PL(q)=2 q^2+3 q^4+4 q^6+q^8+q^{12}-q^{24}\,,
\end{equation}
where the positive terms correctly capture the 10 CP-even and 1 CP-odd invariants, while the negative term indicates there is a syzygy at order 24.

\subsection{Model with \texorpdfstring{$n_N=n_f=2$}{nN=nf=2}}\label{App:HS2x2}

For completeness, we also show the HS for the case $n_N=n_f=2$, which has already been presented in Ref.~\cite{Jenkins:2009dy}. The numerator and denominator are given by
\begin{equation}
\begin{split}
\cN(e,m,n) =& 1+2 e^2 m^2 n^4+m^4 n^4+e^2 m^4 n^4+e^4 m^4 n^4+e^2 m^2 n^6+e^4 m^2 n^6-e^2 m^6 n^6+\\
&-e^4 m^6 n^6-e^2 m^4
   n^8-e^4 m^4 n^8-e^6 m^4 n^8-2 e^4 m^6 n^8-e^6 m^8 n^{12}\,,\\
\cD(e,m,n) =& \left(1-e^2\right) \left(1-e^4\right) \left(1-m^2\right) \left(1-m^4\right) \left(1-n^2\right) \left(1-e^2
   n^2\right) \left(1-m^2 n^2\right)\times \\
   &\left(1-e^2 m^2 n^2\right) \left(1-n^4\right) \left(1-m^2 n^4\right)
   \left(1-e^4 m^2 n^4\right)\,.
\end{split}
\end{equation}
The ungraded HS is
\begin{equation}
\cH(q)=\frac{1+q^6+3 q^8+2 q^{10}+3 q^{12}+q^{14}+q^{20}}{\left(1-q^2\right)^3 \left(1-q^4\right)^5
   \left(1-q^6\right) \left(1-q^{10}\right)}\,.
\end{equation}
The multi-graded PL is given by
\begin{equation}
\begin{split}
\PL(e,m,n)=&e^2+m^2+n^2+e^4+m^4+e^2 n^2+m^2 n^2+n^4+e^2 m^2 n^2+m^2 n^4+\\
&+2 e^2 m^2 n^4+m^4 n^4+e^4 m^2 n^4+e^2 m^4 n^4+e^2 m^2 n^6+e^4 m^4 n^4+\\
&+e^4 m^2 n^6-e^2 m^6 n^6-e^2 m^4 n^8-\cO([emn]^{16})\,.
\end{split}
\end{equation}
As before, the PL is also a non-terminating series for two generations. Therefore, the theory corresponds to a non-complete intersection ring. However, the pure negative order appears at $\cO([emn]^{14})$, the 18 lower order terms in PL are all positive, and they correspond to the \emph{generating} set.

\subsection{Model with \texorpdfstring{$n_N=2,\ n_f=3$}{nN=2, nf=3}}\label{App:HS2x3}

For the case of $n_N=2, n_f=3$, the HS has been calculated in Ref.~\cite{Yu:2021cco}. For completeness, we also provide it here. The numerator and denominator of the multi-graded HS are given by
\begin{align*}
\cN(e,m,n)&= 1-e^2 n^2+e^4 n^4+2 e^2 m^2 n^4+2 e^4 m^2 n^4+2 e^6 m^2 n^4+m^4 n^4+e^2 m^4 n^4+2 e^4 m^4 n^4+\\
&+e^6 m^4 n^4+e^8 m^4 n^4+e^2 m^2 n^6+e^{10} m^2 n^6-e^2 m^4 n^6-e^4 m^4 n^6-3 e^6 m^4 n^6+\\
&-e^8 m^4 n^6-e^{10} m^4 n^6-e^2 m^6 n^6-2 e^4 m^6 n^6-2 e^6 m^6 n^6-2 e^8 m^6 n^6-e^{10} m^6 n^6+\\
&-e^4 m^2 n^8-e^6 m^2 n^8-e^8 m^2 n^8-e^{10} m^2 n^8-e^{12} m^2 n^8-e^2 m^4 n^8-e^4 m^4 n^8+\\
&-3 e^6 m^4 n^8-e^8 m^4 n^8-3 e^{10} m^4 n^8-e^{12} m^4 n^8-e^{14} m^4 n^8-e^4 m^6 n^8+e^6 m^6 n^8+\\
&-e^8 m^6 n^8+e^{10} m^6 n^8-e^{12} m^6 n^8+e^6 m^8 n^8+e^8 m^8 n^8+e^{10} m^8 n^8-e^{10} m^2 n^{10}+\\
&+e^4 m^4 n^{10}+e^6 m^4 n^{10}+2 e^8 m^4 n^{10}+2 e^{12} m^4 n^{10}+e^{14} m^4 n^{10}+e^{16} m^4 n^{10}+\\
&+2 e^6 m^6 n^{10}+3 e^{10} m^6 n^{10}+2 e^{14} m^6 n^{10}-2 e^8 m^6 n^{12}-3 e^{12} m^6 n^{12}-2 e^{16} m^6 n^{12}+\\
&-e^6 m^8 n^{12}-e^8 m^8 n^{12}-2 e^{10} m^8 n^{12}-2 e^{14} m^8 n^{12}-e^{16} m^8 n^{12}-e^{18} m^8 n^{12}+\\
&+e^{12} m^{10} n^{12}-e^{12} m^4 n^{14}-e^{14} m^4 n^{14}-e^{16} m^4 n^{14}+e^{10} m^6 n^{14}-e^{12} m^6 n^{14}+\\
&+e^{14} m^6 n^{14}-e^{16} m^6 n^{14}+e^{18} m^6 n^{14}+e^8 m^8 n^{14}+e^{10} m^8 n^{14}+3 e^{12} m^8 n^{14}+\\
&+e^{14} m^8 n^{14}+3 e^{16} m^8 n^{14}+e^{18} m^8 n^{14}+e^{20} m^8 n^{14}+e^{10} m^{10} n^{14}+e^{12} m^{10} n^{14}+\\
&+e^{14} m^{10} n^{14}+e^{16} m^{10} n^{14}+e^{18} m^{10} n^{14}+e^{12} m^6 n^{16}+2 e^{14} m^6 n^{16}+2 e^{16} m^6 n^{16}+\\
&+2 e^{18} m^6 n^{16}+e^{20} m^6 n^{16}+e^{12} m^8 n^{16}+e^{14} m^8 n^{16}+3 e^{16} m^8 n^{16}+e^{18} m^8 n^{16}+\\
&+e^{20} m^8 n^{16}-e^{12} m^{10} n^{16}-e^{20} m^{10} n^{16}-e^{14} m^8 n^{18}-e^{16} m^8 n^{18}-2 e^{18} m^8 n^{18}+\\
&-e^{20} m^8 n^{18}-e^{22} m^8 n^{18}-2 e^{16} m^{10} n^{18}-2 e^{18} m^{10} n^{18}-2 e^{20} m^{10} n^{18}-e^{18} m^{12} n^{18}+\\
&+e^{20} m^{12} n^{20}-e^{22} m^{12} n^{22}\,,\numberthis\\
\cD(e,m,n)&=\left(1-e^2\right) \left(1-m^2\right) \left(1-n^2\right) \left(1-e^4\right) \left(1-m^4\right)
   \left(1-n^4\right) \left(1-e^2 n^2\right)^2\times \\
   &\left(1-m^2 n^2\right) \left(1-e^6\right) \left(1-e^2
   n^4\right) \left(1-e^4 n^2\right) \left(1-m^2 n^4\right)\times\\
   &\left(1-e^2 m^2 n^2\right) \left(1-e^4 m^2
   n^2\right) \left(1-e^4 m^2 n^4\right) \left(1-e^8 m^2 n^4\right)\,,
\end{align*}
which in the single-graded limit have the following form
\begin{equation}
\begin{split}
\cN(q)&=1+q^2+q^4+2 q^6+6 q^8+10 q^{10}+18 q^{12}+23 q^{14}+28 q^{16}+31 q^{18}+34 q^{20}+32 q^{22}+\\
&+34 q^{24}+31
   q^{26}+28 q^{28}+23 q^{30}+18 q^{32}+10 q^{34}+6 q^{36}+2 q^{38}+q^{40}+q^{42}+q^{44}\,,\\
\cD(q)&=\left(1-q^2\right)^2 \left(1-q^4\right)^5 \left(1-q^6\right)^4 \left(1-q^8\right) \left(1-q^{10}\right)
   \left(1-q^{14}\right)\,.
\end{split}
\end{equation}
The multi-graded PL is given by
\begin{align*}
\PL(e,m,n)&=e^2+m^2+n^2+e^4+m^4+e^2 n^2+m^2 n^2+n^4+e^6+e^4 n^2+e^2 m^2 n^2+e^2 n^4+\\
&+m^2 n^4+e^4 m^2 n^2+e^4 n^4+2 e^2 m^2 n^4+m^4 n^4+3 e^4 m^2 n^4+e^2 m^4 n^4+e^2 m^2 n^6+\\
&+2 e^6 m^2 n^4+2 e^4 m^4 n^4+e^6 n^6+2 e^4 m^2 n^6+e^8 m^2 n^4+e^6 m^4 n^4+2 e^6 m^2 n^6+\\
&-e^2 m^6 n^6-e^2 m^4 n^8+e^8 m^4 n^4+2 e^8 m^2 n^6-e^6 m^4 n^6-2 e^4 m^6 n^6-e^6 m^2 n^8+\\
&-5 e^4 m^4 n^8-2 e^2 m^6 n^8-m^8 n^8+e^{10} m^2 n^6-2 e^6 m^6 n^6-e^8 m^2 n^8-8 e^6 m^4 n^8+\\
&-6 e^4 m^6 n^8-e^2 m^8 n^8-e^6 m^2 n^{10}-2 e^4 m^4 n^{10}-e^2 m^6 n^{10}-\cO([emn]^{20})\,.\numberthis
\end{align*}
The PL is also non-terminating, indicating that the algebraic structure of the flavor invariants is that of a non-complete intersection ring. The pure negative order appears at $\cO([emn]^{20})$, and the \emph{generating} set can be obtained considering invariants up to $\cO([emn]^{18})$ as shown in Ref.~\cite{Yu:2021cco}.

%%%%%%%%%%%%%%%%%%%%%%%%%%%%%%%%%%%%%%%%%%%%%%%%%%%%%%%%%%%%%%%%%%%%
\section{Algorithms} \label{App:Algorithms}

\subsection{Algorithm for the construction of invariants}\label{App:AlgGenInvs}

The flavor invariants can be easily constructed following the invariant graphs. Taking the quark sector as an example, one can refer to the graph in the lower left panel of Fig.~\ref{fig:invariantGraph} to determine the general form of flavor invariants, given by $\Tr(X_u^mX_d^nX_u^k\dots)$ with $X_u\equiv Y_uY_u^{\dagger}$ and $X_d\equiv \neat{Y_{d}^{}Y_{d}^{\dagger}}$. Although this will result in an infinite number of flavor invariants, there are some identities of $n\times n$ matrices that can be utilized to reduce them, such as the Cayley--Hamilton theorem. For instance, for $3\times 3$ matrices, the identity is given by
\begin{equation}
  A^3=A^2\Tr(A)-\frac{1}{2}A\left[\Tr(A)^2-\Tr(A^2)\right]+\frac{1}{6}\left[\Tr(A)^3-3\Tr(A^2)\Tr(A)+2\Tr(A^3)\right]\mathbb{1}_{3\times 3}\,,
  \label{eq:id_a3}
\end{equation}
which will reduce the power of the matrices $X_u$ and $X_d$ to a maximum of 3. Following the discussion in Ref.~\cite{Jenkins:2009dy}, another identity can be derived from the Cayley--Hamilton theorem, which is given by
\begin{align}
        2\Tr(ABAC)&=\Tr(A)^2\Tr(B)\Tr(C)-\Tr(BC)\Tr(A)^2-2\Tr(AB)\Tr(A)\Tr(C)+\nonumber\\
        &-2\Tr(AC)\Tr(A)\Tr(B)+2\Tr(ABC)\Tr(A)+2\Tr(ACB)\Tr(A)+\nonumber\\
        &-\Tr(A^2)\Tr(B)\Tr(C)+2\Tr(AB)\Tr(AC)+\Tr(A^2)\Tr(BC)+\nonumber\\
        &+2\Tr(C)\Tr(A^2B)+2\Tr(B)\Tr(A^2C)-2\Tr(A^2BC)-2\Tr(A^2CB)\,.
        \label{eq:id_abc}
\end{align}
Note that by using the Cayley--Hamilton theorem, we can derive more general identities with 4 or more different matrices entering the theorem.
This particular identity in Eq.~\eqref{eq:id_abc} ensures that the terms present in the trace in the quark sector will not exceed a length of~4. Referring to Ref.~\cite{Jenkins:2009dy} for a detailed explanation, it can be concluded that with the two identities above, there exist only 11 generating invariants in the quark sector of the SM, that can be constructed with the help of the invariant graphs. They are given by
\begin{equation}
  \label{eq:quark_inv}
  \begin{split}
  I_{2,0}&=\Tr(X_u),~
  I_{0,2}=\Tr(X_d),~
  I_{4,0}=\Tr(X_u^2),~
  I_{2,2}=\Tr(X_uX_d),~
  I_{0,4}=\Tr(X_d^2),\\
  I_{6,0}&=\Tr(X_u^3),~
  I_{4,2}=\Tr(X_u^2X_d),~
  I_{2,4}=\Tr(X_uX_d^2),~
  I_{0,6}=\Tr(X_d^3),~\\
  I_{4,4}&=\Tr(X_u^2X_d^2),~
  I_{6,6}^{(-)}=\Im\Tr(\neat{X_{u}^{2}X_{d}^{2}X_{u}^{}X_{d}^{}})\,.
  \end{split}
\end{equation}
These invariants form a \emph{generating} set of all quark invariants, i.e., all flavor invariants in quark sector can be written as a polynomial of these 11 invariants. Furthermore, the first 10 invariants form a \emph{primary} set in the sense that they are algebraically independent invariants which capture all physical parameters in the theory.

Additionally, there exists a separate conjugate graph in the quark sector shown in the lower right panel of Fig.~\ref{fig:invariantGraph}. Because the two graphs are disconnected in the quark sector, invariants can either be built using $X_u,X_d$ or $X_u^*,X_d^*$ without any mixing between them like for the neutrinos. Furthermore, as both matrices $X_u$ and $X_d$ are Hermitian, the identity $\Tr(\neat{X_{u}^{m*}X_{d}^{n*}X_{u}^{k*}X_{d}^{l*}})=\Tr(\neat{X_{d}^{l\dagger}X_{u}^{k\dagger}X_{d}^{n\dagger}X_{u}^{m\dagger}})=\Tr(\neat{X_{d}^{l}X_{u}^{k}X_{d}^{n}X_{u}^{m}})$ can be used to show that all flavor invariants constructed from the conjugate graph are included in the first graph.

To systematically construct flavor invariants in the lepton sector, the walks based on the top graph in Fig.~\ref{fig:invariantGraph} should be explored. As we explained in Sec.~\ref{sec:cons_inv}, all single-trace flavor invariants can be represented as walks on the graph, which can be further denoted by integers. In this regard, our objective is to exhaustively construct walks up to length 26 (correspond the invariants up to order 26) by brute force. These walks can be generated order by order with the extending method. By extending one step further at each vertex, we can generate higher order walks from a given walk, for instance, extending the walk $1232$ leads to
\begin{equation}
  \{1\uline{21}232,\ 12\uline{32}32,\ 12\uline{12}32,\ 123\uline{43}2,\ 123\uline{23}2,\ 1232\uline{32},\ 1232\uline{12}\}\,,
\end{equation}
where the underlined vertices are the ones that have been added. By using the cycling degree of freedom of the walks, for instance, $121232\sim 123212$, we can identify some redundant walks, and the independent walks are determined to be
\begin{equation}
  \{121232,\ 123232,\ 123432\}\,.
\end{equation}
With this algorithm, higher order walks derived from a specific walk will be exhaustively generated. As a result, starting from all possible lowest order walks, the extending algorithm is capable of constructing all walks up to a given order. The initial walks correspond to the order two flavor invariants, which are given by
\begin{equation}
  \{12,\ 23,\ 34,\ 45,\ 56\}\,.
\end{equation}
Running our algorithm up to order 26 with these initial walks results in 516'101 walks. After taking care of the transpose redundancy and conjugate redundancy as explained in Sec.~\ref{sec:cons_inv}, the large list of walks can be reduced. These reduced walks will then be converted into invariants that can be simplified by applying the Cayley--Hamilton theorem, allowing us to effectively eliminate more redundancies.

When using the Cayley--Hamilton theorem and its variation shown in Eq.~\eqref{eq:id_a3} and Eq.~\eqref{eq:id_abc}, we need to confirm that the matrices $A, B$ and $C$ must have the same transformation rules under the flavor group, and can form single trace invariants. We refer to such sequences of flavor matrices as adjoint objects. We note that if one flavor matrix appears consecutively in an invariant for $2n$ times ($n\geq 1$), then the corresponding flavor matrix sequence must be an adjoint object. If it appears $2n+1$ times consecutively, then there must be a $2n$ sub-sequence among them corresponding to adjoint object.
For example, the underlined sequences in the invariant $\Tr(\uwave{\neat{Y_{N}^{}Y_{N}^{\dagger}}}\neat{Y_{N}^{}}\uline{\neat{M_{N}^{*}M_{N}^{}}}\neat{Y_{N}^{\dagger}}\uwave{\neat{Y_{N}^{}Y_{N}^{\dagger}}}\uuline{\neat{Y_{e}^{}Y_{e}^{\dagger}}})$ are adjoint objects. We find two of them are identical, as a result, we can use Eq.~\eqref{eq:id_abc} to eliminate this invariant as it can be represented by other invariants that are already present in our invariant list. In this invariant, the matrix products $\neat{Y_{N}^{}Y_{N}^{\dagger}}$, $\neat{Y_{N}^{}M_{N}^{*}M_{N}^{}Y_{N}^{\dagger}}$ and $Y_eY_e^{\dagger}$ are identified as $A, B$ and $C$ respectively in Eq.~\eqref{eq:id_abc}. It is evident that they share the same adjoint transformation properties, allowing the construction of single trace flavor invariants that accurately match the identity. If we come across an adjoint object with six or more flavor matrices in an invariant, then identity in Eq.~\eqref{eq:id_a3} can be easily applied, and the invariant will be redundant. For example, we can remove the invariant $\Tr(\uwave{\neat{Y_{e}^{}Y_{e}^{\dagger}Y_{e}^{}Y_{e}^{\dagger}Y_{e}^{}Y_{e}^{\dagger}}}\uline{\neat{Y_{N}^{}Y_{N}^{\dagger}}})$ because there is a sequence with six flavor matrices. The identity in Eq.~\eqref{eq:id_a3} cannot eliminate $\Tr(A^3)$. Therefore, our invariant list will include expressions such as $\Tr(Y_eY_e^{\dagger}Y_eY_e^{\dagger}Y_eY_e^{\dagger})$.

By applying these two identities, our invariant list is further reduced, but there are still many more identities that can be used. For instance, setting $C\to AC$ in Eq.~\eqref{eq:id_abc} yields the following identity
\begin{equation}
\label{eq:a2bac}
  \begin{split}
    &3 \Tr(A^2BAC)+3 \Tr(ABA^2C)=2 \Tr(A)^3 \Tr(B) \Tr(C)-2 \Tr(A)^3 \Tr(BC)+\\
    &~~-3 \Tr(A)^2 \Tr(C) \Tr(AB)-3 \Tr(A)^2 \Tr(B) \Tr(AC)+3 \Tr(A)^2 \Tr(ABC)+\\
    &~~+3 \Tr(A)^2 \Tr(ACB)-3 \Tr(A^2) \Tr(A) \Tr(B) \Tr(C)+3 \Tr(A^2) \Tr(A) \Tr(BC)+\\
    &~~+3 \Tr(A) \Tr(C) \Tr(A^2B)+3 \Tr(A) \Tr(B) \Tr(A^2C)-3 \Tr(A) \Tr(A^2BC)+\\
    &~~-3 \Tr(A) \Tr(A^2CB)+\Tr(A^3) \Tr(B) \Tr(C)-\Tr(A^3) \Tr(BC)+\\
    &~~+3 \Tr(A^2B) \Tr(AC)+3 \Tr(AB) \Tr(A^2C)\,,
  \end{split}
\end{equation}
which indicates one of $\Tr(A^2BAC)$ and $\Tr(ABA^2C)$ is redundant, but our algorithm does not eliminate either. We can further remove invariants using more complex identities derived from Cayley--Hamilton theorem at higher orders,  but the identities could be non-trivial and difficult to use. The higher-order variations of the Cayley--Hamilton theorem have been implemented in Refs.~\cite{Darvishi:2023ckq,Darvishi:2024cwe}. However, this approach heavily relies on identifying the building blocks of the invariants, which could be complicated in our theory. Furthermore, while the identities presented in this section are for general matrices, our flavor matrices and their combinations could have special properties. For instance, the invariant may include a Hermitian matrix $Y_eY_e^{\dagger}$ or a symmetric matrix $M_N$, additional identities beyond those for general $3\times3$ matrices could be found in some cases. Therefore, we need a general algorithm to identify all potential identities, which should encompass both regular identities for eliminating redundant invariants and syzygies between non-redundant invariants. The general algorithm becomes computationally expensive when the invariant list is large, with the help of the simple matrix identities, the redundancies have been effectively removed. After we perform CP-even and CP-odd decompositions for the invariants, the number of invariants drops to 8'666. Then we will pass this pre-reduced invariant set to the numerical algorithm to further remove explicit redundancies and get the \emph{generating} set. The whole process is presented by the flow graph in Fig.~\ref{fig:FlowGraphSets}.

\subsection{Algorithm for the construction of a generating set}\label{App:AlgGenSet}

As by definition any invariant in the set can be expressed as a polynomial of the generating invariants, the process of identifying the \emph{generating} set entails eliminating explicit redundancies from the invariant list. We have used matrix identities to reduce the invariant list in advance as we have discussed in App.~\ref{App:AlgGenInvs}. In this section, we will introduce the numerical algorithm that can identify all explicit relations and syzygies. The explicit relations will remove all explicit redundancies, and will result in a \emph{generating} set of invariants, while the syzygies, alongside the explicit relations, will be generated as a byproduct, and their counting serves as a verification of our algorithm when compared with the PL.

Our algorithm is based on the fact that the relations among invariants follow the graded algebra, which means every term in the relations should have the same degree in the spurions. So in order to construct the explicit relations or the syzygies, we need to consider all possible combinations that lead to the specific degree from our invariant list. We create a graded list $\{d_i:\{f_i\}\}$, where $d_i$ is the degree in terms of $(e,m,n)$, and $\{f_i\}$ is a homogeneous list of all (combinations of) invariants at degree $d_i$. We start from the lowest degrees, and the graded list is given by
\begin{equation}
  \{e^2:\{\Tr(\neat{Y_{e}^{}Y_{e}^{\dagger}})\},\ m^2:\{\Tr(\neat{M_{N}^{}M_{N}^{*}})\},\ n^2:\{\Tr(\neat{Y_{N}^{}Y_{N}^{\dagger}})\}\}\,,
\end{equation}
which will be extended degree by degree. At higher degrees $d_i$, the homogeneous list $\{f_i\}$ can be constructed by two possible sources, which includes the invariants at order $d_i$ from our invariant list, and all combinations of invariants $\{f_j\}$ and $\{f_k\}$ in the graded list satisfying $d_jd_k=d_i$. Once the homogeneous list $\{f_i\}$ is formed, the general equation for the polynomial identity should have the form
\begin{equation}
  \sum_{i=1}^{n}c_if_i=0\,,
\end{equation}
where $f_i$ is the element in $\{f_i\}$, $n$ is the length of the homogeneous list, and the coefficient $c_i$ is the parameter we want to determine by numerical method. Since it is a polynomial relation, $c_i$ can be chosen to be integer in general. The invariant $f_i$ (which can be a product of several single-trace invariants) is parameterized in terms of general matrices $Y_{e,N}$ and symmetric matrix $M_N$. In order to find the solutions to the coefficients $\{c_i\}$, a numerical algorithm is introduced. We generate $n$ linear equations of $\{c_i\}$ with $n$ sets of random inputs for the flavor matrices,\footnote{We use the algebraic parameterization in Eq.~\eqref{eq:parameterization_3gen_new} with random integers as inputs. In order to reduce the accidental error from the random numbers, we can generate more than $n$ equations to form a larger linear system, or we can increase the randomness of the inputs} the coefficient matrix $M$ is taken as input by the \texttt{Mathematica} function \textbf{NullSpace}, where the rows represent the set of random inputs and the columns correspond to the homogeneous list $\{f_i\}$. The function will give a list of vectors $V\equiv\{v_i\}$ that forms a basis for the null space of the matrix $M$, as a result, $V$ is the basis solutions of $\{c_i\}$, and the identities are given by
\begin{equation}
   \sum_{i=1}^{n}V_{ji}f_i=0\quad \text{for}\quad j=1,\dots, |V|\,.
\end{equation}

The identities that involve single invariants (not products of invariants) will be called explicit relations, which will be used to remove redundancies from our invariant list. But sometimes it is difficult to use the solution $V$ to determine which invariant should be removed, since the relations of single invariants could be correlated, in such cases, the \texttt{Mathematica} function \textbf{RowReduce} can be applied to generate the row echelon form of $V$, which will be used to disentangle the explicit relations. Here is an illustrative example: we have a homogeneous list consisting of $\{f_i\}=\{s_1,s_2,s_3,p_1,p_2\}$, with $s_{1,2,3}$ as single invariants and $p_{1,2}$ as products of invariants. Suppose they lead to a coefficient matrix $M$ that results in a null space $V$ and can be further reduced to the row echelon form $V'$,
\begin{equation}
M=\begin{pmatrix}
 2 & 1 & -3 & -1 & 1 \\
 -1 & 2 & -1 & -2 & 2 \\
 -3 & 3 & 0 & -3 & 3 \\
 -2 & 4 & -2 & -4 & 4 \\
 -4 & 5 & -1 & -5 & 5 \\
  \end{pmatrix} \xrightarrow{\bf NullSpace}
  V=\begin{pmatrix}
    0 & -1 & 0 & 0 & 1 \\
    0 & 1 & 0 & 1 & 0 \\
    1 & 1 & 1 & 0 & 0 \\
  \end{pmatrix} \xrightarrow{\bf RowReduce}
  V'=\begin{pmatrix}
    1 & 0 & 1 & 0 & 1 \\
    0 & 1 & 0 & 0 & -1 \\
    0 & 0 & 0 & 1 & 1 \\
  \end{pmatrix}\,.
\end{equation}
From the null space matrix $V$, we can find the following identities,
\begin{equation}
 s_2=p_2,\quad s_2=-p_1,\quad s_1=-s_2-s_3\,.
\end{equation}
In order to disentangle the dependence between them, we have to perform additional operations, while for the identities generate by $V'$, they have clear forms,
\begin{equation}
\label{eq:relation}
 s_1=-s_3-p_2,\quad s_2=p_2,\quad p_1=-p_2\,,
\end{equation}
which directly determine that, for instance, $s_1,s_2$ and $p_1$ are redundant. They correspond to the columns that has a leading 1 in the row echelon form $V'$. The first two identities tell us that $s_1,s_2$ should be removed from our invariant list, and $s_3$ should be included in our \emph{generating} set as a generating invariant. 

The first identity can also be written as $s_3=-s_1-p_2$, then $s_1$ and $s_3$ will change their characters, this can be understood as ambiguities in determining generating invariants. The third identity contains only products of invariants and thus is a syzygy. Although this syzygy does not identify any redundant invariant, it does signify the product of invariants $p_1$ is redundant, and we should remove it from our homogeneous list, which results in $\{f_i\}=\{s_3,p_2\}$. It is crucial to remove $p_1$ from the homogeneous list in our algorithm, as keeping it would result in redundant syzygies at higher orders and prevent us from obtaining the correct number of syzygies. For instance, if $p_1$ is not removed, at degree $2d_i$, the syzygy $s_3p_1=-s_3p_2$ will be found. Although it looks trivial in this example, in real analysis of our flavor invariants, the redundant syzygies could have a complicated form, making it difficult to identify them as redundant.

Once the discussion on degree $d_i$ is complete and the reduced graded list $\{d_i:\{f_i\}\}$ has been updated, the analysis is repeated for all other invariants at higher degrees until the scan of all invariants is finished. As a result, all redundancies are removed, and the invariant list is reduced to the \emph{generating} set. While scanning, it is important to keep a record of the number of generating invariants $n_g$ and the number of syzygies $n_s$ at each degree, the difference $n_g-n_s$ should match the corresponding term in the PL.

In the real analysis of our invariants, there are some difficulties, as already mentioned in Sec.~\ref{sec:cons_inv}, with ``redundant syzygies''. They are partially removed by the homogeneous list reduction at each degree as described above, however this reduction does not cover all redundancies. Suppose we have two redundant products $f_a$ and $f_b$ at degree $d_i$, and the syzygies are denoted by the homogeneous linear function $V_a$ and $V_b$ on the homogeneous list $\{f_{i\neq a,b}\}$,
\begin{equation}
  f_a=s_1s_2^2=V_a(\{f_{i\neq a,b}\}),\quad f_b=s_1^2s_2=V_b(\{f_{i\neq a,b}\})\,,
\end{equation}
where $s_1, s_2$ are invariants at degree $d_i/3$. At higher degree, we can easily find a syzygy
\begin{equation}
  s_1V_a(\{f_{i\neq a,b}\})-s_2V_b(\{f_{i\neq a,b}\})=0\,,
\end{equation}
which happens accidentally due to the special forms of $f_a$ and $f_b$. Since it is a linear combination of lower degree syzygies, it must be redundant. Depending on the ordering scheme of the homogeneous list, such redundant syzygies could first appear at different degrees. In our algorithm, the first redundant syzygy appears at order 26, and we find more syzygies than expected from the PL. In order to remove such redundancies, we should record the lower degree syzygies, the linearly independent redundant syzygies can be generated at a specific higher degree, the number of which should be subtracted at given degree when counting the syzygies. Although we did not initially incorporate this into our algorithm, we have verified the validity of this method by checking multiple redundant syzygies at certain degrees. Furthermore, our objective is to determine the \emph{generating} set. It should not pose any issues to run our algorithm without verifying the number of syzygies.

Another issue is related to the complexity of numerical calculations. Our algorithm take random flavor matrices as inputs. When the invariants become highly complex, the matrix calculations will result in extremely large integers. This will generate a complicated linear system that will take considerable time to solve. The \texttt{Mathematica} function \textbf{RowReduce} provides a \textbf{Modulus} option, by setting it to an appropriate large prime number, the massive integers in the linear system are mapped to finite fields, which will considerably reduce the computational complexity, and our algorithm can speed up by several times. There are also some packages, such as \texttt{FiniteFlow}~\cite{Peraro:2019svx}, \texttt{FiniteFieldSolve}~\cite{Mangan:2023eeb}, which are frameworks that can be used to solve the linear system efficiently. Interested readers may refer to these packages.

%%%%%%%%%%%%%%%%%%%%%%%%%%%%%%%%%%%%%%%%%%%%%%%%%%%%%%%%%%%%%%%%%%%%

\section{List of invariants}\label{App:InvariantList}

In this section, we enumerate the 459 generating invariants comprising the \emph{generating} set. We use the walk-based notation to represent these invariants as introduced in Sec.~\ref{sec:cons_inv}. Therefore, a single trace invariant is denoted by an integer. A prefix ``R@'' or ``I@'' means that the real or imaginary part of the trace should be taken. For integer without these prefixes, the single trace itself is inherently real. Consequently, invariants with ``I@'' prefixes are identified as CP-odd invariants, while all the other remaining ones are categorized as CP-even. There are in total 208 CP-even and 251 CP-odd invariants. The complete list we found reads as follows
\begin{adjustwidth}{0cm}{1.7cm}
{
\small
CP-even set:~\{12,\  34,\  23,\  1212,\  3434,\  2323,\  1232,\  2343,\  121212,\  343434,\  232323,\  123232,\  121232,\  234543,\  232343,\  234343,\  123432,\  12123232,\  23234543,\  23234343,\  R@23434543,\  12345432,\  R@12323432,\  12343432,\  12123432,\  2323454543,\  R@2323434543,\  R@2343434543,\  1234545432,\  R@1232345432,\  R@1232323432,\  R@1234345432,\  R@1232343432,\  1212345432,\  1234565432,\  R@1212323432,\  1212343432,\  R@232343454543,\  R@232343434543,\  R@123234545432,\  R@123232345432,\  R@123434545432,\  R@123234345432,\  R@123234543432,\  R@123232343432,\  R@123434345432,\  121234545432,\  R@121232345432,\  R@121232323432,\  R@123234565432,\  R@121234345432,\  R@121232343432,\  R@123434565432,\  121234565432,\  R@23234343454543,\  R@12323234545432,\  R@12323434545432,\  R@12323454543432,\  R@12323234345432,\  R@12323234543432,\  R@12343434545432,\  R@12323434345432,\  R@12323454343432,\  R@12123234545432,\  R@12123232345432,\  R@12323454565432,\  R@12323234565432,\  R@12123434545432,\  R@12123234345432,\  R@12123234543432,\  R@12123232343432,\  R@12323434565432,\  R@12323456543432,\  R@12123434345432,\  R@12343434565432,\  R@12123456545432,\  R@12123234565432,\  R@12123434565432,\  12123456565432,\  R@1232323434545432,\  R@1232323432345432,\  R@1232323454543432,\  R@1232343434545432,\  R@1232345454343432,\  R@1232323434345432,\  R@1232343434543432,\  R@1212323234545432,\  R@1232323454565432,\  R@1212323434545432,\  R@1212323454543432,\  R@1212323234345432,\  R@1212323234543432,\  R@1232343454565432,\  R@1232345434565432,\  R@1232323434565432,\  R@1232323456543432,\  R@1212343434545432,\  R@1212323434345432,\  R@1212323454343432,\  R@1232343434565432,\  R@1232345654343432,\  R@1212323456545432,\  R@1212323454565432,\  R@1212323234565432,\  R@1212343456545432,\  R@1212343454565432,\  R@1212321234345432,\  R@1212323434565432,\  R@1212323456543432,\  R@1212343434565432,\  R@1212323456565432,\  R@1212343456565432,\  R@123232343234545432,\  R@123232343434545432,\  R@123234343454543432,\  R@123232345654543232,\  R@121232323434545432,\  R@121232323432345432,\  R@121232323454543432,\  R@123232343454565432,\  R@123232343234565432,\  R@123232345434565432,\  R@121232343434545432,\  R@121232345454343432,\  R@121232323434345432,\  R@123234343454565432,\  R@123234543434565432,\  R@123232343434565432,\  R@121232343434543432,\  R@123234343456543432,\  R@121232323456545432,\  R@121232323454565432,\  R@121232123434545432,\  R@121232343456545432,\  R@121232343454565432,\  R@121232343212345432,\  R@121232345654345432,\  R@121232345434565432,\  R@121232321234345432,\  R@121232323434565432,\  R@121232323456543432,\  R@121234343456545432,\  R@121234343454565432,\  R@121232123434345432,\  R@121232343434565432,\  R@121232343212343432,\  R@121232345654343432,\  R@121232123456545432,\  R@121232323456565432,\  R@121232123434565432,\  R@121232343456565432,\  R@121232345656543432,\  R@121234343456565432,\  R@12123232343234545432,\  R@12323234323454565432,\  R@12123232343434545432,\  R@12323234343454565432,\  R@12123234343454543432,\  R@12323434345434565432,\  R@12123232345654543232,\  R@12123212345434545432,\  R@12123232123434545432,\  R@12123232343454565432,\  R@12123232343234565432,\  R@12123232345654345432,\  R@12123232345434565432,\  R@12123212343434545432,\  R@12123234343454565432,\  R@12123234343212345432,\  R@12123234565434345432,\  R@12123234543434565432,\  R@12123232123434345432,\  R@12123232343434565432,\  R@12123434345654345432,\  R@12123212343434543432,\  R@12123234343456543432,\  R@12123212345654545432,\  R@12123212343456545432,\  R@12123212343454565432,\  R@12123212345654345432,\  R@12123232123434565432,\  R@12123232343456565432,\  R@12123232345656543432,\  R@12123212343434565432,\  R@12123234343456565432,\  R@12123234565654343432,\  R@12123212343456565432,\  R@1212323234323454565432,\  R@1212321234543434545432,\  R@1212323234343454565432,\  R@1212321234343454543432,\  R@1212323434345434565432,\  R@1212321234345654545432,\  R@1212321234345454565432,\  R@1212321234343456545432,\  R@1212321234343454565432,\  R@1212321234565434345432,\  R@1212323212343434565432,\  R@1212323234343456565432,\  R@1212321234343456543432,\  R@1212323434345656543432,\  R@1212321234345656545432,\  R@1212321234565654345432,\  R@1212321234343456565432,\  R@121232123434345654345432,\  R@121232123434345656545432,\  R@121232123456565434345432,\  R@121232123434345656543432\}\,,

\noindent CP-odd set:~\{I@23434543,\  I@12323432,\  I@2323434543,\  I@2343434543,\  I@1232345432,\  I@1232323432,\  I@1234345432,\  I@1232343432,\  I@1212323432,\  I@121232123232,\  I@232343454543,\  I@232343234543,\  I@232343434543,\  I@232343234343,\  I@234343454343,\  I@123234545432,\  I@123232343232,\  I@123232345432,\  I@123434545432,\  I@123432345432,\  I@123234345432,\  I@123234543432,\  I@123232343432,\  I@123434345432,\  I@123432343432,\  I@121232345432,\  I@121232323432,\  I@123234565432,\  I@121234345432,\  I@121232343432,\  I@123434565432,\  I@121232123432,\  I@23234323454543,\  I@23234343454543,\  I@23234343234543,\  I@23234343454343,\  I@12323234543232,\  I@12323234545432,\  I@12343234545432,\  I@12343232345432,\  I@12323434545432,\  I@12323454543432,\  I@12323234343232,\  I@12323234345432,\  I@12323234543432,\  I@12343434545432,\  I@12343432345432,\  I@12343232343432,\  I@12323434345432,\  I@12323454343432,\  I@12343434543432,\  I@12123234545432,\  I@12123232343232,\  I@12123232345432,\  I@12323454565432,\  I@12323234565432,\  I@12123434545432,\  I@12123432345432,\  I@12123234345432,\  I@12123234543432,\  I@12123232343432,\  I@12343234565432,\  I@12323434565432,\  I@12323456543432,\  I@12123434345432,\  I@12123432343432,\  I@12343434565432,\  I@12123456545432,\  I@12123212345432,\  I@12123234565432,\  I@12123232123432,\  I@12123434565432,\  I@12123212343432,\  I@2323454323454543,\  I@2323434323454543,\  I@2323434345454343,\  I@1232323454543232,\  I@1234543234545432,\  I@1232345434545432,\  I@1232323434545432,\  I@1232323432345432,\  I@1232323454543432,\  I@1234343234545432,\  I@1234343232345432,\  I@1232343434545432,\  I@1232345454343432,\  I@1232323434345432,\  I@1234343454543432,\  I@1232343434543432,\  I@1212323234543232,\  I@1212323234545432,\  I@1232323456543232,\  I@1232323454565432,\  I@1212343234545432,\  I@1212343232345432,\  I@1212323434545432,\  I@1212323454543432,\  I@1212323234343232,\  I@1212323234345432,\  I@1212323234543432,\  I@1234565432345432,\  I@1232343454565432,\  I@1232345654345432,\  I@1232345434565432,\  I@1232323434565432,\  I@1232323456543432,\  I@1212343434545432,\  I@1212343432345432,\  I@1212343232343432,\  I@1212323434345432,\  I@1212323454343432,\  I@1234343234565432,\  I@1232343434565432,\  I@1232345654343432,\  I@1212343434543432,\  I@1234343456543432,\  I@1212321234545432,\  I@1212323456545432,\  I@1212323454565432,\  I@1212323212345432,\  I@1212323234565432,\  I@1212343456545432,\  I@1212343454565432,\  I@1212343212345432,\  I@1212343234565432,\  I@1212321234345432,\  I@1212323434565432,\  I@1212323456543432,\  I@1212323212343432,\  I@1212343434565432,\  I@1212343212343432,\  I@1212321234565432,\  I@1212323456565432,\  I@1212343456565432,\  I@123454323234545432,\  I@123232343234545432,\  I@123234543434545432,\  I@123232343434545432,\  I@123234343454543432,\  I@121232323454543232,\  I@123232345654543232,\  I@121234543234545432,\  I@121232345434545432,\  I@121232323434545432,\  I@121232323432345432,\  I@121232323454543432,\  I@123456543232345432,\  I@123232343454565432,\  I@123232343234565432,\  I@123232345434565432,\  I@121234343234545432,\  I@121234343232345432,\  I@121232343434545432,\  I@121232345454343432,\  I@121232323434345432,\  I@123234343454565432,\  I@123234565434345432,\  I@123234543434565432,\  I@123232343434565432,\  I@121234343454543432,\  I@121232343434543432,\  I@123234343456543432,\  I@121232321234545432,\  I@121232323456543232,\  I@121232323456545432,\  I@121232323454565432,\  I@121234321234545432,\  I@121234565432345432,\  I@121232123434545432,\  I@121232343456545432,\  I@121232343454565432,\  I@121232345654345432,\  I@121232345654543432,\  I@121232345434565432,\  I@121232321234345432,\  I@121232323434565432,\  I@121232323456543432,\  I@121234343456545432,\  I@121234343454565432,\  I@121234343212345432,\  I@121234343234565432,\  I@121232123434345432,\  I@121232343434565432,\  I@121232345654343432,\  I@121234343456543432,\  I@121232123456545432,\  I@121232321234565432,\  I@121232323456565432,\  I@121234321234565432,\  I@121234323456565432,\  I@121232123434565432,\  I@121232343456565432,\  I@121232345656543432,\  I@121234343456565432,\  I@121232123456565432,\  I@12123454323234545432,\  I@12123232343234545432,\  I@12323234323454565432,\  I@12123234543434545432,\  I@12123232343434545432,\  I@12123234343454543432,\  I@12123232345654543232,\  I@12123456543234545432,\  I@12123456543232345432,\  I@12123232123434545432,\  I@12123232343454565432,\  I@12123232343234565432,\  I@12123232345654345432,\  I@12123232345434565432,\  I@12123434321234545432,\  I@12123212343434545432,\  I@12123234343454565432,\  I@12123234565434345432,\  I@12123234565454343432,\  I@12123234543434565432,\  I@12123232343434565432,\  I@12123434345654345432,\  I@12123234343456543432,\  I@12123212345654545432,\  I@12123232345656543232,\  I@12123456565432345432,\  I@12123456543212345432,\  I@12123212343456545432,\  I@12123212343454565432,\  I@12123234565654345432,\  I@12123232123434565432,\  I@12123232343456565432,\  I@12123232345656543432,\  I@12123434321234565432,\  I@12123434323456565432,\  I@12123212343434565432,\  I@12123234343456565432,\  I@12123234565654343432,\  I@12123434345656543432,\  I@12123212345656545432,\  I@12123432123456565432,\  I@12123212343456565432,\  I@1212323234323454565432,\  I@1212345656543232345432,\  I@1212321234345654545432,\  I@1212321234345454565432,\  I@1212321234343456545432,\  I@1212321234343454565432,\  I@1212323456565434345432,\  I@1212323234343456565432,\  I@1212323434345656543432,\  I@1212345656543212345432,\  I@1212321234345656545432,\  I@1212343432123456565432,\  I@1212321234343456565432,\  I@121234565654321234565432\}\,.
}
\end{adjustwidth}
The \emph{generating} set is a union of CP-even set and CP-odd set. We have introduced the notation $\cI_i(\cJ_i$) in the main text to represent the $i$th invariant in the CP-even(CP-odd) set. The invariants in above lists can be easily converted to \texttt{Mathematica} expression with the \textbf{convert} function defined below

\begin{table}[H]
\setlength{\tabcolsep}{5pt}
\renewcommand{\arraystretch}{0.9}
\setlength{\extrarowheight}{2pt}
\centering
\begin{tabular}{m{4.8cm}m{4.94cm}m{4.8cm}}
\begin{tabular}{|c|c|c|c|c|}
\hline
\multirow{2}{*}{Ord.} & \multirow{2}{*}{Degree} & \multicolumn{2}{c|}{$n_g$} & \multirow{2}{*}{$n_s$} \\\cline{3-4}
    & & $n_e$ & $n_o$ & \\ \hline
    & $e^2$ & 1 & 0 & 0 \\
    2 & $m^2$ & 1 & 0 & 0 \\
    & $n^2$ & 1 & 0 & 0 \\ \hline
    & $e^4$ & 1 & 0 & 0 \\
    & $m^4$ & 1 & 0 & 0 \\
    4 & $n^4$ & 1 & 0 & 0 \\ 
    & $e^2n^2$ & 1 & 0 & 0 \\
    & $m^2n^2$ & 1 & 0 & 0 \\ \hline
    & $e^6$ & 1 & 0 & 0 \\
    & $m^6$ & 1 & 0 & 0 \\
    & $n^6$ & 1 & 0 & 0 \\
    \multirow{2}{*}{6} & $e^2n^4$ & 1 & 0 & 0 \\
    & $e^4n^2$ & 1 & 0 & 0 \\
    & $m^2n^4$ & 2 & 0 & 0 \\
    & $m^4n^2$ & 1 & 0 & 0 \\
    & $e^2m^2n^2$ & 1 & 0 & 0 \\ \hline
    & $e^4n^4$ & 1 & 0 & 0 \\
    & $m^2n^6$ & 1 & 0 & 0 \\
   \multirow{2}{*}{8} & $m^4n^4$ & 2 & 1 & 0 \\
    & $e^2m^2n^4$ & 2 & 1 & 0 \\
    & $e^2m^4n^2$ & 1 & 0 & 0 \\
    & $e^4m^2n^2$ & 1 & 0 & 0 \\[-0.05cm] \hline
    & $m^2n^8$ & 1 & 0 & 0 \\
    & $m^4n^6$ & 1 & 1 & 0 \\
    & $m^6n^4$ & 1 & 1 & 0 \\
    10 & $e^2m^2n^6$ & 3 & 2 & 0 \\
    & $e^2m^4n^4$ & 2 & 2 & 0 \\
    & $e^4m^2n^4$ & 3 & 1 & 0 \\
    & $e^4m^4n^2$ & 1 & 0 & 0 \\ \hline
    & $e^6n^6$ & 0 & 1 & 0 \\
    & $m^4n^8$ & 1 & 2 & 0 \\
    & $m^6n^6$ & 1 & 2 & 0 \\
    & $m^8n^4$ & 0 & 1 & 0 \\
    \multirow{2}{*}{12} & $e^2m^2n^8$ & 2 & 3 & 0 \\
    & $e^2m^4n^6$ & 4 & 5 & 0 \\
    & $e^2m^6n^4$ & 1 & 2 & 0 \\
    & $e^4m^2n^6$ & 4 & 3 & 0 \\
    & $e^4m^4n^4$ & 3 & 3 & 0 \\
    & $e^6m^2n^4$ & 1 & 1 & 0 \\[0.01cm] \hline
\end{tabular}
& 
\begin{tabular}{|c|c|c|c|c|}
\hline
\multirow{2}{*}{Ord.} & \multirow{2}{*}{Degree} & \multicolumn{2}{c|}{$n_g$} & \multirow{2}{*}{$n_s$} \\\cline{3-4}
    & & $n_e$ & $n_o$ & \\ \hline
    & $m^4n^{10}$ & 0 & 1 & 0 \\
    & $m^6n^8$ & 1 & 2 & 0 \\
    & $m^8n^6$ & 0 & 1 & 0 \\
    & $e^2m^2n^{10}$ & 1 & 2 & 0 \\
    & $e^2m^4n^8$ & 4 & 7 & 0 \\
    & $e^2m^6n^6$ & 3 & 5 & 0 \\
    14 & $e^2m^8n^4$ & 0 & 1 & 0 \\
    & $e^4m^2n^8$ & 4 & 5 & 0 \\
    & $e^4m^4n^6$ & 6 & 8 & 0 \\
    & $e^4m^6n^4$ & 2 & 3 & 0 \\
    & $e^6m^2n^6$ & 2 & 4 & 0 \\
    & $e^6m^4n^4$ & 1 & 2 & 0 \\
    & $e^8m^2n^4$ & 1 & 0 & 0 \\ \hline
    & $m^4n^{12}$ & 0 & 1 & 0 \\
    & $m^6n^{10}$ & 0 & 1 & 0 \\
    & $m^8n^8$ & 0 & 1 & 1 \\
    & $e^2m^2n^{12}$ & 0 & 1 & 0 \\
    & $e^2m^4n^{10}$ & 3 & 5 & 0 \\
    & $e^2m^6n^8$ & 3 & 5 & 0 \\
    & $e^2m^8n^6$ & 1 & 2 & 0 \\
    \multirow{2}{*}{16} & $e^4m^2n^{10}$ & 2 & 4 & 0 \\
    & $e^4m^4n^8$ & 8 & 13 & 1 \\
    & $e^4m^6n^6$ & 5 & 8 & 0 \\
    & $e^4m^8n^4$ & 0 & 2 & 0 \\
    & $e^6m^2n^8$ & 3 & 5 & 0 \\
    & $e^6m^4n^6$ & 5 & 8 & 0 \\
    & $e^6m^6n^4$ & 1 & 2 & 0 \\
    & $e^8m^2n^6$ & 1 & 2 & 0 \\
    & $e^8m^4n^4$ & 1 & 1 & 0 \\ \hline
    & $m^8n^{10}$ & 0 & 0 & 3 \\
    & $m^{10}n^8$ & 0 & 0 & 3 \\
    & $e^2m^4n^{12}$ & 1 & 2 & 1 \\
    & $e^2m^6n^{10}$ & 1 & 2 & 6 \\
    \multirow{2}{*}{18} & $e^2m^8n^8$ & 1 & 1 & 6 \\
    & $e^2m^{10}n^6$ & 0 & 0 & 1 \\
    & $e^4m^2n^{12}$ & 1 & 2 & 0 \\
    & $e^4m^4n^{10}$ & 6 & 9 & 4 \\
    & $e^4m^6n^8$ & 6 & 9 & 4 \\
    & $e^4m^8n^6$ & 2 & 3 & 0 \\ \hline
\end{tabular}
&
\begin{tabular}{|c|c|c|c|c|}
\hline
\multirow{2}{*}{Ord.} & \multirow{2}{*}{Degree} & \multicolumn{2}{c|}{$n_g$} & \multirow{2}{*}{$n_s$} \\\cline{3-4}
    & & $n_e$ & $n_o$ & \\ \hline
    & $e^6m^2n^{10}$ & 2 & 4 & 1 \\
    & $e^6m^4n^8$ & 9 & 11 & 2 \\
    & $e^6m^6n^6$ & 6 & 7 & 0 \\
    \multirow{2}{*}{18} & $e^6m^8n^4$ & 0 & 1 & 0 \\
    & $e^8m^2n^8$ & 2 & 3 & 1 \\
    & $e^8m^4n^6$ & 3 & 5 & 0 \\
    & $e^8m^6n^4$ & 1 & 1 & 0 \\
    & $e^{10}m^2n^6$ & 0 & 1 & 0 \\ \hline
    & $m^6n^{14}$ & 0 & 0 & 1 \\
    & $m^8n^{12}$ & 0 & 0 & 9 \\
    & $m^{10}n^{10}$ & 0 & 0 & 8 \\
    & $m^{12}n^8$ & 0 & 0 & 4 \\
    & $e^2m^4n^{14}$ & 0 & 0 & 2 \\
    & $e^2m^6n^{12}$ & 0 & 0 & 21 \\
    & $e^2m^8n^{10}$ & 0 & 0 & 32 \\
    & $e^2m^{10}n^8$ & 0 & 0 & 14 \\
    & $e^2m^{12}n^6$ & 0 & 0 & 1 \\
    & $e^4m^4n^{12}$ & 2 & 3 & 14 \\
    & $e^4m^6n^{10}$ & 2 & 2 & 32 \\
    20 & $e^4m^8n^8$ & 2 & 1 & 21 \\
    & $e^4m^{10}n^6$ & 0 & 0 & 2 \\
    & $e^6m^2n^{12}$ & 1 & 1 & 2 \\
    & $e^6m^4n^{10}$ & 6 & 7 & 14 \\
    & $e^6m^6n^8$ & 7 & 7 & 12 \\
    & $e^6m^8n^6$ & 3 & 2 & 1 \\
    & $e^8m^2n^{10}$ & 1 & 2 & 3 \\
    & $e^8m^4n^8$ & 6 & 8 & 5 \\
    & $e^8m^6n^6$ & 3 & 5 & 1 \\
    & $e^8m^8n^4$ & 0 & 1 & 0 \\
    & $e^{10}m^2n^8$ & 0 & 1 & 1 \\
    & $e^{10}m^4n^6$ & 1 & 2 & 0 \\ \hline
    & $m^6n^{16}$ & 0 & 0 & 1 \\
    & $m^8n^{14}$ & 0 & 0 & 10 \\
    & $m^{10}n^{12}$ & 0 & 0 & 18 \\
    \multirow{2}{*}{22} & $m^{12}n^{10}$ & 0 & 0 & 9 \\
    & $m^{14}n^8$ & 0 & 0 & 2 \\
    & $e^2m^4n^{16}$ & 0 & 0 & 2 \\
    & $e^2m^6n^{14}$ & 0 & 0 & 34 \\
    & $e^2m^8n^{12}$ & 0 & 0 & 76 \\\hline
\end{tabular}
\end{tabular}
\caption{Number of generating invariants $n_g$, divided into the number of CP-even and CP-odd generating invariants ($n_e$ and $n_o$ respectively), as well as the number of syzygies $n_s$ at each degree and total order. Note that the difference $n_g-n_s$ can be mapped to the coefficient of the multi-graded PL at each degree up to order 26.}
\label{tab:graded_count_1}
\end{table}

\begin{table}[H]
\setlength{\tabcolsep}{5pt}
\setlength{\extrarowheight}{2pt}
\centering
\begin{tabular}{m{5.3cm}m{5.3cm}}
\begin{tabular}{|c|c|c|c|c|}
\hline
\multirow{2}{*}{Ord.} & \multirow{2}{*}{Degree} & \multicolumn{2}{c|}{$n_g$} & \multirow{2}{*}{$n_s$} \\\cline{3-4}
    & & $n_e$ & $n_o$ & \\ \hline
    & $e^2m^{10}n^{10}$ & 0 & 0 & 55 \\
    & $e^2m^{12}n^8$ & 0 & 0 & 12 \\
    & $e^4m^6n^{12}$ & 0 & 0 & 103 \\ 
    & $e^4m^4n^{14}$ & 0 & 0 & 27 \\
    & $e^4m^8n^{10}$ & 0 & 0 & 109 \\
    & $e^4m^{10}n^8$ & 0 & 0 & 39 \\
    & $e^4m^{12}n^6$ & 0 & 0 & 2 \\
    & $e^6m^4n^{12}$ & 1 & 1 & 47 \\
    & $e^6m^2n^{14}$ & 0 & 0 & 2 \\
    22& $e^6m^6n^{10}$ & 2 & 0 & 85 \\
    & $e^6m^8n^8$ & 2 & 0 & 37 \\
    & $e^6m^{10}n^6$ & 0 & 0 & 3 \\
    & $e^8m^2n^{12}$ & 0 & 0 & 6 \\
    & $e^8m^4n^{10}$ & 2 & 3 & 26 \\
    & $e^8m^6n^8$ & 5 & 4 & 18 \\
    & $e^8m^8n^6$ & 2 & 1 & 0 \\
    & $e^{10}m^2n^{10}$ & 0 & 0 & 4 \\
    & $e^{10}m^4n^8$ & 2 & 2 & 5 \\
    & $e^{10}m^6n^6$ & 1 & 2 & 1 \\ \hline
    & $e^{12}n^{12}$ & 0 & 0 & 1 \\
    & $m^8n^{16}$ & 0 & 0 & 11 \\
    & $m^{10}n^{14}$ & 0 & 0 & 20 \\
    \multirow{2}{*}{24} & $m^{12}n^{12}$ & 0 & 0 & 21 \\
    & $m^{14}n^{10}$ & 0 & 0 & 5 \\
    & $m^{16}n^8$ & 0 & 0 & 1 \\
    & $e^2m^4n^{18}$ & 0 & 0 & 1 \\
    & $e^2m^6n^{16}$ & 0 & 0 & 33 \\ \hline
\end{tabular}
&
\raisebox{0.3cm}{
\begin{tabular}{|c|c|c|c|c|}
\hline
\multirow{2}{*}{Ord.} & \multirow{2}{*}{Degree} & \multicolumn{2}{c|}{$n_g$} & \multirow{2}{*}{$n_s$} \\\cline{3-4}
    & & $n_e$ & $n_o$ & \\ \hline
    & $e^2m^8n^{14}$ & 0 & 0 & 110 \\
    & $e^2m^{10}n^{12}$ & 0 & 0 & 116 \\
    & $e^2m^{12}n^{10}$ & 0 & 0 & 48 \\
    & $e^2m^{14}n^8$ & 0 & 0 & 5 \\
    & $e^4m^{4}n^{16}$ & 0 & 0 & 31 \\
    & $e^4m^{6}n^{14}$ & 0 & 0 & 174 \\
    & $e^4m^{8}n^{12}$ & 0 & 0 & 284 \\
    & $e^4m^{10}n^{10}$ & 0 & 0 & 162 \\
    & $e^4m^{12}n^8$ & 0 & 0 & 33 \\
    & $e^6m^2n^{16}$ & 0 & 0 & 1 \\
    & $e^6m^4n^{14}$ & 0 & 0 & 87 \\
    & $e^6m^6n^{12}$ & 0 & 0 & 261 \\
    \multirow{2}{*}{24} & $e^6m^8n^{10}$ & 0 & 0 & 226 \\
    & $e^6m^{10}n^8$ & 0 & 0 & 59 \\
    & $e^6m^{12}n^6$ & 0 & 0 & 2 \\
    & $e^8m^2n^{14}$ & 0 & 0 & 8 \\
    & $e^8m^4n^{12}$ & 0 & 0 & 94 \\
    & $e^8m^6n^{10}$ & 0 & 0 & 134 \\
    & $e^8m^{10}n^6$ & 0 & 0 & 2 \\
    & $e^8m^8n^8$ & 1 & 0 & 49 \\
    & $e^{10}m^6n^8$ & 2 & 0 & 17 \\
    & $e^{10}m^8n^6$ & 1 & 0 & 0 \\
    & $e^{10}m^2n^{12}$ & 0 & 0 & 10 \\
    & $e^{10}m^4n^{10}$ & 0 & 0 & 32 \\
    & $e^{12}m^2n^{10}$ & 0 & 0 & 3 \\
    & $e^{12}m^4n^8$ & 0 & 1 & 3 \\ \hline
\end{tabular}
}
\end{tabular}
\caption{Tab.~\ref{tab:graded_count_1} continued. We only show the results up to order 24. There is no generating invariant at higher orders, only syzygies can be found. At order 26, the numbers of syzygies at each degree can be read off from the multi-graded PL shown in Eq.~\eqref{eq:gradedPL}. At higher orders, the PL loses its ability to explain the correct number of generating invariants and the number of syzygies. Please refer to the text around Eq.~\eqref{eq:PL_order28} for further details.}
\label{tab:graded_count_2}
\end{table}
\newpage

\begin{snugshade}
\begin{verbatim}
convert[walk_]:=Block[{map,head,num,c,t,ct},
  map={12->ct@Ye,21->Ye,23->Yn,32->ct@Yn,34->c@Mn,43->Mn,54->c@Yn,45->t@Yn,
  65->t@Ye,56->c@Ye}/.{c->Conjugate,t->Transpose,ct->ConjugateTranspose};
  head=Switch[Head@walk,R,Re@Tr@#&,I,Im@Tr@#&,_,Tr];
  num=IntegerDigits@Cases[{walk},_Integer,2][[1]];
  head[Dot@@FromDigits/@Partition[num,2,1,1]/.map]];
\end{verbatim}
\end{snugshade}
\noindent For example, evaluating \textbf{convert}[12] in \texttt{Mathematica} will generate the expression of a CP-even flavor invariant $\cI_1=\Tr(\neat{Y_{e}^{\dagger}Y_{e}^{}})=\Tr(X_e)$, and evaluating \textbf{convert}[I@23434543] will generate the expression of a CP-odd flavor invariant $\cJ_{1}=\Im\Tr(\neat{Y_{N}^{} M_{N}^{*}M_{N}^{} M_{N}^{*} Y_{N}^{T} Y_{N}^{*} M_{N}^{} Y_{N}^{\dagger}})$. 

We have summarized the number of generating invariants, divided into the number of CP-even and CP-odd generating invariants, as well as the number of syzygies at each degree and total order in Tabs.~\ref{tab:graded_count_1} and \,\ref{tab:graded_count_2}. Although the numbers in these tables at every degree are fixed, there are still degrees where ambiguity arises in determining the form of the invariants. This ambiguity occurs when our algorithm identifies multiple invariants appearing linearly within a single explicit relation. Such explicit relations can simply arise from the Cayley--Hamilton theorem as shown in Eqs.~\eqref{eq:id_abc} and \eqref{eq:a2bac}. We also discuss the ambiguity in App.~\ref{App:AlgGenSet} around Eq.~\eqref{eq:relation}. Let us now present a specific example. At degree $e^4n^4$, according to the summary table, there is only one generating invariant. However, there exists another invariant $\cI = \Tr(\neat{Y_{e}^{}Y_{e}^{\dagger} Y_{N}^{}Y_{N}^{\dagger} Y_{e}^{}Y_{e}^{\dagger} Y_{N}^{}Y_{N}^{\dagger}})$, which is not included in the \emph{generating} set. We can find the following explicit relation
\begin{equation}
2\cI+4 \cI_{18}- \cI_1^2\cI_3^2+\cI_1^2\cI_6 +4  \cI_1 \cI_3 \cI_7-4 \cI_1 \cI_{12}-2 \cI_7^2+\cI_3^2 \cI_4-\cI_4 \cI_6-4 \cI_3 \cI_{13}=0\,.
\end{equation}
In this relation, two invariants appearing linearly, namely $\cI$ and $\cI_{18}$, are observed. Both of these invariants are of degree $e^4n^4$. The ambiguity arises because it is possible to select either of them as the generating invariant, rendering the other one redundant.

We find that our \emph{generating} set contains some subsets that can be used as \emph{generating} set of other theories. For example, for the theory with $n_N=n_f=2$, the \emph{generating} set can be formed with the invariants
\begin{equation}
\label{eq:gen_nf2}
\text{Gen.}(n_N=n_f=2):~\{\cI_1, \cI_2, \cI_3, \cI_4, \cI_5, \cI_6, \cI_7, \cI_8, \cI_{14}, \cI_{17}, \cI_{22}, \cI_{35}, \cJ_1, \cJ_2, \cJ_5, \cJ_7, \cJ_{28}, \cJ_{31}\}\,,
\end{equation}
which is also shown in Ref.~\cite{Jenkins:2009dy} with a different convention. For the theory with $n_N=2, n_f=3$, the \emph{generating} set is given by
\begin{equation} \label{eq:GenSetnN2}
\begin{split}
\text{Gen.}&(n_N=2, n_f=3):~\{\cI_1, \cI_2, \cI_3, \cI_4, \cI_5, \cI_6, \cI_7, \cI_8, \cI_9, \cI_{12}, \cI_{13}, \cI_{14}, \cI_{17}, \cI_{18}, \cI_{22}, \cI_{25},\\
&\cI_{34}, \cI_{35}, \cI_{54}, \cI_{79}, \cJ_1, \cJ_2, \cJ_5, \cJ_7, \cJ_9, \cJ_{10}, \cJ_{26}, \cJ_{28}, \cJ_{29}, \cJ_{31}, \cJ_{32}, \cJ_{68}, \cJ_{70}, \cJ_{72}, \\
&\cJ_{132}, \cJ_{133}, \cJ_{134}, \cJ_{195}\}\,.
\end{split}
\end{equation}
The above \emph{generating} set has already been shown in Ref.~\cite{Yu:2021cco}, but the commutation notation is equivalently represented by taking imaginary part in our notation. 

We can also easily identify that the primary invariants shown in Eq.~\eqref{eq:PrimSetnuSM} correspond to the invariants
\begin{equation}
\text{Primary set}:~\{\cI_1, \cI_2, \cI_3, \cI_5, \cI_6, \cI_7, \cI_8, \cI_9, \cI_{12}, \cI_{13}, \cI_{15}, \cI_{23}, \cI_{25}, \cI_{34}, \cI_{35}, \cI_{47}, \cI_{50}, \cI_{54}, \cI_{65}, \cI_{79}, \cI_{91}\}\,.
\end{equation}
In the Dirac limit, the \emph{generating} set will be reduced to have only 11 invariants, which are given by
\begin{equation}
\text{Gen.(Dirac limit)}:~\{\cI_1, \cI_3, \cI_4, \cI_6, \cI_7, \cI_9, \cI_{11}, \cI_{12}, \cI_{13}, \cI_{18}, \cJ_{10}\}\,.
\end{equation}
There is a one-to-one correspondence between these invariants and the invariants in the quark sector as shown in Eq.~\eqref{eq:quark_inv}. For completeness, we also show the walk notations of generating flavor invariants in the quark sector, they are given by
\begin{equation}
\begin{split}
\text{Gen.(Quark sector):~}\{&78, 89, 7878, 8989, 7898, 787878, 898989, 789898, \\
&787898, 78789898, \text{I}@787898789898\}\,.
\end{split}
\end{equation}
These walks are based on the graph shown in left bottom panel of Fig.~\ref{fig:invariantGraph}.

\section{The Hironaka decomposition}\label{app:Hironaka}
As defined in Sec.~\ref{sec:PlethysticProgram}, the \emph{generating} set $\{\cI_1,...,\cI_m\}$ allows us to write any invariant $\cI'$ of the ring as a polynomial of the generating invariants:
\begin{equation}
    \cI^{\prime} = P\(\cI_1,...,\cI_m\) \, .
\end{equation}The Hironaka decomposition refines this last equation. This decomposition comes from the Cohen-Macaulay property~\cite{Derksen:2015Inv, Sturmfels:2008Inv, procesi2007lie} which only holds for reductive groups. In this cases it is possible to construct two finite sets of invariants: the set of primary invariants $\{\theta_1,\dots,\theta_k\}$ which is algebraically independent, and the set of secondary invariants $\{\eta_1,\dots,\eta_r\}$, such that any invariant $\cI'$ can be written as
\begin{equation} \label{eq:Hironaka}
    \cI'=\sum_{i=1}^{r}\,\eta_i\, P_i(\theta_1,\dots,\theta_k)\, ,
\end{equation}where $P_i(\theta_1,\dots,\theta_k)$ is a polynomial in the primary invariants. This decomposition is related to the HS in the following way. If $z_j$ is the degree of $\theta_j$, and $s_i$ the degree of $\eta_i$, then
\begin{equation} \label{eq:HironakaHS}
    \cH(q) = \frac{\cN(q)}{\cD(q)} =\dfrac{\sum\limits_{i=1}^{r}\,q^{s_i}}{\prod\limits_{j=1}^{k}\,(1-q^{z_j})}\, .
\end{equation}
Note that, secondary invariants can be a product of several generating invariants. Furthermore, for a given group the Hironaka decomposition is not unique and the degrees of the primary and secondary invariants can be different.

%%%%%%%%%%%%%%%%%%%%%%%%%%%%%%%%%%%%%%%%%%%%%%%%%%%%%%%%%%%%%%%%%%%%
\section{Hilbert's Nullstellensatz}\label{App:theorem}

Hilbert’s Nullstellensatz~\cite{eisenbud2013commutative,atiyah2018introduction,Cox:2015ode}, a fundamental result in algebraic geometry, establishes a profound connection between polynomial equations and the geometry of algebraic varieties. The traditional formulation of Hilbert's Nullstellensatz often involves a polynomial ring and its associated ideals. Consider the polynomial ring $R=k[x_1, x_2, \ldots, x_n]$ in $n$ variables over the field $k$ (a mathematical structure that generalizes the concept of numbers). This ring consists of polynomials in the variables $x_1, x_2, \ldots, x_n$ with coefficients in $k$. We will now introduce the fundamental mathematical concepts required for presenting Hilbert's Nullstellensatz.

\begin{itemize}
\item Ideal

An ideal $I$ in the polynomial ring $R$ is a subset of polynomials, which satisfies:
\begin{equation}
    \begin{aligned}
        & \text{(1)} && 0\in I. \\
        & \text{(2)} && \text{If $f,g\in I$, then $f+g\in I$.} \\
        & \text{(3)} && \text{If $f\in I$ and $g\in R$, then $fg\in I$.}
    \end{aligned}
\end{equation}

\item Variety

  Given an ideal $I$, the variety $V(I)$ is the set of common zeros of all polynomials in $I$. Formally, a point $(a_1, a_2, \ldots, a_n)$ lies in the variety $V(I)$ if and only if every polynomial in $I$ evaluates to zero at that point:
\begin{equation}
V(I) = \{(a_1, a_2, \ldots, a_n) \mid f(a_1, a_2, \ldots, a_n) = 0 \text{ for all } f \in I\}\,.
\end{equation}

\item Radical of an Ideal

The radical of an ideal $I$, denoted by $\sqrt{I}$, is the set of all polynomials $g$ such that some power of $g$ belongs to $I$. Mathematically, $\sqrt{I}$ is defined as:
\begin{equation}
  \sqrt{I} = \{ g \mid g^k \in I \text{ for some } k \geq 1 \}\,.
\end{equation}
\end{itemize}

Hilbert's Nullstellensatz asserts that for any algebraically closed field $k$, there is a bijective correspondence between the points of a variety $V(I)$ and the radical ideals $\sqrt{I}$ defining that variety. Formally, this correspondence is expressed as:
\begin{equation}
  \label{eq:nullstellensatz}
  \text{Ideal}(V(I)) = \sqrt{I}\,,
\end{equation}
where $\text{Ideal}(V(I))$ denotes the ideal of polynomials vanishing on the variety $V(I)$.

In a more polynomial-centric language, Hilbert's Nullstellensatz can also be formulated differently. If a polynomial $p$ vanishes on the variety $V(I)$, it belongs to $\text{Ideal}(V(I))$, and, by Hilbert's Nullstellensatz as shown in Eq.~\eqref{eq:nullstellensatz}, it also belongs to $\sqrt{I}$. According to the definition of $\sqrt{I}$, there exists $k \geq 1$ such that $p^k\in I$, thus can be expressed as:
\begin{equation}
p^k = f_1p_1 + f_2p_2 + \dots + f_mp_m\,,
\end{equation}
where $f_i \in R$ and $p_i$ are the defining polynomials of $I$. This equation essentially states that if $p$ is vanishing under the common zeros of the defining polynomials of the ideal $I$, then some $k$-th power of the polynomial $p$ can be expressed as a combination of these defining polynomials.

%%%%%%%%%%%%%%%%%%%%%%%%%%%%%%%%%%%%%%%%%%%%%%%%%%%%%%%%%%%%%%%%%%%%
\section{CPC conditions for \texorpdfstring{$n_N=n_f=2$}{nN=nf=2}}\label{App:CPC2Gen}

\subsection{Minimal CPC set for \texorpdfstring{$n_N=n_f=2$}{nN=nf=2}}
The simplified model with two generations of fermions serves as a good example for the algebraic studies. In Ref.~\cite{Jenkins:2009dy}, the authors find the following 6 CP-odd invariants in the \emph{generating} set
\begin{equation}
\label{eq:cpodd2gen}
    \begin{split}
        J_1 & = \Im\text{Tr}\left(M_N Y_N^{\dagger } Y_N Y_N^{\dagger } Y_e Y_e^{\dagger } Y_N M_N^*\right) \sim \cJ_2\\
        J_2 & = \Im\text{Tr}\left(M_N^* M_N Y_N^{\dagger } Y_N M_N^* Y_N^T Y_N^* M_N\right) \sim \cJ_1\\
        J_3 & = \Im\text{Tr}\left(M_N^* M_N Y_N^{\dagger } Y_e Y_e^{\dagger } Y_N M_N^* Y_N^T Y_N^* M_N\right) \sim \cJ_7\\
        J_4 & = \Im\text{Tr}\left(M_N Y_N^{\dagger } Y_N Y_N^{\dagger } Y_e Y_e^{\dagger } Y_N M_N^* Y_N^T Y_N^*\right) \sim \cJ_5\\
        J_5 & = \Im\text{Tr}\left(M_N M_N^* M_N Y_N^{\dagger } Y_e Y_e^{\dagger } Y_N M_N^* Y_N^T Y_e^* Y_e^T Y_N^*\right) \sim \cJ_{31}\\
        J_6 & = \Im\text{Tr}\left(M_N Y_N^{\dagger } Y_N Y_N^{\dagger } Y_e Y_e^{\dagger } Y_N M_N^* Y_N^T Y_e^* Y_e^T Y_N^*\right) \sim \cJ_{28}\\
    \end{split}
\end{equation}
which we have translated to our notation. Although the \emph{generating} set is small, it is still difficult to find the common zeros of the polynomials based on usual methods. However, the invariants can also be considered as ideals in the polynomial ring of the theory. In this section, we will analyze these ideals with the software package \texttt{Macaulay2}~\cite{M2} based on the parameterization in Eq.~\eqref{eq:parameterization_2gen_new}. To simplify the notation, we take $r_{11}=a,r_{21}=c,c_{12}=b+p\,i$ and $c_{22}=d+q\,i$. Therefore, the polynomial ring is defined as $R:=\mathbb{Q}[y_e,y_\mu,m_1,m_2,a,b,c,d,p,q]$ and all CPV effects can be characterized by the ideal $I$ defined by the six CP-odd invariants, i.e., $I\equiv\langle J_1, \dots, J_6\rangle$. The vanishing set denoted by $V(I)$ captures all the CPC conditions. The problem of finding common zeros is equivalent to finding the irreducible components of the ideal.

According to Hilbert's Nullstellensatz, the ideal of all polynomials that vanish on the common zero set $V(I)$ is the radical of the ideal $\sqrt{I}$, which can be calculated by the \textbf{radical} function in \texttt{Macaulay2}. The CPC conditions are captured by the minimal primes of the radical
\begin{align*}
&\{\langle q,p\rangle,\ \langle q,a\rangle,\ \langle p,c\rangle,\ \langle c,a\rangle,\ \langle c,b\rangle,\ \langle d,a\rangle,\ \langle d,b\rangle,\ \langle m_1,a\rangle,\ \langle m_1,c\rangle,\ \langle m_2,a\rangle,\ \langle m_2,c\rangle,\ \\
&\langle m_2,m_1\rangle,\ \langle m_1,y_e-y_{\mu}\rangle,\ \langle m_2,y_e-y_{\mu}\rangle,\ \langle m_1,d\,p-b\,q\rangle,\ \langle m_2,d\,p-b\,q\rangle,\ \numberthis \label{eq:CPCconditions2gen}\\
&\langle m_1-m_2,y_e-y_{\mu}\rangle,\ \langle y_e-y_{\mu},a\,b+c\,d\rangle,\ \langle y_e-y_{\mu},a\,p+c\,q\rangle,\ \\
&\langle m_1-m_2,a\,b\,c^2-a^2\,c\,d+b^2\,c\,d-a\,b\,d^2+c\,d\,p^2-a\,b\,q^2\rangle\}\,.
\end{align*}
These solutions have been simplified with the physical spectrum. For instance, all masses are taken to be non-negative. The CPC conditions can be obtained by setting the generators of the ideals to 0. For instance, the first ideal in the above set indicates that there is one condition $p=q=0$ that can lead to CPC, which is just the trivial solution of vanishing phases. In the algebraic geometry picture, all these conditions are fundamental objects, and they correspond to points, lines, surfaces, etc. In addition, each of these conditions has a connection to the special spectrum and enlarged symmetries of the theory. 

As we explore the CPC conditions using the \texttt{Macaulay2} package, we also find that without introducing $J_3$, we can still derive the CPC conditions in Eq.~\eqref{eq:CPCconditions2gen}. This suggests that $J_3$ must be redundant when determining the CPC conditions. This is indeed observed by the syzygy approach based on Hilbert's Nullstellensatz. The redundant $J_3$ in our notation is $\cJ_7$ (refer to Eq.~\eqref{eq:cpodd2gen} for the mapping). If we use the \emph{generating} set identified in our paper, as shown in Eq.~\eqref{eq:gen_nf2}, we can find the following syzygy:
\begin{equation}
2 \cJ_7^2=(\cI_2^2-\cI_5) \cJ_2^2 +2 \cJ_1 \cJ_{31}\,.
\end{equation}
Therefore, $\cJ_7$ vanishes given that $\cJ_1=\cJ_2=0$. We also attempted to find similar syzygies for other CP-odd invariants at their square order in the \emph{generating} set. However, no other syzygy could be found, and we also observed that at higher orders, the syzygies are not easy to solve. Thus, we can conclude that the minimal CPC set is given by $\{J_1,J_2,J_4,J_5,J_6\}$ up to square order based on Hilbert's Nullstellensatz.

By analyzing the polynomial ring, it is also possible to generate the conditions leading to a special spectrum, which are not necessarily the CPC conditions. For instance, the conditions leading to unphysical phase of $p$ or $q$ can be obtained by the elimination of variables, and the relevant function in \texttt{Macaulay2} is called \textbf{eliminate}. By eliminating the CP-odd variable $p$ or $q$, one can find the following conditions
\begin{equation}
\begin{split}
&\text{unphysical $p$:~}\{\langle q\rangle,\ \langle c\rangle,\ \langle m_1\rangle,\ \langle m_2\rangle,\ \langle m_1-m_2\rangle,\ \langle y_e-y_\mu\rangle,\ \langle d,b\rangle,\ \langle d,a\rangle\}\,,\\
&\text{unphysical $q$:~}\{\langle p\rangle,\ \langle a\rangle,\ \langle m_1\rangle,\ \langle m_2\rangle,\ \langle m_1-m_2\rangle,\ \langle y_e-y_\mu\rangle,\ \langle d,b\rangle,\ \langle c,b\rangle\}\,,
\end{split}
\end{equation}
where the unphysical conditions such as $\langle m_1+m_2\rangle$ is removed. The above conditions can also be calculated with the more physical parameterization in Eq.~\eqref{eq:parameterization_2gen}. They are given as follows
\begin{equation}
\begin{split}
\text{unphysical $\phi$:~}&\{\langle \sin\varphi\rangle,\ \langle \sin\varphi-1\rangle,\ \langle \sin\varphi+1\rangle,\ \langle \sin\alpha\rangle,\ \langle \sin\alpha-1\rangle,\ \langle \sin\alpha+1\rangle,\ \\
&\langle m_1\rangle,\ \langle m_2\rangle,\ \langle m_1-m_2\rangle,\ \langle y_1-y_2\rangle\}\,,\\
\text{unphysical $\varphi$:~}&\{\langle\sin\phi\rangle,\ \langle\sin\alpha\rangle,\ \langle\sin\alpha-1\rangle,\ \langle\sin\alpha+1\rangle,\ \langle\sin\theta\rangle,\ \langle\sin\theta-1\rangle,\ \\
&\langle\sin\theta+1\rangle,\ \langle y_e-y_\mu\rangle,\ \langle m_1-m_2\rangle,\ \langle y_1\rangle,\ \langle y_2\rangle,\ \langle y_1-y_2\rangle\}\,.
\end{split}
\end{equation}
By exploring these special conditions and their combinations, one can obtain all of the special spectra with enlarged symmetries that can be used to remove phases in the theory.

\subsection{Pseudo-real couplings}

There are some highly non-trivial conditions in the solution list. For instance, the last one shows that the mass degeneracy of $m_1=m_2$ and a vanishing of a specific combination of the matrix elements of $Y_N$ can lead to CPC.
One example that solves these conditions is
\begin{equation}
\label{eq:special_param}
m_1=m_2=1,\  y_e=1,\  y_\mu =2,\  a=6,\  b=2,\  c=3,\  d=4,\  p=8,\  q=5,\  
\end{equation}
which corresponds to the following flavor matrices:
\begin{equation}
\label{eq:special_mat}
Y_e=\begin{pmatrix}
    1 & 0\\
    0 & 2
\end{pmatrix},\quad
Y_N=\begin{pmatrix}
    6 & 2+8i\\
    3 & 4+5i
\end{pmatrix},\quad
M_N=\begin{pmatrix}
    1 & 0\\
    0 & 1
\end{pmatrix}.
\end{equation}
With this setup, we can check that all CP-odd invariants are vanishing, and there is no CP violation. However there are two complex numbers in $Y_N$, that cannot be made real by field redefinitions. Such a scenario of CP conservation in the presence of irremovable complex parameters was previously noted in Ref.~\cite{Ivanov:2015mwl,Trautner:2016ezn,Bonnefoy:2021tbt} as pseudo-real couplings in the context of Three Higgs Doublet Models and toy models with complicated discrete symmetries. This can be understood as follows. Since $M_N$ has degenerate eigenvalues, there is an $O(2)$ freedom for the field redefinition of $N$, while, because $Y_e$ has non-degenerate eigenvalues, there is only a rephasing freedom for the field $L$. By applying these field redefinitions, we find that the phases in $Y_N$ can not be removed. However, we can find the following field redefinition that can map $Y_N$ to $Y_N^*$\footnote{For the choice of parameters shown in Eq.~\eqref{eq:special_param}, we find $\theta = 2\arctan(3),\, \alpha = \pi+\arctan(4/3),\, \beta = 3\pi/2$.}
\begin{equation} \label{eq:gCP}
\begin{pmatrix}
e^{i\alpha} & 0\\
0 & e^{i\beta}
\end{pmatrix}
\begin{pmatrix}
a & b+p i\\
c & d+q i
\end{pmatrix}
\begin{pmatrix}
\cos\theta & \sin\theta\\
\sin\theta & -\cos\theta
\end{pmatrix}=
\begin{pmatrix}
a & b-p i\\
c & d-q i
\end{pmatrix}.
\end{equation}
Note that $M_N$ and $Y_e$ are real and diagonal in this basis, thus they are also mapped to their complex conjugate under this field redefinition. This indicates that the Lagrangian is symmetric under CP transformations up to a field redefinition, which is referred to as generalized CP symmetry~\cite{Grimus:2003yn,Feruglio:2012cw}. Hence, even though there are irremovable phases present in the theory, CP is still conserved for this set of parameters. These special CPC conditions are correctly captured by the CP-odd flavor invariants, as they are all vanishing in these cases. In other words, all CPC conditions, no matter how special they may appear, can be obtained by setting CP-odd generating invariants to zero.

Note, that for a specific value of the rotation angle that solves Eq.~\eqref{eq:gCP} for the explicit example in Eq.~\eqref{eq:special_param}, a discrete symmetry is defined that leaves the Lagrangian invariant. Furthermore, the generalized CP transformation that imposes Eq.~\eqref{eq:gCP}, is of order 2, since both the rephasing matrix and the rotation matrix fulfill $A A^* = \mathbb{1}$, indicating a trivial flavor symmetry. There are additional cases in Eq.~\eqref{eq:CPCconditions2gen} that result in pseudo-real couplings. For example, one can verify that the ideal $\langle y_e-y_\mu, a\,p+c\,q\rangle$ corresponds to another discrete symmetry that leads to pseudo-real couplings. In general, if the condition involves the phases non-trivially, meaning the phases can be set to certain constrained non-zero values, the pseudo-real couplings must arise in such scenarios.

%%%%%%%%%%%%%%%%%%%%%%%%%%%%%%%%%%%%%%%%%%%%%%%%%%%%%%%%%%%%%%%%%%%%
\bibliographystyle{apsrev4-1_title}
\bibliography{biblio.bib}

\end{document}